\documentclass[pra, twocolumn, superscriptaddress, aps]{revtex4-1}
\usepackage{amsmath,amsfonts, amssymb, amsthm, dsfont}
\usepackage{yfonts}
\usepackage{bm}
\usepackage{mathrsfs}
\usepackage{graphicx}
\usepackage{subfigure}
\usepackage{verbatim}
\usepackage{pst-all}
\usepackage{leftidx}

\usepackage{tikz}
	\usetikzlibrary{calc}
\usetikzlibrary{arrows,shapes, positioning, matrix}
\usetikzlibrary{decorations.markings}
\newcommand*{\halfway}{0.5*\pgfdecoratedpathlength+.5*4pt}\tikzstyle arrowstyle=[scale=1]
\tikzstyle arrowstyle=[scale=1]
\tikzstyle directed=[postaction={decorate,decoration={markings,
    mark=at position .15 with {\arrow[arrowstyle]{stealth}}}}]
\tikzstyle string=[thick,postaction={decorate},decoration={markings,
    mark=at position \halfway with {\arrow[arrowstyle]{stealth}}}]
\tikzstyle string2=[thick,postaction={decorate},decoration={markings,
    mark=at position .5 with {\arrow[arrowstyle]{stealth}}}]
\tikzstyle dual_string=[dashed,postaction={decorate,decoration={markings,
    mark=at position 1 with {\arrow[arrowstyle]{stealth}}}}]
\tikzstyle trivial=[dashed, thick]

\tikzstyle dw=[thick,postaction={decorate,decoration={markings,
    mark=at position 1 with {\arrow[arrowstyle]{stealth}}}}]
\tikzstyle group=[]

\newcommand{\di}{\mathrm{d}}

\renewcommand{\vec}[1]{{\mathbf #1}}
\newcommand{\ket}[1]{|#1\rangle}
\newcommand{\bra}[1]{\langle#1|}

\renewcommand{\vr}{{\vec{r}}}

\renewcommand{\ol}[1]{\overline{#1}}
\newcommand{\comments}[1]{}
\newcommand{\Ref}[1]{Ref.~\onlinecite{#1}}
\newcommand{\Refs}[1]{Refs.~\onlinecite{#1}}

\newcommand{\mb}[1]{\mathbf{#1}}

\renewcommand{\cal}[1]{\mathcal{#1}}
\newcommand{\coho}[1]{\textswab{#1}}
\newcommand{\cohosub}[1]{\scalebox{0.7}{\textswab{#1}}}

\def\doubleunderline#1{\underline{\underline{#1}}}

\newcommand{\ag}[2]{#1_\mb{#2}}

\makeatletter
\newcommand{\vast}{\bBigg@{3}}
\newcommand{\Vast}{\bBigg@{4}}
\makeatother

\renewcommand{\t}[1]{\tilde{#1}}

\newcommand{\eq}[1]{(\ref{#1})}
\newcommand{\eqn}[1]{eqn.~(\ref{#1})}

\renewcommand{\>}{\rangle}

\newcommand{\al}{\alpha}
\newcommand{\bt}{\beta}
\newcommand{\del}{\delta}

\newcommand{\ga}{\gamma}

\newcommand{\cF}{ {\cal F} }

\newcommand{\cO}{ {\cal O} }

\newcommand{\cY}{ {\cal Y} }

\newcommand{\bpm}{\begin{pmatrix}}
\newcommand{\epm}{\end{pmatrix}}
\newcommand{\bmm}{\begin{matrix}}
\newcommand{\emm}{\end{matrix}}

\newcommand{\tikzpicmidline}[1]{\begin{tikzpicture}[baseline={($ (current bounding box) - (0,0pt) $)}]#1\end{tikzpicture}}

\makeatletter
\def\l@subsubsection#1#2{}
\makeatother

\begin{document}

\psset{arrowsize=2.5pt 1.2, arrowinset=0.3}
\newpsobject{psstring}{psline}{linearc=2pt}
\newpsobject{psid}{psline}{linestyle=dotted,dotsep=2pt}
\newpsobject{pspsi}{psline}{doublecolor=lightgray, linecolor=blue, doubleline=true, linewidth=0.8pt}
\newpsobject{pssigma}{psline}{linewidth=1.5pt}

\title{Exactly Solvable Models for Symmetry-Enriched Topological Phases}
\author{Meng Cheng}
\affiliation{Department of Physics, Yale University, New Haven, CT
06511-8499, USA}
\affiliation{Station Q, Microsoft Research, Santa Barbara, California 93106-6105, USA}
\author{Zheng-Cheng Gu}
\affiliation{Department of Physics, The Chinese University of Hong Kong, Shatin, New Territories, Hong Kong}
\affiliation{Perimeter Institute for Theoretical Physics, Waterloo, ON N2L 2Y5, Canada}
\author{Shenghan Jiang}
\affiliation{Department of Physics, Boston College, Chestnut Hill, MA 02467}
\author{Yang Qi}
\thanks{Current address: Department of physics, Massachusetts Institute of Technology, Cambridge, MA 02139, USA}
\affiliation{Perimeter Institute for Theoretical Physics, Waterloo, ON N2L 2Y5, Canada}
\affiliation{Institute for Advanced Study, Tsinghua University, Beijing 100084, China}

\date{\today}

\begin{abstract}

	We construct fixed-point wave functions and exactly solvable commuting-projector Hamiltonians for a large class of bosonic symmetry-enriched topological (SET) phases, based on the concept of equivalent classes of symmetric local unitary transformations. We argue that for onsite unitary symmetries, our construction realizes all SETs free of anomaly, as long as the underlying topological order itself can be realized with a commuting-projector Hamiltonian. We further extend the construction to anti-unitary symmetries (e.g. time-reversal symmetry), mirror-reflection symmetries, and to anomalous SETs on the surface of three-dimensional symmetry-protected topological phases. Mathematically, our construction naturally leads to a generalization of group extensions of unitary fusion categories to anti-unitary symmetries.
\end{abstract}

\maketitle

\section{Introduction}

Interplay between global symmetry and topological order has been an exciting research direction in recent years. It is by now well appreciated that symmetries play very important roles in our understanding of gapped phases of quantum many-body systems, even in the absence of spontaneous symmetry breaking. The classification of gapped quantum systems often becomes much richer in the presence of symmetries. For instance, an otherwise trivial phase (i.e., adiabatically connected to an atomic product state) can split into distinct gapped phases when symmetries are taken into account, called symmetry-protected topological (SPT) phases. Eminent examples of SPT phases include time-reversal-invariant topological insulators and superconductors in both two- and three-dimensional free fermion systems~\cite{Bernevig2006, Bernevig_PRL2006, Kane2005a, Kane2005b, Fu_PRL07, Moore_PRB07, Roy_PRB2009, kitaev2009, schnyder2008}, whose theoretical predictions and experimental discoveries have generated intense research interest in the past decade. Very recently, it has been realized that SPT phases also exist in interacting bosonic systems~\cite{XieScience2012}, e.g., the Haldane phase in spin chains~\cite{Gu2009, Haldane1,Haldane2}.

On the other hand, if a two-dimensional (2D) gapped phase exhibits an intrinsic topological order, characterized by quasiparticle excitations with fractional braiding and exchange statistics, symmetry can act in a nontrivial way on the quasiparticle excitations, leading to the notion of symmetry-enriched topological (SET) phases. Specifically, quasiparticle excitations can carry fractionalized quantum numbers under the global symmetry, a phenomenon known as symmetry fractionalization. For example, quasiholes in fractional quantum Hall (FQH) states have fractional electric charges~\cite{Laughlin83}. Such fractionalization has long been regarded as a signature of the underlying topological order. Another well-studied topologically ordered phase of matter, gapped quantum spin liquids (QSL) in frustrated magnets~\cite{BalentsSLReview, SavaryBalentsReview}, also exhibits symmetry fractionalization~\cite{wen2002psg, essin2013, LuBFU}. In fact, a defining feature of QSLs is the existence of a spin-$1/2$ spinon excitation~\cite{KivelsonPRB1987}, which transforms projectively under the $\mathrm{SO}(3)$ spin rotation symmetry and oftentimes under space-time symmetries as well~\cite{wen2002psg}.

Aside from fractionalizations, symmetries can also transform one type of quasiparticles into another. It was recently realized that extrinsic defects of such symmetries can harbor exotic zero modes, giving rise to topologically protected degeneracies and non-Abelian braiding transformations. By now many examples of non-Abelian defects in Abelian parent states have been found, including ``genons'' in bilayer quantum Hall systems~\cite{barkeshli2012a, barkeshli2013genon, teo2013b}, parafermion zero modes in FQH/superconductor heterostructures~\cite{clarke2013, lindner2012, cheng2012} and lattice dislocations or disinclinations in certain exactly solvable lattice models~\cite{bombin2010, you2012, you2013, teo2013}. The non-Abelian defects can potentially be exploited in topological quantum information processing to enhance the computational power~\cite{nayak2008}.

A further motivation for the study of SETs comes from a remarkable connection to three-dimensional (3D) SPT phases~\cite{VishwanathPRX2013}: when the 3D phase has a boundary, the nontrivial bulk SPT order manifests as anomalous symmetry transformations on the boundary degrees of freedom. As a result, a symmetry-preserving gapped boundary must exhibit topological order, and the symmetry has to be implemented in a way that can not be consistently realized in truly 2D systems, i.e., the SET is said to be anomalous. Due to the bulk-boundary correspondence, the study of anomalous surface topological order has become an essential tool in classifying and characterizing 3D SPT phases~\cite{VishwanathPRX2013, wang2013, ChenASF2014, cho2014, kapustin2014, bonderson2013, wang2013b, metlitski2013, chen2014b, Fidkowski13, WangPRB2014, metlitski2014, WangPRX2016, Seiberg_unpub, Witten_unpub}. Identifying anomalous SETs also has important implications for the classification of SETs in two dimensions~\cite{wang2013, Hermele_unpub, Qi_unpub1, Qi_unpub2}.

Theoretically, a number of different approaches have been developed to understand and classify SPT and SET phases~\cite{wen2002psg, Wang2007, LevinPRB2012, essin2013, lu2013, mesaros2013, HungPRB2013a, SET2, HermelePRB2014, ChenASF2014,  Tarantino_arxiv, ZhenPRB2015, KapustinSPT1, freed2016}. We will closely follow the classification scheme developed in \Refs{SET1, ChenASF2014, Tarantino_arxiv}, based on the mathematical framework of tensor category theory.

In this paper, we construct exactly solvable lattice models for bosonic SET phases, based on the concept of equivalent class of symmetric local unitary transformations~\cite{ChenPRB2010, GuPRB2015}. The motivation for the work is three-fold: first, exactly solvable models (with commuting-projector Hamiltonians) provide valuable insights into the general structure of the ground-state wave functions, since they represent the fixed-point state of the quantum phase under wave-function renormalization, and may shed light on the search for microscopic realizations of such phases. One can also study excitations in the model and understand the symmetry actions concretely.  Second, constructions of fixed-point wave functions for quantum phases imply that these states have exact tensor-network representations. Therefore, these states can in principle be efficiently targeted in numerical algorithms based on tensor-network states~\cite{SJiangTPS2015X}. Lastly, investigation of the fixed-point wave functions constructed from equivalent class of symmetric local unitary transformations provides an independent derivation of the classification of SET phases.

In particular, we believe our construction provides a possible framework to classify non-chiral bosonic SETs with space-time symmetries, which are relevant to most material realizations of such phases including TIs and QSLs. Unlike the case of onsite unitary symmetries~\cite{SET1, Tarantino_arxiv}, there still lacks a systematic framework for the classification of SET phases protected by space-time symmetries, which is needed to study possible symmetry fractionalization patterns in 2D QSLs. So far, progress has been made with the help of other onsite unitary symmetries, especially the spin-rotational symmetry~\cite{Hermele_unpub,Qi_unpub1, Qi_unpub2}. However, these methods do not apply to systems without any onsite unitary symmetries, e.g., materials with strong spin-orbit couplings~\cite{LiYbMgGaO2015}.

More specifically, we focus on symmetry-enriched phases in a large class of 2D topological phases, known broadly as the quantum doubles. The defining feature of a quantum-double phase is that there exists a representative ``fixed-point'' (i.e., zero correlation length) wave function with a commuting-projector parent Hamiltonian. Well-known examples of quantum-double models include discrete gauge theories~\cite{Kitaev97,YTHuTQD, BuerschaperAP2014}, string-net models~\cite{Levin05a, LinPRB2014}, and doubled Chern-Simons theory~\cite{Freedman04a}.
Physically, it is known that all topologically ordered states with gappable boundaries belong to this class~\cite{kitaev2012}, which certainly implies vanishing of chiral central charges, but in fact stronger than just that. Due to the fixed-point nature, these states admit natural tensor-network representations with relatively small bond dimensions~\cite{GuPRB2009, BuerschaperPRB2009, QDPEPS}. In the following, we will loosely refer to topological phases which can arise from quantum doubles as being ``non-chiral.''

For an onsite unitary symmetry group, we show that all (non-anomalous and non-chiral) SETs, at least within the classification scheme introduced in \Ref{SET1}, can be realized in our construction. In fact, our construction in this case can be understood naturally as ``ungauging'' the Levin-Wen model for the gauged SET state: because the symmetry group is onsite and unitary, one can always gauge the symmetry for the SET state (i.e., by coupling to lattice gauge fields). If the topological order of the SET state is a quantum double, one can show that the gauged model remains so. Starting from the string-net construction of the gauged model, one can apply a duality transformation~\cite{levin2012} which then ``ungauges'' the symmetry to get the SET state. Since every non-anomalous SET state is ``gaugable'', such a procedure can always be carried through to produce a string-net construction of the SET state.

For anti-unitary and mirror symmetries, we conjecture that our construction is also general enough to represent all the non-anomalous {and non-chiral} SETs.  We further extend these ideas to construct fixed-point wave functions for anomalous surface topological orders of 3D SPT phases described by group cohomology models~\cite{VishwanathPRX2013, wang2013, ChenASF2014, cho2014, kapustin2014}. The general mathematical structure underlying our construction is a generalization of group extensions of unitary fusion categories~\cite{ENO2009, Turaev}. More precisely, the self-consistent conditions derived from equivalent class of symmetric local unitary transformations are weaker than the group extensions of unitary fusion categories, which allows us to consider much more general types of symmetry actions, including anti-unitary symmetry, reflection symmetry, and anomalous symmetry. Essentially, our fixed-point-wave-function constructions produce Hamiltonian-type~\cite{DanFreed2014} topological phases which do not necessarily admit topologically invariant actions in arbitrary space-time manifold (known as Lagrangian type).

Once the general formalism is laid out, we present an extensive list of examples, namely, symmetry-enriched Abelian gauge theories, in Sec. \ref{sec:segt}. In particular, we construct all non-anomalous SETs in this family where symmetries do not permute quasiparticles. Using our construction, we also derive a sufficient and necessary condition for a pattern of symmetry fractionalization to be non-anomalous. A similar obstruction-vanishing condition was obtained for unitary onsite symmetries in Refs.~\cite{ChenASF2014,SET1}. The novelty of our approach is the applicability to anti-unitary and spatial symmetries, although the computation is only explicitly carried out for Abelian gauge theories so far.

We then analyze the example of $\mathbb{Z}_2$ toric code SET with unitary/anti-unitary $\mathbb{Z}_2$ symmetry in Sec. \ref{sec:z2tc-unitary}, showing explicitly the symmetry actions on quasiparticles. Aside from symmetry fractionalization, the $\mathbb{Z}_2$ symmetry can also permute the $e$ and $m$ particles in the toric code, known as an electro-magnetic duality (EMD) symmetry. To the best of our knowledge, our model provides the first onsite realization of the EMD symmetry in the $\mathbb{Z}_2$ toric code with commuting-projector Hamiltonians~\footnote{We note that solvable models where such EMD symmetry is realized as an onsite symmetry were constructed in \Ref{SET1} and \Ref{Fidkowski13}, by gauging the fermion parity symmetry in topological superconductors. However they are not commuting-projector models. In addition, the well-known plaquette model introduced in Ref. [\onlinecite{WenPRL2003}] realizes $e\leftrightarrow m$ exchange by the lattice translation symmetry. }.

\section{Fixed-Point Wave Functions}
\label{sec:fpwf}

In the following we outline the construction of fixed-point wave functions for SET phases, inspired by the string-net construction as well as group cohomology models of SPT phases.

\subsection{String-net states and local relations}
\begin{figure}
  \centering
  \includegraphics{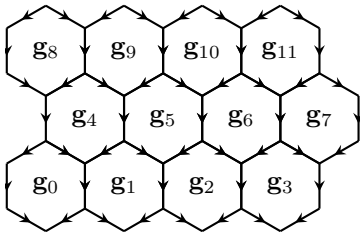}
  \caption{\label{fig:graph}A trivalent-graph lattice. The arrows on the links show the branching structure. Each plaquette is decorated by a group element $\mb g_i\in G$.}
\end{figure}

The fixed-point wave functions can be defined on any trivalent graph, as shown in Fig.~\ref{fig:graph}. We shall assume that the edges of the graph are directed, and the directions are such that the arrows are two-in-one-out or one-in-two-out on each vertex%
. This is called a branching structure. As we will discuss later in this section, the wave function does not depend on the particular choice of the branching structure, although the branching structure is explicitly used in the construction.

Let $G$ be a finite global symmetry group. We will consider both onsite symmetries (unitary and anti-unitary) and mirror symmetries, so $G$ is equipped with two $\mathbb{Z}_2$ gradings: $p:G\rightarrow \{0,1\}$ , where $p(\mb{g})=1$ means $\mb{g}$ is an orientation-reversing operation (i.e., mirror reflection), and  $q: G\rightarrow\{0,1\}$ where $q(\mb{g})=1$ means $\mb{g}$ corresponds to an anti-unitary operation.

Let us now specify the Hilbert space of the model. Each edge of the graph is associated with a $n$-dimensional Hilbert space, and an orthonormal basis is denoted by $\ket{a}$, where $a$ is drawn from a label set of order $n$. They can be thought as different types of strings that occupy the edges, with a unique ``vacuum'' label $0$ (sometimes denoted by $I$) corresponding to no string. We will denote the label set by $\mathcal{C}_{G}$, for reasons that will become clear shortly.

Three strings meet at a vertex. Whether three string types $a$, $b$ and $c$ are allowed to meet or not is determined by the fusion rule $N^{ab}_c$, which is a non-negative integer. If $N^{ab}_c>0$, $a,b$ and $c$ can meet at a vertex:
\begin{equation}
	\begin{tikzpicture}[baseline={($ (current bounding box) - (0,0pt) $)}]
		\draw (0, .5) node [above] {$a$}--(.5, 0) [string];
		\draw (1, .5) node [above] {$b$}--(.5, 0) [string];
		\draw (.5, 0)--(.5, -.5) node [below] {$c$} [string];
	\end{tikzpicture}
	\label{}
\end{equation}
When $N^{ab}_c>1$, one has to include additional local degrees of freedom at each vertex. We will assume $N^{ab}_c$ only takes values in $\{0,1\}$ to simplify the discussions.
In addition, each label $a$ has a unique ``dual'' label $\bar{a}$ such that $N^{a\bar{a}}_0=1$. We associate to each label $a$ a positive number $d_a$, called the quantum dimension of $a$, which satisfy $d_a d_b=\sum_c N^{ab}_c d_c$.

To account for the symmetry, we add a spin degree of freedom in the center of each plaquette, whose basis $\ket{\mb{g}}$ are labeled by the element $\mb{g}$ of the symmetry group $G$. For each edge we can then associate a group element $\bar{\mb g}_0\mb{g}_1$ (here $\bar{\mb g}$ denotes the inverse of group element $\mb g$):
\begin{equation}
	\begin{pspicture}[shift=-0.5](0,-0.6)(1,0.6)
            \scriptsize
			\rput(0.25, 0.0){$\mb{g}_0$}
			\rput(0.8, 0.0){$\mb{g}_1$}
            \psline(0.5,-0.5)(0.5,0.5)
			\rput(0.9,-0.4){${\bar{\mb{g}}_0\mb{g}_1}$}
            \psline{<-}(0.5, -0.1)(0.5,0.0)
			\psline[linestyle=dashed, dash=3pt 2pt]{->}(0.1, 0.2)(0.9, 0.2)
        \end{pspicture}
	\label{eq:dw}
\end{equation}
We say that the there is a $\bar{\mb{g}}_0\mb{g}_1$ domain wall on the edge. In the symmetry-enriched wave function, domain walls decorated with different group elements have different sets of labels. We require that the set of labels have a $G$-graded structure, in the following sense: labels are be organized into $|G|$ different sectors $\cal{C}_\mb{g}$, where each sector $\mathcal C_\mb{g}$ contains labels allowed on a $\mb g$ domain wall. Following the notations in Ref.~[\onlinecite{SET1}], we denote labels in $\mathcal C_\mb{g}$ by $a_\mb{g}$. Furthermore, the fusion rules must respect the $G$-grading: $N_{a_\mb{g} b_\mb{h}}^{c_\mb{k}}=\delta_{\mb{k}, \mb{gh}}N_{a_\mb{g}, b_\mb{h}}^{c_\mb{gh}}$, so one has $\mathcal{C}_\mb{g}\times \mathcal{C}_\mb{h}\in \mathcal{C}_\mb{gh}$. In particular, the $\mathcal{C}_\mb{1}$ sector is closed under fusion. A useful fact that follows from the $G$-graded fusion rules is the total quantum dimensions of each sector must be equal: $\mathcal{D}_\mb{g}^2=\sum_{a_\mb{g}\in \cal{C}_\mb{g}}d_{a_\mb{g}}^2=\mathcal{D}_\mb{1}^2$~\cite{SET1}. We define the total quantum dimension $\mathcal{D}^2=\sum_{a_\mb{g}\in \cal{C}_G}d_{a_\mb{g}}^2=|G|\mathcal{D}_\mb{1}^2$.


The ground-state wave function is a superposition of string-net states (i.e., string states on the lattice that satisfy the branching rules). A defining feature of the string-net wave function is that the amplitudes for different string-net states satisfy a set of local relations:
\begin{enumerate}
  \item The wave function is invariant under local deformation of strings,
  \begin{equation}
    \label{eq:locrel1}
    \Psi\left(\begin{tikzpicture}[baseline={($ (current bounding box) - (0,0pt) $)}]
      \draw (0, .8) node [above] {\small $a_{\mb g}$}--(0, -.8) [string];
      \draw (0, .4) -- (1, .4) -- (1, -.4)--(0, -.4) [trivial];
      \node [group] at (0, 0) [left] {\small $\mb g_0$};
      \node [group] at (.5, 0) {\small $\mb g_0\mb g$};
    \end{tikzpicture}\right)=
    \Psi\left(\begin{tikzpicture}[baseline={($ (current bounding box) - (0,0pt) $)}]
      \draw (0, .8) node [above] {$a_{\mb g}$}--(0, .4)
      --(1, .4)--(1, -.4)--(0, -.4)--(0, -.8) [string];
      \draw (0, .4) -- (0, -.4) [trivial];
      \node [group] at (0, 0) [left] {$\mb g_0$};
    \end{tikzpicture}\right).
  \end{equation}
  Here, the graph in the parentheses represents a local patch of the string state. We notice that the group element in the right plaquette, $\mb g_0\mb g$, is determined from the group element in the left plaquette and the grading $\mb{g}$ on the domain wall $a_{\mb g}$. Therefore, without causing ambiguity, the label of the group element in the right plaquette can be omitted. We will follow this convention in the rest of the paper.
  \item The wave function is invariant up to a normalization factor, under the creation/annihilation of bubbles:
  \begin{equation}
    \label{eq:locrel2}
	\begin{split}
    \Psi\left(\begin{tikzpicture}[baseline={($ (current bounding box) - (0,0pt) $)}]
      \draw (0, .6)--(0, .3) [string]
	  node [midway, right] {$a_\mb{hk}$};
      \draw (0, .3) ..controls(-.3, .2)..(-.3, .1)
      --(-.3, -.1) node [midway,left] {$\ag bh$}
      ..controls(-.3, -.2).. (0, -.3) [string];
      \draw (0, .3) ..controls(.3, .2)..(.3, .1)
      --(.3, -.1) node [midway,right] {$\ag ck$}
      ..controls(.3, -.2).. (0, -.3) [string];
      \draw (0, -.3)--(0, -.6) [string]
	  node [midway, right] {$a_{\mb{hk}}^\prime$};
      \node [group] at (-.3, .2) [above left] {$\mb g_0$};
    \end{tikzpicture}\right)
    &=\delta_{aa^\prime}\sqrt{\frac{d_{b_{\mb h}}d_{c_{\mb k}}}{d_{a_{\mb hk}}}}
    \Psi\left(\begin{tikzpicture}[baseline={($ (current bounding box) - (0,0pt) $)}]
    \draw (0, .6)--(0, .3) [string]
	node [midway, right] {$a_\mb{hk}$};
    \draw (0, .3) ..controls(-.3, .2)..(-.3, .1)
    --(-.3, -.1)
    ..controls(-.3, -.2).. (0, -.3) [string];
    \draw (0, .3) ..controls(.3, .2)..(.3, .1)
    --(.3, -.1)
    ..controls(.3, -.2).. (0, -.3) [trivial];
    \draw (0, -.3)--(0, -.6) [string];
    \node [group] at (-.3, .2) [above left] {$\mb g_0$};
    \end{tikzpicture}\right)\\
    &=\delta_{aa^\prime}\sqrt{\frac{d_{b_{\mb h}}d_{c_{\mb k}}}{d_{a_{\mb hk}}}}
    \Psi\left(\begin{tikzpicture}[baseline={($ (current bounding box) - (0,0pt) $)}]
    \draw (0, .6)--(0, .3) [string]
	node [midway, right] {$a_\mb{hk}$};
    \draw (0, .3) ..controls(-.3, .2)..(-.3, .1)
    --(-.3, -.1)
    ..controls(-.3, -.2).. (0, -.3) [trivial];
    \draw (0, .3) ..controls(.3, .2)..(.3, .1)
    --(.3, -.1)
    ..controls(.3, -.2).. (0, -.3) [string];
    \draw (0, -.3)--(0, -.6) [string];
    \node [group] at (-.3, .2) [above left] {$\mb g_0$};
    \end{tikzpicture}\right).
\end{split}
  \end{equation}

  In Eqs.~\eqref{eq:locrel1} and \eqref{eq:locrel2}, the dashed lines denote strings carrying the vacuum label ``0''. Hence, these two moves alter labels on the edges in a way that appear to change the shape of the strings, if edges carrying label ``0'' are treated as vacuum. However, the underlying lattice, and the degrees of freedom on it, are left unchanged. This is in contrast to the generalized symmetric local unitary transformations we introduce in Appendix~\ref{fixwv}, which truly change the underlying lattice, and consequently the number of degrees of freedom. These two types of moves differ by a normalization factor, if a local bubble is added or removed, as explained in Appendix~\ref{fixwv}.

  \item The wave function transforms as the following under the so-called $F$ moves:
	\begin{equation}
    \label{eq:fmove}
    \Psi\left(\begin{tikzpicture}[baseline={($ (current bounding box) - (0,0pt) $)}, scale=0.8]
		\draw (0, 0) node [above] {$\ag ag$} --(.3, -.5) [string];
		\draw (.6, 0) node [above] {$\ag bh$} --(.3, -.5) [string];
		\draw (.3, -.5)--(.6, -1) [string]
		      node [midway, left] {$\ag e{gh}$};
		\draw (1.2, 0) node [above] {$\ag ck$} --(.6, -1) [string];
		\draw (.6, -1)--(.9, -1.5) [string] node [below] {$\ag d{ghk}$};
		\node [group] at (0, -1.3) {$\mb g_0$};
		\end{tikzpicture}\right)
		=\sum_{f_\mb{hk}\in \mathcal{C}_\mb{hk}} {}^{\mb g_0}
    [\mathcal F^{\ag ag\ag bh\ag ck}_{\ag d{ghk}}]_{\ag e{gh}\ag f{hk}}
    \Psi\left(\begin{tikzpicture}[baseline={($ (current bounding box) - (0,0pt) $)}, scale=0.8]
		\draw (0, 0) node [above] {$\ag ag$} --(.6, -1) [string];
		\draw (.6, 0) node [above] {$\ag bh$} --(.9, -.5) [string];
		\draw (.9, -.5)--(.6, -1) [string]
		      node [midway, right] {$\ag f{hk}$};
		\draw (1.2, 0) node [above] {$\ag ck$} --(.9, -.5) [string];
		\draw (.6, -1)--(.9, -1.5) [string] node [below] {$\ag d{ghk}$};
		\node [group] at (0, -1.3) {$\mb g_0$};
		\end{tikzpicture}\right).
  \end{equation}
 Here, $\cal F$ is a generalized $F$ symbol, in which $\mb g_0$ labels the group element in the left-most plaquette.
 As before, the elements in other plaquettes can be determined from $\mb g_0$ and the gradings on the edges. The gradings of edges labeled by $\ag d{ghk}$, $\ag e{gh}$, and $\ag f{hk}$ can be determined from those of $\ag ag$, $\ag bh$, and $\ag ck$ from fusion, hence, we can omit the grading labels of these sectors and use a simplified notation  ${}^{\mb g_0}[\cal F^{\ag ag\ag bh\ag ck}_{d}]_{ef}$ for the generalized $F$ symbol. We will often view ${}^{\mb g_0}[\cal F^{\ag ag\ag bh\ag ck}_{d}]$ as a matrix, with $e,f$ being the two indices.
\end{enumerate}

The local moves are defined in terms of a set of data $d_a, {}^{\mb g_0}[\cal F^{\ag ag\ag bh\ag ck}_{d}]_{ef}$. They need to satisfy several consistency conditions, which will be given in the following.

First of all, to preserve the norm of the wave function, we require that the $F$ moves are unitary:
  \begin{equation}
  \Big({}^{\mb{g}_0}[\mathcal{F}^{\ag ag\ag bh\ag ck}_{\ag d{ghk}}]\Big)^{-1} = \Big({}^{\mb{g}_0}[\mathcal{F}^{\ag ag\ag bh\ag ck}_{\ag d{ghk}}]\Big)^{\dagger}.
	  \label{eqn:F-unitarity1}
  \end{equation}
  Obviously, the following associativity relation of fusion rules needs to be satisfied:
\begin{align}
\label{NNNN}
\sum_{e_{\mb g \mb h}} N^{a_{\mb g}b_{\mb h}}_{e_{\mb g\mb h}} N^{e_{\mb g \mb h}c_{\mb k}}_{d_{\mb g \mb h \mb k}}=\sum_{f_{\mb h \mb k}} N^{b_{\mb h }c_{\mb k}}_{f_{\mb h \mb k}}N^{a_{\mb g}f_{\mb h \mb k}}_{d_{\mb g \mb h \mb k}}.
\end{align}

Another unitarity condition comes from considering a local move similar (but inequivalent) to the $F$ move with one of the lines bent down. Leaving the details to Appendix~\ref{fixwv}, the condition reads
\begin{equation}
	\sum_{f_{\mb{ghk}}} d_{f_{\mb{ghk}}} {}^{\mb{g}_0}[\cal F^{c_\mb{g}e'_{\mb{h}}b_\mb{k}}_f]_{ad} {}^{\mb{g}_0}[\cal F^{c_\mb{g}e_{\mb{h}}b_\mb{k}}_f]^*_{ad}=\frac{d_{a_\mb{gh}} d_{d_{\mb{hk}}} }{d_{e_\mb{h}}}\delta_{e_\mb{h}e'_\mb{h}}.
	\label{eqn:F-unitarity2}
\end{equation}

Self-consistency of local moves requires that any two sequences of moves starting from and ending with the same string-net states must yield the same total amplitudes. This can be achieved by imposing the so-called pentagon equations on the $\mathcal F$ symbols:
\begin{multline}
  \label{eq:pentagon4}
	{}^{\mb g_0}[\cal F^{\ag f{gh}\ag ck\ag dl}_e]_{mq}
	{}^{\mb g_0}[\cal F^{\ag ag\ag bh\ag q{kl}}_e]_{fp}=\\
	\sum_{n_{\mb{hk}}\in \cal{C}_\mb{hk}}{}^{\mb g_0}[\cal F^{\ag ag\ag bh\ag ck}_m]_{fn}
	{}^{\mb g_0}[\cal F^{\ag ag\ag n{hk}\ag dl}_e]_{mp}
	{}^{\mb g_0\mb g}[\cal F^{\ag bh\ag ck\ag dl}_p]_{nq}
\end{multline}

Another slightly more technical condition is that the diagrammatic rules we have defined for string-net states should be isotopy-invariant, i.e., one has the ability to introduce and remove bends in lines. One can show that this leads to the condition
\begin{equation}
	d_{a_\mb{g}}=\big|{}^{\mb{g}_0}[\cal F^{a_\mb{g}\ol{a_\mb{g}}a_\mb{g}}_{a_\mb{g}}]_{0,0}\big|^{-1}.
	\label{eqn:isotopy-invariance}
\end{equation}
The isotopy invariance is completely analogous to the usual diagrammatic calculus of fusion categories, and we refer the readers to \Refs{Kitaev06a, Bonderson07b} for more details.

The structure of fixed-point wave function defined in this section is modeled on the well-known string-net construction of quantum doubles of {unitary fusion categories(UFC)~\cite{Levin05a}. In particular, notice that the sector $\mathcal{C}_\mb{1}$ is closed under fusion, so if we restrict all group elements to be $\mb{1}$, the consistency conditions Eqs. \eqref{eqn:F-unitarity1}, \eqref{eqn:F-unitarity2}, \eqref{eq:pentagon4} and \eqref{eqn:isotopy-invariance} define $\mathcal{C}_\mb{1}$ as a UFC~\cite{Kitaev06a}. We should note however that the equivalence classes of wave functions under local unitary transformations produce a weaker set of axioms than those of UFCs, which in a sense can be thought as a ``Hamiltonian-type'' UFC~\cite{ChenPRB2010, GuPRB2015}. This strongly suggests that the topological order of the system is identical to the quantum double of $\cal{C}_\mb{1}$ if we ignore the symmetry. In the rest of the paper, we will use $\mathcal{Z}(\cal{C})$ to denote the topological order realized by the quantum double of a UFC $\cal{C}$.

Although our construction explicitly uses a branching structure on the trivalent graph, the wave function obtained in such a construction is actually independent of the choice of the branching structure. On one hand, a branching structure can be induced from an ordering of the vertices of the graph, by assigning the orientation of each edge according to the ordering. For the usual quantum double (i.e., $G$ is trivial), if the category $\cal C$ satisfies the so-called sphericity condition, it has been shown that the wave function is invariant under the reordering of vertices~\cite{BW}, on the same trivalent graph. We believe that a similar conclusion holds for the present construction as long as $\cal C_G$ is spherical in a suitable sense. On the other hand, as explained in Appendix~\ref{fixwv}, the wave function is invariant under the generalized symmetric local unitary (gSLU) transformations, which can add or remove vertices on the graph. Thus, using the gSLU transformations, one can change the branching structure by first removing the vertices, and then adding them back, with a different branching structure. Such processes of removing and adding vertices can be used to relate any two branching structures~\cite{branching}.

We notice that the fixed-point wave function constructed using the generalized $F$ symbols can be viewed as the string-net construction of a unitary multifusion category~\cite{multifusion}. Instead of the $G$-graded structure, one can also view the labels on the edges as having a double-graded structure $a_{\mb g,\mb h}$, where $\mb g$ and $\mb h$ are the group elements on the two sides of the domain wall, respectively. Then the labels form a multifusion category, and the generalized $F$ symbols are the $F$ symbols of the multifusion category, satisfying the pentagon equation in Eq.~\eqref{eq:pentagon4}.

The $F$ symbols have gauge redundancies. Physically, we can consider the following local unitary transformation on the state:
\begin{equation}
	\Psi\left(\begin{tikzpicture}[baseline={($ (current bounding box) - (0,0pt) $)}]
		\draw (0, .5) node [above] {$\ag ag$}--(.5, 0) [string];
		\draw (1, .5) node [above] {$\ag bh$}--(.5, 0) [string];
		\draw (.5, 0)--(.5, -.5) node [below] {$\ag c{gh}$} [string];
		\node [group] at (-.2, -.2) {$\mb g_0$};
	\end{tikzpicture}
	\right)\rightarrow
	{}^{\mb g_0}[v^{\ag ag\ag bh}_c]
	\Psi\left(\begin{tikzpicture}[baseline={($ (current bounding box) - (0,0pt) $)}]
		\draw (0, .5) node [above] {$\ag ag$}--(.5, 0) [string];
		\draw (1, .5) node [above] {$\ag bh$}--(.5, 0) [string];
		\draw (.5, 0)--(.5, -.5) node [below] {$\ag c{gh}$} [string];
		\node [group] at (-.2, -.2) {$\mb g_0$};
	\end{tikzpicture}\right).
	\label{eqn:gauge-F}
\end{equation}
${}^{\mb g_0}v^{\ag ag\ag bh}_c$ are $\mathrm{U}(1)$ phase factors. Again, here in ${}^{\mb g_0}v^{\ag ag\ag bh}_c$ we omit the $G$ grading of $\ag c{gh}$, since it can be inferred from the gradings of $\ag ag$ and $\ag bh$. In order for Eq. \eqref{eqn:gauge-F} to be a symmetric local unitary transformation, ${}^{\mb{g}_0}[v^{\ag ag\ag bh}_c]$ also needs to satisfy a symmetry condition, which we will postpone to Sec. \ref{sec:sym-act-F}.

As a result, $\cal{F}$ symbols which are related through the following gauge transformations should yield the same SET phases:
\begin{equation}
	{}^{\mathbf{g}_0}[\cal{F}^{a_\mb{g}b_\mb{h}c_\mb{k}}_{d}]_{ef}\rightarrow \frac{{}^{\mb{g}_0\mb{g}}[{v}^{b_\mb{h}c_\mb{k}}_f]\, {}^{\mb{g}_0}[v^{a_\mb{g}f_\mb{hk}}_d]}{{}^{\mb{g}_0}[v^{a_\mb{g}b_\mb{h}}_e]\, {}^{\mb{g}_0}[v^{e_\mb{gh}c_\mb{k}}_d]}
{}^{\mathbf{g}_0}[\cal{F}^{a_\mb{g}b_\mb{h}c_\mb{k}}_{d}]_{ef}.
	\label{eq:v-equiv}
\end{equation}

To summarize, we have defined a fixed-point ground-state wave function for a SET phase using local moves. This construction generalizes the usual string-net wave functions in two aspects: first, the fusion rules now have a $G$-graded structure; second, the F symbols depend explicitly on the group element in the left-most domain. One should also notice that so far the symmetry has not entered the discussion. In fact, the rules we have defined so far are not enough to uniquely determine the wave function on the plane or on a sphere (i.e there is unstable $|G|$-fold ground-state degeneracy on a sphere). In the next section, we will complete the theory by eliminating the dependence of F symbols on the left-most group element using the symmetry.

\subsection{Symmetry action on the $F$ symbols}
\label{sec:sym-act-F}

The consistency conditions in Eqs.~\eqref{eq:locrel1}-\eqref{eq:fmove} ensure the existence of a fixed-point wave function $\Psi$. However, the wave function needs to be symmetric under the symmetry group $G$, which yields additional conditions on the input data. More precisely, we assume that the wave function on the sphere is invariant under $G$, or forms a one-dimensional representation.

We will show that the symmetry condition relates the $F$ symbol ${}^{\mb g_0}[\cal F^{\ag ag\ag bh\ag ck}_d]_{ef}$ to ${}^{\mb 1}[\cal F^{\ag ag\ag bh\ag ck}_d]_{ef}$. The former can be viewed as the result of the $\mb g_0$ action of the latter. In this subsection, we discuss the form of the $G$ action for different types of symmetry operations. We find it convenient to define
\begin{equation}
	{}^{\mb 1}[\cal F^{\ag ag\ag bh\ag ck}_d]_{ef} =[F^{\ag ag\ag bh\ag ck}_d]_{ef}, {}^{\mb 1}[v^{\ag ag\ag bh}_c]\equiv u^{\ag ag\ag bh}_c.
	\label{}
\end{equation}
The $F$ symbols $[F^{\ag ag\ag bh\ag ck}_d]_{ef}$ then satisfy a twisted pentagon equation,
\begin{multline}
	\label{eq:pentagon-t}
[F^{\ag f{gh}\ag ck\ag dl}_e]_{mq}
	[ F^{\ag ag\ag bh\ag q{kl}}_e]_{fp}\\
=\sum_{n_{\mb{hk}}\in \cal{C}_\mb{hk}}[ F^{\ag ag\ag bh\ag ck}_m]_{fn}
	[F^{\ag ag\ag n{hk}\ag dl}_e]_{mp}
	{}^{\mb g}[F^{\ag bh\ag ck\ag dl}_p]_{nq}.
\end{multline}
where ${}^{\mb g}[F^{\ag bh\ag ck\ag dl}_p]_{nq}$ schematically denotes that there is a nontrivial $\mb{g}$ action on the $F$ symbols. The detailed forms of the action for different types of symmetry operations will be determined below.

As we will see, when $G$ is an onsite unitary symmetry group,  the $F$ symbols defined in Eq.~\eqref{eq:fmove} is independent of $\mb g_0$, and Eq. \eqref{eq:pentagon-t} becomes the usual pentagon equation of $F$ for the $G$-graded fusion category $\mathcal{C}_G$. In this case, what we have defined is called a $G$-extension of the UFC $\cal{C}_\mb{1}$~\cite{ENO2009}. The mathematical classification of such extensions has been obtained in \Ref{ENO2009}. Remarkably, \Ref{ENO2009} showed that the equivalence classes of $G$-extensions of $\cal{C}_\mb{1}$ are in one-to-one correspondence with the (non-anomalous) symmetry-enriched topological orders in the double of $\cal{C}_\mb{1}$ (for a summary of the mathematical results, see Appendix \ref{sec:eno}). Therefore, our construction can represent all SETs in $\cal{Z}(\cal{C}_\mb{1})$ with a unitary finite symmetry group $G$.

\subsubsection{Onsite symmetry}
First, we consider an onsite unitary symmetry operation $\mb g_0$. Such a symmetry operation acts on group elements in all plaquettes: $\ket{\mb{g}_i}\rightarrow \ket{\mb{g}_0\mb{g}_i}$, while leaving all the edge labels unchanged. To get a wave function invariant under $\mb g_0$, we demand that the symmetry action commutes with the $F$ move in Eq.~\eqref{eq:fmove}: (the symmetry action obviously commutes with the other two types of moves in Eqs.~\eqref{eq:locrel1} and \eqref{eq:locrel2}),
\begin{equation}
  \label{eq:unitary-gF}
  \begin{tikzpicture}[baseline={($ (current bounding box) - (0,0pt) $)}]
    \matrix (m) [matrix of math nodes,row sep=2em, column sep=6em,minimum width=2em]
    {
    \Psi\left(\begin{tikzpicture}[baseline={($ (current bounding box) - (0,0pt) $)}]
		\draw (0, 0) node [above] {$\ag ag$} --(.3, -.5) [string];
		\draw (.6, 0) node [above] {$\ag bh$} --(.3, -.5) [string];
		\draw (.3, -.5)--(.6, -1) [string]
		      node [midway, left] {$\ag e{gh}$};
		\draw (1.2, 0) node [above] {$\ag ck$} --(.6, -1) [string];
		\draw (.6, -1)--(.9, -1.5) [string] node [below] {$\ag d{ghk}$};
		\node [group] at (0, -1.3) {$\mb 1$};
		\end{tikzpicture}\right) &
    \Psi\left(\begin{tikzpicture}[baseline={($ (current bounding box) - (0,0pt) $)}]
		\draw (0, 0) node [above] {$\ag ag$} --(.6, -1) [string];
		\draw (.6, 0) node [above] {$\ag bh$} --(.9, -.5) [string];
		\draw (.9, -.5)--(.6, -1) [string]
		      node [midway, right] {$\ag f{hk}$};
		\draw (1.2, 0) node [above] {$\ag ck$} --(.9, -.5) [string];
		\draw (.6, -1)--(.9, -1.5) [string] node [below] {$\ag d{ghk}$};
		\node [group] at (0, -1.3) {$\mb 1$};
		\end{tikzpicture}\right)\\
    \Psi\left(\begin{tikzpicture}[baseline={($ (current bounding box) - (0,0pt) $)}]
		\draw (0, 0) node [above] {$\ag ag$} --(.3, -.5) [string];
		\draw (.6, 0) node [above] {$\ag bh$} --(.3, -.5) [string];
		\draw (.3, -.5)--(.6, -1) [string]
		      node [midway, left] {$\ag e{gh}$};
		\draw (1.2, 0) node [above] {$\ag ck$} --(.6, -1) [string];
		\draw (.6, -1)--(.9, -1.5) [string] node [below] {$\ag d{ghk}$};
		\node [group] at (0, -1.3) {$\mb g_0$};
		\end{tikzpicture}\right) &
    \Psi\left(\begin{tikzpicture}[baseline={($ (current bounding box) - (0,0pt) $)}]
		\draw (0, 0) node [above] {$\ag ag$} --(.6, -1) [string];
		\draw (.6, 0) node [above] {$\ag bh$} --(.9, -.5) [string];
		\draw (.9, -.5)--(.6, -1) [string]
		      node [midway, right] {$\ag f{hk}$};
		\draw (1.2, 0) node [above] {$\ag ck$} --(.9, -.5) [string];
		\draw (.6, -1)--(.9, -1.5) [string] node [below] {$\ag d{ghk}$};
		\node [group] at (0, -1.3) {$\mb g_0$};
		\end{tikzpicture}\right)\\};
    \path[-stealth]
      (m-1-1) edge node [left] {$\mb g_0$} (m-2-1)
              edge node [above]
			  {$^{1}[\cal F^{\ag ag\ag bh\ag ck}_d]_{ef}$} (m-1-2)
      (m-2-1.east|-m-2-2) edge node [below]
              {${}^{\mb g_0}[\cal F^{\ag ag\ag bh\ag ck}_d]_{ef}$} (m-2-2)
      (m-1-2) edge node [right] {$\mb g_0$} (m-2-2);
  \end{tikzpicture}
\end{equation}
This implies that the $F$ symbol is independent of $\mb g_0$,
\begin{equation}
  \label{eq:unitary-gaction}
  {}^{\mb g_0}[\cal F^{\ag ag\ag bh\ag ck}_d]_{ef}
  ={}^1[\cal F^{\ag ag\ag bh\ag ck}_d]_{ef}.
\end{equation}
Similarly, we find that the gauge transformations are also independent of $\mb{g}_0$:
\begin{equation}
	{}^{\mb g_0}[v^{\ag ag\ag bh}_c]=u^{\ag ag\ag bh}_c.
	\label{eq:g0v-u}
\end{equation}

Second, we consider an onsite anti-unitary symmetry operation $\mb{g}_0$. In this case, the action of $\mb g_0$ not only transforms all group elements $\mb g_i\rightarrow \mb g\mb g_i$, but also complex conjugate the amplitude. The condition that the following diagram commutes,
\begin{equation}
  \label{eq:au-gF}
  \begin{tikzpicture}[baseline={($ (current bounding box) - (0,0pt) $)}]
    \matrix (m) [matrix of math nodes,row sep=2em, column sep=8em,minimum width=2em]
    {
    \Psi\left(\begin{tikzpicture}[baseline={($ (current bounding box) - (0,0pt) $)}]
		\draw (0, 0) node [above] {$\ag ag$} --(.3, -.5) [string];
		\draw (.6, 0) node [above] {$\ag bh$} --(.3, -.5) [string];
		\draw (.3, -.5)--(.6, -1) [string]
		      node [midway, left] {$\ag e{gh}$};
		\draw (1.2, 0) node [above] {$\ag ck$} --(.6, -1) [string];
		\draw (.6, -1)--(.9, -1.5) [string] node [below] {$\ag d{ghk}$};
		\node [group] at (0, -1.3) {$\mb 1$};
		\end{tikzpicture}\right) &
    \Psi\left(\begin{tikzpicture}[baseline={($ (current bounding box) - (0,0pt) $)}]
		\draw (0, 0) node [above] {$\ag ag$} --(.6, -1) [string];
		\draw (.6, 0) node [above] {$\ag bh$} --(.9, -.5) [string];
		\draw (.9, -.5)--(.6, -1) [string]
		      node [midway, right] {$\ag f{hk}$};
		\draw (1.2, 0) node [above] {$\ag ck$} --(.9, -.5) [string];
		\draw (.6, -1)--(.9, -1.5) [string] node [below] {$\ag d{ghk}$};
		\node [group] at (0, -1.3) {$\mb 1$};
		\end{tikzpicture}\right)\\
    \Psi^\ast\left(\begin{tikzpicture}[baseline={($ (current bounding box) - (0,0pt) $)}]
		\draw (0, 0) node [above] {$\ag ag$} --(.3, -.5) [string];
		\draw (.6, 0) node [above] {$\ag bh$} --(.3, -.5) [string];
		\draw (.3, -.5)--(.6, -1) [string]
		      node [midway, left] {$\ag e{gh}$};
		\draw (1.2, 0) node [above] {$\ag ck$} --(.6, -1) [string];
		\draw (.6, -1)--(.9, -1.5) [string] node [below] {$\ag d{ghk}$};
		\node [group] at (0, -1.3) {$\mb g_0$};
		\end{tikzpicture}\right) &
    \Psi^\ast\left(\begin{tikzpicture}[baseline={($ (current bounding box) - (0,0pt) $)}]
		\draw (0, 0) node [above] {$\ag ag$} --(.6, -1) [string];
		\draw (.6, 0) node [above] {$\ag bh$} --(.9, -.5) [string];
		\draw (.9, -.5)--(.6, -1) [string]
		      node [midway, right] {$\ag f{hk}$};
		\draw (1.2, 0) node [above] {$\ag ck$} --(.9, -.5) [string];
		\draw (.6, -1)--(.9, -1.5) [string] node [below] {$\ag d{ghk}$};
		\node [group] at (0, -1.3) {$\mb g_0$};
		\end{tikzpicture}\right)\\};
    \path[-stealth]
      (m-1-1) edge node [left] {$\mb g_0$} (m-2-1)
              edge node [above]
			  {$^{1}[\cal F^{\ag ag\ag bh\ag ck}_d]_{ef}$} (m-1-2)
      (m-2-1.east|-m-2-2) edge node [below]
              {${}^{\mb g_0}[\cal F^{\ag ag\ag bh\ag ck}_d]_{ef}^\ast$} (m-2-2)
      (m-1-2) edge node [right] {$\mb g_0$} (m-2-2);
  \end{tikzpicture}
\end{equation}
implies that the $\mb g_0$ action on the $F$ symbol is the complex conjugation,
\begin{equation}
  \label{eq:au-gaction}
  {}^{\mb g_0}[\cal F^{\ag ag\ag bh\ag ck}_d]_{ef}
  ={}^1[\cal F^{\ag ag\ag bh\ag ck}_d]_{ef}^\ast.
\end{equation}
The action on the gauge transformation is similar:
\begin{equation}
	{}^{\mb g_0}[v^{\ag ag\ag bh}_c]=(u^{\ag ag\ag bh}_c)^*.
	\label{eq:g0v-au}
\end{equation}

\subsubsection{Mirror symmetry}
We now consider $\mb{g}_0$ a mirror reflection operation. Other point-group operations can be generally composed out of reflections.

Due to the branching structure, there may seem to be different ways to position the mirror axis. We choose the mirror reflection according to the following convention:
\begin{equation}
 \begin{tikzpicture}[baseline={($ (current bounding box) - (0,0pt) $)}]
	 \draw (0, 0) node [above] {$\ag ag$} -- (0.5, {-sqrt(3)/2}) [string];
	 \node [group] at (0.0, {-sqrt(3)/2+0.25}) {\scalebox{0.8}{$\mb{g}_1$}};
	 \node [group] at (0.7, {-sqrt(3)/4+0.2}) {\scalebox{0.8}{$\mb{g}_1\mb{g}$}};
		\draw[dashed, thick] (0.0, {-sqrt(3)/2}) -- (0.75, {-sqrt(3)/4}) [dw];
		\draw (-1, 0.7) -- (1,0.7);
		\draw (0.5, {1.4+sqrt(3)/2})  -- (0, 1.4) node [below] {$\ag ag$} [string];
		\draw[dashed, thick] (0.0, {1.4+sqrt(3)/2}) -- (0.75, {1.4+sqrt(3)/4}) [dw];
	\end{tikzpicture}
	\label{}
\end{equation}
Heuristically, it means that the strings $a_\mb{g}$ transform as if they are pseudo-vectors, which is consistent with the intuitive interpretation that they are like ``symmetry flux lines''. Therefore, we choose a branching structure that transforms as a pseudo-vector under the mirror symmetry. Since the construction is independent of the branching structure, such a choice is always possible.

We again demand that the F move commutes with the symmetry action, as shown in the following diagram:
\begin{equation}
  \label{eq:mirror-gF}
  \begin{tikzpicture}[baseline={($ (current bounding box) - (0,0pt) $)}]
    \matrix (m) [matrix of math nodes,row sep=2em, column sep=8em,minimum width=2em]
    {
    \Psi\left(\begin{tikzpicture}[baseline={($ (current bounding box) - (0,0pt) $)}]
		\draw (0, 0) node [above] {$\ag ag$} --(.3, -.5) [string];
		\draw (.6, 0) node [above] {$\ag bh$} --(.3, -.5) [string];
		\draw (.3, -.5)--(.6, -1) [string]
		      node [midway, left] {$\ag e{gh}$};
		\draw (1.2, 0) node [above] {$\ag ck$} --(.6, -1) [string];
		\draw (.6, -1)--(.9, -1.5) [string] node [below] {$\ag d{ghk}$};
		\node [group] at (0, -1.3) {$\mb 1$};
		\end{tikzpicture}\right) &
    \Psi\left(\begin{tikzpicture}[baseline={($ (current bounding box) - (0,0pt) $)}]
		\draw (0, 0) node [above] {$\ag ag$} --(.6, -1) [string];
		\draw (.6, 0) node [above] {$\ag bh$} --(.9, -.5) [string];
		\draw (.9, -.5)--(.6, -1) [string]
		      node [midway, right] {$\ag f{hk}$};
		\draw (1.2, 0) node [above] {$\ag ck$} --(.9, -.5) [string];
		\draw (.6, -1)--(.9, -1.5) [string] node [below] {$\ag d{ghk}$};
		\node [group] at (0, -1.3) {$\mb 1$};
		\end{tikzpicture}\right)\\
    \Psi\left(\begin{tikzpicture}[baseline={($ (current bounding box) - (0,0pt) $)}]
		\draw (.3, .5) -- (0, 0) node [below] {$\ag ag$} [string];
		\draw (.3, .5) -- (.6, 0) node [below] {$\ag bh$}[string];
		\draw (.6, 1) -- (.3, .5) [string]
		      node [midway, left] {$\ag e{gh}$};
		\draw (.6, 1) -- (1.2, 0) node [below] {$\ag ck$}  [string];
		\draw (.9, 1.5) node [above] {$\ag d{ghk}$} -- (.6, 1)[string] ;
		\node [group] at (0, 1.3) {$\mb g_0$};
		\end{tikzpicture}\right) &
    \Psi\left(\begin{tikzpicture}[baseline={($ (current bounding box) - (0,0pt) $)}]
		\draw (.6, 1) -- (0, 0) node [below] {$\ag ag$}  [string];
		\draw (.9, .5) -- (.6, 0) node [below] {$\ag bh$} [string];
		\draw (.6, 1) -- (.9, .5) [string]
		      node [midway, right] {$\ag f{hk}$};
		\draw (.9, .5) -- (1.2, 0) node [below] {$\ag ck$}  [string];
		\draw (.9, 1.5) [string] node [above] {$\ag d{ghk}$} -- (.6, 1);
		\node [group] at (0, 1.3) {$\mb g_0$};
		\end{tikzpicture}\right)\\};
    \path[-stealth]
      (m-1-1) edge node [left] {$\mb g_0$} (m-2-1)
              edge node [above]
			  {$^{1}[\cal F^{\ag ag\ag bh\ag ck}_d]_{ef}$} (m-1-2)
      (m-2-1.east|-m-2-2) edge node [below]
	  {${}^{\mb g_0}[\tilde{\cal F}^{\ag ag\ag bh\ag ck}_d]_{ef}$} (m-2-2)
      (m-1-2) edge node [right] {$\mb g_0$} (m-2-2);
  \end{tikzpicture}
\end{equation}
We need to evaluate the ``dual'' F move denoted by $\tilde{F}$ in the diagram. By stacking the diagrams of $\tilde{\cal F}$ on top of those of ${\cal F}$, we can easily derive the following relation:
\begin{equation}
\sum_f{}^{\mb{g}_0}[\tilde{\cal F}^{\ag ag\ag bh\ag ck}_d]_{ef} {}^{\mb{g}_0}[{\cal F}^{\ag ag\ag bh\ag ck}_d]_{e'f}=\delta_{ee'},
	\label{}
\end{equation}
or in matrix form, ${}^{\mb{g}_0}[\tilde{\cal F}^{\ag ag\ag bh\ag ck}_d]\,{}^{\mb{g}_0}[{\cal F}^{\ag ag\ag bh\ag ck}_d]^{\mathsf{T}}=1$. Because ${}^{\mb{g}_0}[{\cal F}^{\ag ag\ag bh\ag ck}_d]$ is unitary, it follows that ${}^{\mb{g}_0}[\tilde{\cal F}^{\ag ag\ag bh\ag ck}_d]_{ef}={}^{\mb{g}_0}[{\cal F}^{\ag ag\ag bh\ag ck}_d]_{ef}^\ast$. So we find that the mirror action on $\mathcal{F}$ symbols is the same as that of an anti-unitary symmetry:
\begin{equation}
  \label{eq:mirror-action}
  {}^{\mb g_0}[\cal F^{\ag ag\ag bh\ag ck}_d]_{ef}
  ={}^1[\cal F^{\ag ag\ag bh\ag ck}_d]_{ef}^\ast.
\end{equation}

Using the two $\mathbb Z_2$ gradings $p(\mb g)$ and $q(\mb g)$ we introduced at the beginning of this section, the symmetry transformations in Eqs.~\eqref{eq:unitary-gaction}, \eqref{eq:au-gaction}, and \eqref{eq:mirror-action} can be unified into the following form,
\begin{equation}
  \label{eq:sg-action}
  {}^{\mb g_0}[\cal F^{\ag ag\ag bh\ag ck}_d]_{ef}
  ={}^1[\cal F^{\ag ag\ag bh\ag ck}_d]_{ef}^{s(\mb g_0)},
\end{equation}
where $s(\mb g_0)=1$ if $p(\mb g_0)q(\mb g_0)=1$, and $s(\mb g_0)=\ast$ if $p(\mb g_0)q(\mb g_0)=-1$.

\subsubsection{Anomalous symmetry}
\label{sec:anomalous-symmetry}

Finally, we discuss anomalous symmetry actions, which can be used to study anomalous SET states that can only exist on the surface of a 3D SPT state. On a symmetry-preserving surface of a nontrivial 3D SPT state, the symmetry cannot be realized in an onsite fashion in terms of degrees of freedoms on the 2D surface. Due to the anomalous symmetry action, a symmetry-preserving surface state (i.e., no spontaneous symmetry breaking) is either gapless or gapped by an anomalous SET state. In this section, we define a generalized form of $G$-extension of a UFC to study such anomalous SET states, realized on the surface of 3D group-cohomology SPT models. We only consider onsite symmetries (unitary or anti-unitary) in this section, and will comment on possible generalizations to mirror symmetries in the end.

First, we outline how our construction can be adopted to study the surface topological order of a 3D SPT state. We will focus on those 3D SPT states within the so-called group-cohomology classification~\cite{chen2013}. These SPT phases can be realized in exactly-solvable commuting-projector models. For this reason one can decouple the boundary degrees of freedom from the bulk, in the sense that the boundary can be formally treated as a stand-alone two-dimensional system,  but the bulk SPT state leaves its fingerprint in how the global symmetry acts on the boundary degrees of freedom. If the bulk SPT state is nontrivial, the symmetry action on the boundary is ``anomalous'', in a way that can not be realized as a truly onsite symmetry in the 2D lattice model. In our construction, we will view the plaquette spin degrees of freedom as coming from a 3D SPT state (after the bulk has been traced out). Notice that by construction the symmetry only acts on the spin degrees of freedom in the plaquettes. We now derive the precise form of the anomalous symmetry transformation.

\begin{figure}[htbp]
  \centering
  \includegraphics{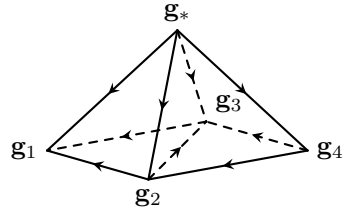}
  \label{fig:conf3d}
  \caption{A triangulation of a 3D bulk, with a 2D surface. The bulk is represented by only one vertex, carrying a group element $\mb g_\ast$. The vertices carrying $\mb g_1,\dots \mb g_4$ belong to the 2D surface.}
\end{figure}

According to \Ref{chen2013}, each cohomology class of $\cal H^4[G, \mathrm U(1)]$ describes a distinct 3D SPT phase, of which a fixed-point wave function can be constructed using a representative 4-cocycle $\beta$ of the class. The wave function can be defined on any triangulation of the 3D spatial manifold. For the convenience of studying the surface state, we choose the following minimal triangulation of the bulk, by adding a single vertex to the bulk and connecting the bulk vertex to all surface vertices. The branching structure is chosen such that all additional links point from the bulk vertex to the surface. As shown in Fig.~\ref{fig:conf3d}, there is a one-to-one correspondence between the tetrahedra in the bulk and the triangles on the surface.

The basis states in the Hilbert space are labeled by assigning a group element to every vertex in the triangulation. In particular, we denote the group element on the bulk vertex as $\mb g_\ast$. The fixed-point wave function is a superposition of all basis states, with the following phase factor associated to each tetrahedron:
\begin{equation}
  \label{eq:3d-fpwf}
  \Psi\left(
  \begin{tikzpicture}[baseline={($ (current bounding box) - (0,0pt) $)}]
    \draw (-.25, -.25) node [below] {$\mb g_0$} --
          (.5, .5) node [above] {$\mb g_2$} [string];
    \draw (.5, .5) -- (1.25, -.25) node [below] {$\mb g_1$} [string];
    \draw (-.25, -.25) -- (1.25, -.25) [string];
	\end{tikzpicture},\mb g_\ast
  \right)=\beta(\mb g_\ast,\mb g_\ast^{-1}\mb g_0,
  \mb g_0^{-1}\mb g_1, \mb g_1^{-1}\mb g_2).
\end{equation}

The states on the surface are in fact dual to group-element configurations on the trivalent graph used in Sec.~\ref{sec:fpwf}, where a vertex on the surface corresponds to a plaquette of the trivalent graph, and the branching structure on the triangulated surface is canonically induced from the one on the trivalent graph, according to the rule in Eq.~\eqref{eq:dw}. We can then write the phase factors for a given group configuration on the trivalent graph:
\begin{equation}
  \label{eq:3d-fpwf2}
  \Psi\left(
  \begin{tikzpicture}[baseline={($ (current bounding box) - (0,0pt) $)}]
    \draw (-.25, -.25) node [below] {$\mb g_0$} --
          (.5, .5) [dual_string];
    \draw (.5, .5) -- (1.25, -.25) [dual_string];
    \draw (-.25, -.25) -- (1.25, -.25) [dual_string];
		\draw (0, .5) node [above] {$\mb g$}--(.5, 0) [string];
	  \draw (1, .5) node [above] {$\mb h$}--(.5, 0) [string];
		\draw (.5, 0)--(.5, -.5) node [below] {$\mb{gh}$} [string];
	\end{tikzpicture},\mb g_\ast
  \right)=\beta(\mb g_\ast,\mb g_\ast^{-1}\mb g_0,
  \mb g, \mb h).
\end{equation}
Plugging this phase factor into Eq.~\eqref{eq:fmove}, we find that this wave function has a nontrivial phase factor associated with an $F$ move,
\begin{gather}
  \label{eq:fmove-spt}
  \Psi\left(\begin{tikzpicture}[baseline={($ (current bounding box) - (0,0pt) $)}]
  \draw (0, 0) node [above] {$\mb g$} --(.3, -.5) [string];
  \draw (.6, 0) node [above] {$\mb h$} --(.3, -.5) [string];
  \draw (.3, -.5)--(.6, -1) [string];
  \draw (1.2, 0) node [above] {$\mb k$} --(.6, -1) [string];
  \draw (.6, -1)--(.9, -1.5) [string];
  \node [group] at (0, -1.3) {$\mb g_0$};
  \end{tikzpicture},\mb g_\ast\right)
  ={}^{\mb g_0}[\cal F^{\mb{ghk}}](\mb g_\ast)
  \Psi\left(\begin{tikzpicture}[baseline={($ (current bounding box) - (0,0pt) $)}]
  \draw (0, 0) node [above] {$\mb g$} --(.6, -1) [string];
  \draw (.6, 0) node [above] {$\mb h$} --(.9, -.5) [string];
  \draw (.9, -.5)--(.6, -1) [string];
		\draw (1.2, 0) node [above] {$\mb k$} --(.9, -.5) [string];
  \draw (.6, -1)--(.9, -1.5) [string];
  \node [group] at (0, -1.3) {$\mb g_0$};
  \end{tikzpicture},\mb g_\ast\right),\\
  \label{eq:F=b4}
  {}^{\mb g_0}[\cal F^{\mb{ghk}}](\mb g_\ast)
  =\frac{\beta(\mb g_\ast,\mb g_\ast^{-1}\mb g_0\mb g,\mb h,\mb k)
         \beta(\mb g_\ast,\mb g_\ast^{-1}\mb g_0,\mb g,\mb{hk})}
        {\beta(\mb g_\ast,\mb g_\ast^{-1}\mb g_0,\mb g,\mb h)
         \beta(\mb g_\ast,\mb g_\ast^{-1}\mb g_0,\mb{gh},\mb k)}.
\end{gather}
Using the cocycle condition $\di\beta=1$, the $F$ symbol in Eq.~\eqref{eq:F=b4} can be simplified as
\begin{equation}
  \label{eq:F=bb}
  {}^{\mb g_0}[\cal F^{\mb{ghk}}](\mb g_\ast)
  =\frac{\beta(\mb g_\ast^{-1}\mb g_0,\mb{g,h,k})^{s(\mb g_\ast)}}
  {\beta(\mb g_0,\mb{g,h,k})}.
\end{equation}

Next, we consider how the symmetry acts on the wave function. Similar to the discussion in Eqs.~\eqref{eq:unitary-gF} and \eqref{eq:au-gF}, we compare the $F$ moves before and after a symmetry transformation that changes the group element in the left-most plaquette from $\mb1$ to $\mb g_0$. However, the symmetry also acts in the bulk, and changes $\mb g_\ast$ to $\mb g_0\mb g_\ast$. Comparing the two $F$ moves, we get
\begin{equation}
  \label{eq:F=Fbeta}
  \frac{{}^{\mb g_0}[\cal F^{\mb{ghk}}](\mb g_0\mb g_\ast)}
  {[\cal F^{\mb{ghk}}](\mb g_\ast)^{s(\mb g_0)}}
  =\frac{\beta(\mb{1,g,h,k})^{s(\mb g_0)}}
  {\beta(\mb g_0,\mb{g,h,k})}.
\end{equation}
Using the coboundary equivalence, we can choose a gauge, such that $\beta(\mb{1,g,h,k})=+1$. In this case, the result in Eq.~\eqref{eq:F=Fbeta} is simplified to
\begin{equation}
  \label{eq:F=Fbeta2}
  {}^{\mb g_0}[\cal F^{\mb{ghk}}](\mb g_0\mb g_\ast)=
  \beta(\mb g_0,\mb{g,h,k})^{-1}
  [\cal F^{\mb{ghk}}](\mb g_\ast)^{s(\mb g_0)}.
\end{equation}

The additional phase factor appearing in Eq.~\eqref{eq:F=Fbeta2} reflects the anomalous nature of the symmetry action on the surface of a nontrivial 3D SPT, and cannot be gauged away by any redefinition of the symmetry action. In fact, such redefinitions can only account for trivial phase factors, which are the coboundary of a 3-cochain.

To demonstrate this, we consider a general symmetry action which generates a nontrivial phase factor on each vertex of the trivalent graph,
\begin{equation}
  \label{eq:gact-v}
  \mb g_0:\vast|\begin{tikzpicture}[baseline={($(current bounding box)-(0,0pt)$)}]
    \draw (0, 0) node [above] {$\ag ag$} -- (.3, -.5) [string];
    \draw (.6, 0) node [above] {$\ag bh$} -- (.3, -.5) [string];
    \draw (.3, -.5) -- (.3, -1) node [below] {$\ag c{gh}$} [string];
    \node [group] at (-.3, -.7) {$\mb 1$};
  \end{tikzpicture}\vast\rangle
  \rightarrow \omega(\mb g_0, \mb{g, h})
    \vast|\begin{tikzpicture}[baseline={($(current bounding box)-(0,0pt)$)}]
    \draw (0, 0) node [above] {$\ag ag$} -- (.3, -.5) [string];
    \draw (.6, 0) node [above] {$\ag bh$} -- (.3, -.5) [string];
    \draw (.3, -.5) -- (.3, -1) node [below] {$\ag c{gh}$} [string];
    \node [group] at (-.3, -.7) {$\mb g_0$};
  \end{tikzpicture}\vast\rangle,
\end{equation}
where $\omega$ is an arbitrary 3-cochain.  We also assume that the symmetry group is onsite and unitary, but the following discussion can be easily generalized to antiunitary and mirror symmetries by adding the complex conjugation $s(\mb g)$ at appropriate places.

Plugging this new definition of the symmetry action into the $F$ move in Eq.~\eqref{eq:fmove}, we see that after symmetry actions, the $F$ symbols acquire additional phase factors, comparing to the results in Sec.~\ref{sec:sym-act-F},
\begin{equation}
  \label{eq:redef-sym-act-F0}
  {}^{\mb g_0}[\cal F^{\ag ag\ag bh\ag ck}_d]_{ef}
  = \frac{\omega(\mb g_0,\mb{g, hk})
    \omega(\mb g_0\mb g,\mb{h,k})}
  {\omega(\mb g_0,\mb{g,h})
  \omega(\mb g_0, \mb{gh, k})}
  [F^{\ag ag\ag bh\ag ck}_d]_{ef}
\end{equation}
The phase factor can be rearranged into the following form,
\begin{equation}
  \label{eq:redef-sym-act-F}
  {}^{\mb g_0}[\cal F^{\ag ag\ag bh\ag ck}_d]_{ef}
  = \frac{d\omega(\mb{1,g,h,k})}
    {d\omega(\mb g_0,\mb{g,h,k})}
  [F^{\ag ag\ag bh\ag ck}_d]_{ef}.
\end{equation}

Added to the symmetry transformation in Eq.~\eqref{eq:F=Fbeta}, such a redefinition changes the cocycle $\beta$ to $\beta d\omega$. Therefore, a redefinition of the symmetry action on the surface can change $\beta$ by a coboundary term, but cannot alter its cohomology class. This is consistent with our claim that a symmetry action with a nontrivial $\beta$ is anomalous, and thus cannot be realized in a purely 2D system.

We can now carry through the construction of the fixed-point wavefunction for the surface SET. In previous sections, Eq.~\eqref{eq:sg-action} ensures that the wave function is invariant under global symmetry actions. With the anomalous symmetry transformation, to make sure that the wave function is symmetric the extended $F$ symbols ${}^{\mb g_0}[\cal F^{\ag ag\ag bh\ag ck}_d]_{ef}$ have to acquire an additional phase factor under symmetry actions:
\begin{equation}
  \label{eq:beta-gaction}
  {}^{\mb g_0}[\cal F^{\ag ag\ag bh\ag ck}_d]_{ef}
  =\beta(\mb{g}_0,\mb{g,h,k})
  {}^1[\cal F^{\ag ag\ag bh\ag ck}_d]_{ef}^{s(\mb g_0)}.
\end{equation}
The diagrams in Eq.~\eqref{eq:unitary-gF}, \eqref{eq:au-gF}, and \eqref{eq:mirror-gF} commute because the relative phases for $F$ move in Eq. \eqref{eq:beta-gaction} are exactly canceled out by the same phase factors from anomalous symmetry transformations.

Furthermore, we notice that the extra phase factor in Eq.~\eqref{eq:F=Fbeta2} exactly cancels the phase factor in Eq.~\eqref{eq:beta-gaction}. Consequently, if we put the 2D SET fixed-point wave function satisfying the anomalous symmetry transformation in Eq.~\eqref{eq:beta-gaction} on the surface (meaning that the group elements in the SET state are actually part of the 3D SPT fixed-point state), all phase factors cancel out and we obtain a symmetric wave function.
 This way, we explicitly demonstrate that the anomalous symmetry transformation discussed in Sec.~\ref{sec:anomalous-symmetry} can be used to study anomalous SET states realized on the surface of the corresponding 3D SPT state.

Applying the general twisted pentagon equation in Eq.~\eqref{eq:pentagon4}, we obtain the following ``obstructed'' pentagon equation:
	\begin{multline}
	\label{eq:pentagon-obs}
[F^{\ag f{gh}\ag ck\ag dl}_e]_{mq}
  	[ F^{\ag ag\ag bh\ag q{kl}}_e]_{fp}=\\
  \beta(\mb{g,h,k,l})
  \sum_{n_{\mb{hk}}\in \cal{C}_\mb{hk}}[ F^{\ag ag\ag bh\ag ck}_m]_{fn}
  	[F^{\ag ag\ag n{hk}\ag dl}_e]_{mp}
    [F^{\ag bh\ag ck\ag dl}_p]_{nq}^{s(\mb g)}
\end{multline}
The solution of this equation describes anomalous SET states that can only be constructed together with a nontrivial 3D bulk if we demand that the symmetry action is onsite, belonging to the SPT state corresponding to $\beta\in\cal H^4[G, \mathrm U(1)]$. It is worth noticing that at a heuristic level, Eq. \eqref{eq:pentagon-obs} resembles the pentagon equation of symmetry defects in \Ref{ChenASF2014}, if we think of $a_\mb{g}$'s as representing symmetry defects to some extent. However, one should not confuse these two equations, since Eq. \eqref{eq:pentagon-obs} applies to the input data to our generalized ``string-net'' type construction, while \Ref{ChenASF2014} discussed the actual physical defects in a symmetry-enriched topological phase.

We note that it is fairly well-established that 3D SPT states protected by onsite symmetries, either unitary or anti-unitary, are partially classified by $\cal H^4[G,\mathrm U(1)]$. It has been conjectured that the classification takes a similar form if $G$ contains mirror-reflection symmetries~\cite{ZXLiu_mirrorSPT}, where mirror reflections act on the U(1) coefficients by complex conjugations.

\subsection{Parent Hamiltonians}
\label{sec:parent_ham}
We now briefly describe how to construct a parent Hamiltonian for the fixed-point wave function, generalizing the Levin-Wen Hamiltonians. The Hamiltonian takes the following form:
\begin{equation}
	H=-\sum_v Q_v - \sum_e Q_e - \sum_p B_p.
	\label{}
\end{equation}
Here $Q_v, Q_e$ and $B_p$ are all commuting projectors.

The vertex terms $Q_v$ ensure that fusion rules are obeyed at each vertex:
\begin{equation}
	Q_v\vast|
	\begin{tikzpicture}[baseline={($ (current bounding box) - (0,0pt) $)}, scale=0.9]
		\draw (0, .5) node [above] {$a$}--(.5, 0) [string];
		\draw (1, .5) node [above] {$b$}--(.5, 0) [string];
		\draw (.5, 0)--(.5, -.5) node [right] {$c$} [string];
	\end{tikzpicture}
	\vast\rangle
	=N_{ab}^c\vast|
	\begin{tikzpicture}[baseline={($ (current bounding box) - (0,0pt) $)}, scale=0.9]
		\draw (0, .5) node [above] {$a$}--(.5, 0) [string];
		\draw (1, .5) node [above] {$b$}--(.5, 0) [string];
		\draw (.5, 0)--(.5, -.5) node [right] {$c$} [string];
	\end{tikzpicture}
	\vast\rangle.
	\label{}
\end{equation}
And the edge terms $Q_e$ enforce the $G$-grading structure:
\begin{equation}
	Q_e\vast|
 \begin{tikzpicture}[baseline={($ (current bounding box) - (0,3.5pt) $)}]
		\draw (0, 0) node [right] {$\ag ag$} -- (0, -1.0) [string];
		\node [group] at (-0.3, -0.4) {\scalebox{0.8}{$\mb{g}_1$}};
		\node [group] at (0.35, -0.4) {\scalebox{0.8}{$\mb{g}_2$}};
		\draw[dashed, thick] (-0.5, -0.7) -- (0.5, -0.7) [dw];
		\end{tikzpicture}
		\vast\rangle
		=\delta_{\mb{g},\overline{\mb{g}}_1\mb{g}_2}\vast|
 \begin{tikzpicture}[baseline={($ (current bounding box) - (0,3.5pt) $)}]
		\draw (0, 0) node [right] {$\ag ag$} -- (0, -1.0) [string];
		\node [group] at (-0.3, -0.4) {\scalebox{0.8}{$\mb{g}_1$}};
		\node [group] at (0.35, -0.4) {\scalebox{0.8}{$\mb{g}_2$}};
		\draw[dashed, thick] (-0.5, -0.7) -- (0.5, -0.7) [dw];
		\end{tikzpicture}
		\vast\rangle.
	\label{}
\end{equation}

The vertex and edge projectors are fairly straightforward to define. The most important part of the construction is the plaquette terms, which take the following form:
\begin{equation}
	B_p = \frac{1}{\mathcal{D}^2}\sum_{\mb{g}\in G}\ket{\mb{g}_p\mb{g}}\bra{\mb{g}_p}\sum_{s_\mb{g}\in \mathcal{C}_\mb{g}}d_{s_\mb{g}}B_p^{s_\mb{g}}.
	\label{eqn:Bp}
\end{equation}
As in the Levin-Wen construction, here $B_p^{s_\mb{g}}$ has the following graphic representation: imagine adding a loop of $s_\mb{g}$ to the plaquette, and fuse the loop onto the edges using the local moves defined in.
\begin{equation}
	B_p^s
	\vast|\,
	 \begin{tikzpicture}[baseline={($ (current bounding box) - (0,1.5pt) $)}]
		 \draw[thick] ({sqrt(3)/4},0.25) -- (0,0.5) -- ({-sqrt(3)/4},0.25) -- ({-sqrt(3)/4},-0.25) -- (0,-0.5) -- ({sqrt(3)/4},-0.25) -- ({sqrt(3)/4},0.25);
	\draw[thick] ({sqrt(3)/4},0.25) -- ({1.7*sqrt(3)/4},1.7*0.25);
	\draw[thick] (0,0.5) -- (0,1.7*0.5);
	\draw[thick] ({-sqrt(3)/4},0.25) -- ({-1.7*sqrt(3)/4},1.7*0.25);
	\draw[thick] ({-sqrt(3)/4},-0.25) -- ({-1.7*sqrt(3)/4},-1.7*0.25);
	\draw[thick] (0,-0.5) -- (0,-1.7*0.5);
	\draw[thick] ({sqrt(3)/4},-0.25) -- ({1.7*sqrt(3)/4},-1.7*0.25);
\end{tikzpicture}\,
	\vast\rangle
	=
	\vast|\,
	 \begin{tikzpicture}[baseline={($ (current bounding box) - (0,1.5pt) $)}]
		 \draw[thick] ({sqrt(3)/4},0.25) -- (0,0.5) -- ({-sqrt(3)/4},0.25) -- ({-sqrt(3)/4},-0.25) -- (0,-0.5) -- ({sqrt(3)/4},-0.25) -- ({sqrt(3)/4},0.25);
\draw [rounded corners, thick] ({0.8*sqrt(3)/4},0.8*0.25) -- (0,0.8*0.5) -- ({-0.8*sqrt(3)/4},0.8*0.25) -- ({-0.8*sqrt(3)/4},-0.8*0.25) -- (0,-0.8*0.5) -- ({0.8*sqrt(3)/4},-0.8*0.25) -- cycle;
	\draw[thick] ({sqrt(3)/4},0.25) -- ({1.7*sqrt(3)/4},1.7*0.25);
	\draw[thick] (0,0.5) -- (0,1.7*0.5);
	\draw[thick] ({-sqrt(3)/4},0.25) -- ({-1.7*sqrt(3)/4},1.7*0.25);
	\draw[thick] ({-sqrt(3)/4},-0.25) -- ({-1.7*sqrt(3)/4},-1.7*0.25);
	\draw[thick] (0,-0.5) -- (0,-1.7*0.5);
	\draw[thick] ({sqrt(3)/4},-0.25) -- ({1.7*sqrt(3)/4},-1.7*0.25);
	\node  at (0.2, 0) {\scalebox{0.8}{$s$}};
\end{tikzpicture}\,
	\vast\rangle
	\label{eq:bp}
\end{equation}
The right-hand side of this equation contains a loop carrying a topological charge $s$ running inside of the hexagon, and it is a graphic representation of a superposition of different configurations on the hexagon. The precise form of this superposition can be computed by deforming the diagram using the combination of basic moves, including the $F$ move in Eq.~\eqref{eq:fmove}, the $H$ move, which is a variation of $F$ move discussed in Appendix~\ref{fixwv} , and the elimination of bubbles in Eq.~\eqref{eq:locrel2}, as shown in Fig.~\ref{fig:bpmoves}.

\begin{figure}[htbp]
	\subfigure[Step1]{\includegraphics[width=0.3\columnwidth]{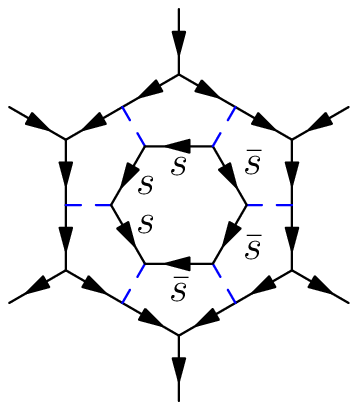}}~~~~~~
  \subfigure[Step2]{\includegraphics[width=0.3\columnwidth]{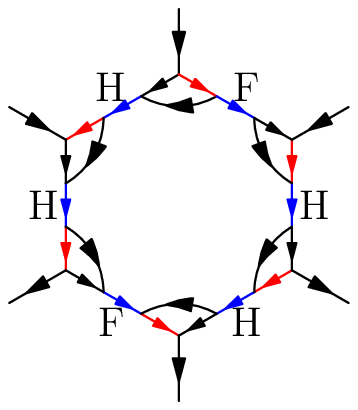}}~~~~~~
  \subfigure[Step3]{\includegraphics[width=0.3\columnwidth]{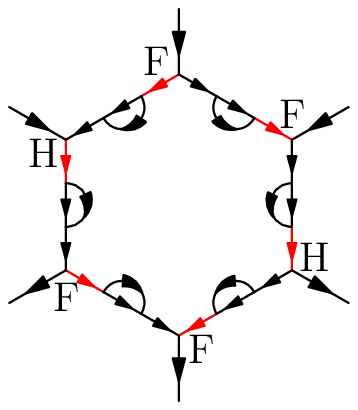}}
  \caption{\label{fig:bpmoves} Steps of deforming the right-hand side of Eq.~\eqref{eq:bp} to its left-hand side. (a) Step 1: the initial configuration containing an inner loop carrying a topological charge $s$. The inner loop consists of counter-propagating segments carrying charges $s$ and $\bar{s}$, respectively. The dashed line are vacuum strings carrying charge $0$. The blue color indicates the locations where the $F$ moves will be applied to obtain the next configuration. (b) Step 2: the second configuration is obtained after applying six $F$ moves and $H$ moves (see Appendix~\ref{fixwv}), located at the links marked by the blue color. The letter denotes whether an $F$ move or an $H$ move is performed. The red links mark the locations of the $F$ moves leading to the next configuration. (c) Step 3: the third configuration is obtained through six $F$ moves and $H$ moves located at the red links. The letter denotes whether an $F$ move or an $H$ move is performed. Finally, this configuration is changed into the one on the left-hand side of Eq.~\eqref{eq:bp}, by eliminating the bubbles using the move in Eq.~\eqref{eq:locrel2}.}
\end{figure}

In summary, in this section, we explicitly construct 2D SET states using fixed-point wave functions, which are ground states of commuting-projector Hamiltonians. The fixed-point wave function is based on a set of data describing a generalized $G$-extension of the UFC $\cal C_\mb{1}$, including the generalized $\cal F$ symbols.

When the symmetry group $G$ is unitary, the generalized $\mathcal{F}$ symbols reduce to the usual F symbols, and the structure of the fixed-point wave function is essentially given by the $G$-extension $\cal{C}_G$ of $\mathcal{C}_\mb{1}$. The parent Hamiltonian can then be understood as ``ungauging'' the parent Hamiltonian of the string-net model with $\cal{C}_G$ as the input data. In other words, because of the $G$-grading of $\cal{C}_G$, its quantum double $\cal{Z}(\cal{C}_G)$ can be thought as a gauge theory. Strings with labels in the $\cal{C}_\mb{g}$ sectors with $\mb{g}\neq \mb{1}$ play the role of gauge connections of $G$, and in the SET phase they are forced to align with the $G$ domain walls of the plaquette spins, so in the language of a gauge theory, the $G$ connections become pure gauges. This is a generalization of the duality between SPT phases and Dijkgraaf-Witten topological gauge theories~\cite{levin2012}.

Using the formalism of anyon models, the ungauging procedure can be understood in terms of anyon condensation. Among the quasiparticles in $\cal{Z}(\cal{C}_G)$, there is a subset whose fusion rules and braiding statistics are isomorphic to those of $\text{Rep}(G)$, i.e., the category of irreducible linear representations of $G$. The bosons in $\text{Rep}(G)$ can be condensed, driving a topological phase transition from the gauged theory to the SET phase~\cite{SET1}. The effect of the condensation of $\text{Rep}(G)$ on the topological order can be analyzed algebraically~\cite{bais2009, eliens2013, kong2014, burnell2012}. Very roughly speaking, all gauge fluxes become confined by the condensation. Microscopically, the confinement corresponds precisely to ``ungauging'' the strings with nontrivial gradings to domain walls. Furthermore, because quasiparticles in $\text{Rep}(G)$ are local excitations after the condensation (carrying $G$ charges), the remaining quasiparticles in $\cal{Z}(\cal{C}_G)$ are re-organized: for example, those that can be transformed into each other by fusing with particles in $\text{Rep}(G)$ are now identified as the same type of anyons, and sometimes a quasiparticle needs to split into several distinct types of quasiparticles. In Sec. \ref{sec:z2tc-unitary} we will see very explicitly how the condensation works.

When $G$ contains anti-unitary or mirror-reflection symmetries, the generalized $\cal F$ symbols satisfy instead the twisted pentagon equation in Eq.~\eqref{eq:pentagon-t}. We conjecture that the corresponding generalized $G$-extension $\cal C_G$ also classifies the space-time SET phases of $\cal Z(\cal C)$.

In the rest of this paper, the framework presented in this section will be used to study SET phases of different topological orders. We begin with a basic example, with a trivial $\cal C_\mb{1}=\mathrm{Vec}$. This means that there is no topological order to begin with, and we are therefore classifying 2D SPT states. In this case, each sector $\cal C_\mb{g}$ contains only one simple object with $d=1$, and hence all objects in $\cal C_G$ can be labeled simply by group elements. Therefore, the $F$ symbols can be viewed as 3-cochain, the pentagon equation in Eq.~\eqref{eq:pentagon-t} becomes the cocycle conditions in group cohomology, and the gauge equivalence in Eq.~\eqref{eq:v-equiv} becomes the coboundary equivalence. Therefore, different fixed-point wave functions are classified by $\cal H^3[G, \mathrm U(1)]$, where the $G$ action on U(1) is specified by the function $s(\mb g)$ in Eq.~\ref{eq:sg-action}. For onsite symmetries, this result reproduces the group-cohomology classification of 2D SPT states, and for mirror-reflection symmetry, this is consistent with the conjecture that mirror-SPT states are also classified by group cohomologies, where mirror-reflections act as anti-unitary operations on the U(1) coefficients.

\section{Symmetry-Enriched Gauge Theories}
\label{sec:segt}

In this section we will study symmetry-enriched (untwisted) discrete gauge theories $D(N)$, where the gauge group $N$ is Abelian. Applying the formalism in Sec. \ref{sec:fpwf}, we first outline how to construct general SETs in $D(N)$ in which symmetries do not permute charges with fluxes. Then for the cases where symmetries do not permute any anyons (except that space-time symmetry operations that reverse orientation, such as time-reversal and mirror-reflection symmetries, have to map anyons to ones with opposite topological spins), we explicitly write all the data necessary for the construction and show that all non-anomalous symmetry fractionalization classes can be realized. We further derive a sufficient and necessary condition for a symmetry fractionalization class to be non-anomalous.

We first review briefly the topological order in $D(N)=\cal{Z}(\text{Vec}_N)$. The underlying UFC $\text{Vec}_N$ is defined as follows: the labels are group elements of $N$ and the fusion rules are given by group multiplications. In particular, the $F$ symbols are all trivial:
\begin{equation}
	[F^{abc}_{a\times b\times c}]_{a\times b,b\times c}=1\: a,b,c\in N.
	\label{}
\end{equation}
Anyons in the discrete gauge theory are labeled as dyons $([a], \pi_a)$, where the ``magnetic flux'' is a conjugacy class $[a]$ of $N$, and the ``electric charge'' is an irreducible representation $\pi_a$ of the centralizer group $C_a$. Since $N$ is Abelian, each conjugacy class is a singleton $\{a\}$. The centralizer group $C_a$ is always $N$, whose irreducible one-dimensional representations are given by the characters $\lambda:N\rightarrow \mathrm{U}(1)$. So, we can simply label the dyons as a pair $(a, \lambda)$. The characters form a group, called the character group $\hat{N}$ which is isomorphic to $N$. So, the anyons in $D(N)$ form a fusion group $N\times\hat{N}$.

We further assume that symmetry operations permute anyons in a simple way: they can permute different types of gauge fluxes arbitrarily as long as the fusion rules are preserved. In other words, the permutations are automorphisms of the group $N$. Their actions on gauge charges can then be deduced, since their braiding statistics with gauge fluxes have to be invariant  under the symmetry (complex conjugated when the symmetry is anti-unitary). In particular, we do not consider the so-called ``electro-magnetic duality''(EMD) symmetry in this section, which permutes charges with fluxes. We will, however, study an example of the EMD symmetry in $D(\mathbb{Z}_2)$ in Sec. \ref{sec:z2tc-unitary}.

This limited form of symmetry actions is sufficient to include the aforementioned natural action of time-reversal and mirror-reflection symmetries. In terms of the $G$-extension of the UFC $\cal C$ described in Sec.~\ref{sec:fpwf}, such SET phases can be described an extension in which all objects in $\cal C_G$ have quantum dimensions equal to $1$. In the following we refer to $\cal C_G$ with this property as being Abelian. As it turns out,  $\cal C_G$ can be thought as a group $\mathcal{G}$ (with multiplication given by fusion), and importantly $N$ is a normal subgroup of $\mathcal{G}$. The reason is that because of the $G$-grading, for any $n\in N$ and $x\in \mathcal{G}$, $xnx^{-1}$ has to be in $N$ regardless of the grading of $x$, i.e., $N$ is invariant under conjugation.  The problem of determining the fusion rules of $\cal C_G$ becomes the problem of finding group extensions. Physically, since symmetries only permute among charges (fluxes), for Abelian gauge theories such permutations have to be uniquely induced from a group automorphism in order to preserve the fusion rules (for non-Abelian gauge groups, such permutations are induced from outer automorphisms), and therefore $\cal C_G$ must be group extensions. However, one should notice that we do not need to assume that $\cal C_G$ has Abelian group multiplication; in fact in general they do not. Lattice models of this kind of symmetry-enriched gauge theory were considered also in \Ref{Tarantino_arxiv}.

\subsection{Group extension}
\label{sec:group-ext}

We start with some general remarks on $G$ extensions of a finite Abelian group $N$.  Such group extensions correspond to SETs of $D(N)$ where $G$ does not permute charges with fluxes, but permutations among fluxes and charges are allowed.

Let us now briefly review mathematically how to classify group extensions. Given an Abelian group $N$ and another finite group $G$, to specify a group extension we first need to pick a homomorphism $\varphi: G\rightarrow \text{Aut}(N)$. Here $\text{Aut}(N)$ is the automorphism group of $N$, i.e., all permutations of elements of $N$ preserving the group multiplications. Then possible extensions are classified by $[\nu]\in\cal{H}^2_\varphi[G, N]$.
More explicitly, we can parametrize the elements of $\cal{G}$ as $a_\mb{g}$ where $a\in N, \mb{g}\in G$, with the group multiplication law given by
\begin{equation}
	a_\mb{g}\times b_\mb{h}=[a\varphi_\mb{g}(b)\nu(\mb{g,h})]_{\mb{gh}}.
	\label{}
\end{equation}
Associativity of group multiplication in $\cal{G}$ requires
 \begin{equation}
	 \varphi_\mb{g}[\nu(\mb{h,k})]\nu(\mb{g,hk})
   =\nu(\mb{g,h})\nu(\mb{gh,k}),
	 \label{eq:fusion-nu}
 \end{equation}
 which is of course, the twisted $2$-cocycle condition. In other words, given $\varphi$ and $\nu$, we have explicitly constructed the multiplication table of the group $\mathcal{G}$.

Physically, $\varphi$ and $\nu$ determine how symmetry acts on anyons in $D(N)$.  For each $\mb{g}\in G$, we have an automorphism of $N$ denoted by $\varphi_\mb{g}$. $\varphi_\mb{g}$ induces canonically a permutation action $\rho_\mb{g}$ on the anyons in $D(N)$, as $\rho_\mb{g}\big( (a, \lambda)\big)= (\varphi_\mb{g}(a), \lambda')$, where the new character $\lambda'$ is defined by $\lambda'(b)=\lambda^{s(\mb{g})}\big(\varphi_\mb{g}(b) \big), b\in N$. One can check that this transformation preserves all fusion rules as well anyon braiding statistics. We notice that the factor $s(\mb g)$ ensures that the anyons $(a,\lambda)$ and $(\varphi(a),\lambda^\prime)$ has the same (opposite) topological spin, if $s(\mb g)=1$ or $\ast$, respectively.

To illustrate, let us consider an example with $N=\mathbb{Z}_n$ and $G=\mathbb{Z}_2$. We will denote the elements of $\mathbb{Z}_n$ by $j=0,1,\dots, n-1$ and the multiplication is $j\times k = (j+k)\,\text{mod }n$. Choose the symmetry action on $N$ to be $\varphi_\mb{g}(j)=-j$ for $j\in N$. The extension of $\mathbb{Z}_n$ by $\mathbb{Z}_2$ with the given action is actually isomorphic to the dihedral group $\mathbb{D}_{2n}$. We can also obtain the action of the symmetry on the quantum double $D(\mathbb{Z}_n)$. Label the quasiparticles by $(j,q)$ where $j$ is the flux and $q$ is the charge (corresponding to a character $\chi_q(j)=e^{\frac{2\pi iqj}{n}}$). Under the symmetry, we have
\begin{equation}
	(j, q)\rightarrow (-j, -q).
	\label{}
\end{equation}
This is in fact the charge conjugation symmetry in $D(\mathbb{Z}_n)$.

On the other hand, $\nu$ accounts for fractionalization of $G$ on the charges of $N$. We will come back to this point later.

From now on in this section, for clarity we will focus on the case of $\varphi=1$, i.e., gauge fluxes are not permuted by $G$ at all, while $G$ acts on gauge charges as $\lambda^\prime = \lambda^{s(\mb g)}$, to illustrate our approach.

\subsection{$F$ symbols and group cohomology classes}
\label{sec:cohomology}

Next, we discuss possible $F$ symbols compatible with the fusion rule given by the group structure of $\cal G$.
Because the all objects are Abelian, the $F$ symbol $[F^{\ag ag\ag bh\ag ck}_{a_\mb{g}\times b_\mb{h}\times \ag ck}]_{\ag ag\times \ag bh, \ag bh\times \ag ck}$ only depends on $\ag ag$, $\ag bh$, and $\ag ck$, and we will write it as $ \omega(\ag ag,\ag bh,\ag ck)$, where $\ag ag$, $\ag bh, \ag ck\in \cal G$. Furthermore, the unitarity condition implies that $\omega(\ag ag, \ag bh, \ag ck)$ is a U(1) phase factor. Therefore, the $F$ symbol can be viewed as a 3-cochain.

In terms of the 3-cochain $\omega(\ag ag, \ag bh, \ag ck)$, the twisted pentagon equation in Eq.~\eqref{eq:pentagon-t} becomes the following cocycle equation,
\begin{equation}
	\label{eq:cocycle3}
	\begin{split}
		d\omega &(\ag ag,\ag bh,\ag ck,\ag dl)\\
	&=\frac{\omega(\ag ag,\ag bh,\ag ck)
	\omega(\ag ag,\ag bh\times\ag ck,\ag dl)
	{}^{\mb g}\omega(\ag bh,\ag ck,\ag dl)}
	{\omega(\ag ag\times\ag bh,\ag ck,\ag dl)
	\omega(\ag ag,\ag bh, \ag ck\times\ag dl)}\\
	&=1.
	\end{split}
	\end{equation}
Similarly, the phase factor $[u^{\ag ag\ag bh}_{\ag ag\times \ag bh}]$ defined in Eq.~\eqref{eqn:gauge-F} only depends on the first two elements, and can be viewed as a 2-cochain $[u^{\ag ag\ag bh}_c]=u(\ag ag, \ag bh)$. Correspondingly, the gauge equivalence condition in Eq.~\eqref{eq:v-equiv} becomes the following coboundary equivalence:
\begin{equation}
	\label{eq:coboundary3}
	\omega\rightarrow \omega\cdot du, \quad
	du(\ag ag,\ag bh,\ag ck)=
	\frac{{}^{\mb g}u(\ag bh,\ag ck)u(\ag ag,\ag bh\times\ag ck)}
	{u(\ag ag,\ag bh)u(\ag ag\times\ag bh, \ag ck)}.
\end{equation}

In the above two equations, ${}^\mb{g}\omega$ and ${}^\mb{g} v$ denote the symmetry actions discussed in Sec.~\ref{sec:sym-act-F}, i. e., time-reversal and mirror-reflection symmetries act by complex conjugation. Therefore, the $F$ symbols are classified by the twisted third group cohomology $\cal H^3[\cal G, \mathrm U(1)]$, with the nontrival group actions on $\mathrm{U}(1)$.

\subsection{Symmetry fractionalization and anomaly}
\label{sec:qn-sf}

In this section, we have seen that without any anyon permutations, such extensions are classified by a second group cohomology class $[\nu]\in\cal H^2[G, N]$ specifying the group extension, and then a third group cohomology class $[\omega]\in\cal H^3[\cal G, \mathrm{U}(1)]$ for the $F$ symbols. Now, we explain how these two pieces of data together can encode all possible symmetry fractionalization classes.

Generally speaking, symmetry fractionalization in a topological phase is classified by $\cal H^2_{\rho}[G, \cal A]$~\cite{SET1}, where $\cal A$ is the fusion group of the Abelian anyons, and the action of $G$ on $\cal A$ is indicated explicitly by the subscript $\rho$. In the Abelian gauge theory $D(N)$, $\cal A=N\times\hat N$, where $N$ and $\hat N$ denote gauge fluxes and charges, respectively. According to our assumptions in Sec.~\ref{sec:group-ext}, $G$ acts trivially on $N$, and acts on $\hat N$ as $\rho_{\mb g}(\lambda)=\lambda^{s(\mb g)}$. Since the actions we consider also factor through, the second group cohomology $\cal H^2[G, N\times\hat N]$ can be decomposed to $\cal H^2[G, N]\times\cal H^2_\rho[G, \hat N]$. In other words, a symmetry fractionalization class $\coho{w}\in\cal H^2_\rho[G, N\times\hat N]$ can be separated into two parts, $\coho{w}=\coho{w}_e\times\coho{w}_m$, where $\coho{w}_e\in\cal H^2[G, N]$, and $\coho{w}_m\in\cal H^2_\rho[G, \hat N]$. Physically, $\coho{w}_e$ and $\coho{w}_m$ represent symmetry fractionalization classes of gauge charges and fluxes, respectively. As we shall explain below, these two symmetry fractionalization classes are encoded differently in the data of $\cal C_G$: $\coho{w}_e$ is encoded in the group extension (hence the same notation), while $\coho{w}_m$ is encoded in the third group-cohomology class $[\omega]\in\cal H^3[\cal G, \mathrm{U}(1)]$.

As introduced in Sec.~\ref{sec:group-ext}, $\nu$ is used to determine the group extension. Therefore, it appears in the fusion rule of $\cal C_G$, as shown in Eq.~\eqref{eq:fusion-nu}. This indicates that, when fusing two domain walls carrying group elements $\mb f$ and $\mb g$, respectively, the gauge flux on the fused domain wall is changed by $\nu(\mb f, \mb g)$. It is well known that such fusion rules reflect the symmetry fractionalization class of the gauge charges, so we should identify $\coho{w}_e\equiv \nu$.

The symmetry fractionalization of the gauge flux, $\coho{w}_m$, is encoded in the cocycle $\omega\in\cal H^3[\cal G, \mathrm U(1)]$. In particular, we consider $3$-cocycles that take the following form:
\begin{equation}
	\omega(a_\mb{g},b_\mb{h},c_\mb{k})=\chi_c(\mb{g,h})\alpha(\mb{g,h,k}).
	\label{eq:w=ca}
\end{equation}
As shown in Appendix~\ref{sec:param3}, the 3-cocycle condition $\di\omega=1$ implies the following properties of $\chi$: 1) $\chi$ is a character on $N$, i.e., $\chi_a(\mb{g,h})\chi_b(\mb{g,h})=\chi_{a\times b}(\mb{g,h})$. 2) $\chi_a$ satisfies the cocycle condition $\chi_a(\mb g,\mb h)\chi_a(\mb{gh},\mb k)=\chi_a^{s(\mb g)}(\mb h, \mb k)\chi_a(\mb g, \mb{hk})$. Therefore, $\chi$ can be viewed as a cocycle in $\cal H^2_\rho[G, \hat N]$, as $\coho{w}_m$ is. When $G$ is unitary, we can give a further argument for the identification of $\chi$ with the fractionalization class of gauge fluxes, by gauging the symmetry group $G$ and analyzing the anyon content of the gauged theory. The details of the argument can be found in Appendix \ref{sec:frac_gauge_theory}. We believe this is true for anti-unitary and mirror symmetries as well.

To construct a 2D SET, the twisted pentagon equation Eq.~\eqref{eq:pentagon-t} should be satisfied, which reduces to the $3$-cocycle condition $d\omega=1$ in $\mathcal{G}$. For cocycles of the form given in Eq.~\eqref{eq:w=ca}, this is equivalent to
\begin{equation}
	\chi_{\nu(\mb{k,l})}(\mb{g,h})=\di\alpha(\mb{g,h,k,l}).
	\label{eqn:anomaly-free}
\end{equation}
The proof of these assertions can be found in Appendix~\ref{sec:param3}.
For later reference, we mention that $\chi_{\nu(\mb{k,l})}(\mb{g,h})$ is generally a $4$-cocycle, and the above condition says the $4$-cocycle belongs to the trivial cohomology class in $\cal{H}^4[G, \mathrm{U}(1)]$.

For onsite unitary symmetries, Eq. \eqref{eqn:anomaly-free} is equivalent to the obstruction-vanishing condition obtained in \Ref{SET1} and \Ref{ChenASF2014} applied to $D(N)$. To see this, recall that the gauge anomaly for a symmetry fractionalization class is captured by the so-called obstruction class $O$~\cite{SET1,ChenASF2014}, which is a $4$-cocycle: $[O]\in \mathcal{H}^4[G, \mathrm{U}(1)]$. For $D(N)$, $O$ is given by
\begin{equation}
	O(\mb{g,h,k,l})=R^{\cohosub{w}(\mb{g,h}),\cohosub{w}(\mb{k,l})}.
	\label{eqn:obstruction}
\end{equation}
Here $R$ is the R symbol of the anyon theory for $D(N)$~\cite{Kitaev06a}, see below for an explicit expression. As shown in \Ref{SET1} and \Ref{ChenASF2014}, the symmetry can be gauged (hence the SET is free of gauge anomaly) if and only if $O$ belongs to a trivial cohomology class in $\cal H^4[G, \mathrm{U}(1)]$. For $D(N)$, one can choose $R^{(a, \lambda), (a',\lambda')}=\lambda(a')$. Using $\coho{w}=\coho{w}_e\times\coho{w}_m$ we can write
\begin{equation}
	R^{\cohosub{w}(\mb{g,h}),\cohosub{w}(\mb{k,l})}=[\coho{w}_m(\mb{g,h})]\big(\coho{w}_e(\mb{k,l})\big)=\chi_{\nu(\mb{k,l})}(\mb{g,h}).
	\label{}
\end{equation}
Therefore, Eq. \eqref{eqn:anomaly-free} is exactly the same condition as the vanishing of the obstruction class Eq. \eqref{eqn:obstruction}.

On the contrary, when the 4-cocycle on the left-hand side of Eq.~\eqref{eqn:anomaly-free} is a nontrivial cohomology class in $\cal H^4[G,\mathrm U(1)]$, Eq. \eqref{eqn:anomaly-free} has no solutions. Thus our construction can not represent such a fractionalization class parametrized by $\nu$ and $\chi$. In fact, for such fractionalization classes, one can use the obstructed pentagon equation in Eq.~\eqref{eq:pentagon-obs}, which takes the following form for the 3-cocyles in Eq.~\eqref{eq:w=ca}:
\begin{equation}
	\chi_{\nu(\mb{k,l})}(\mb{g,h})=\beta(\mb{g,h,k,l})\di\alpha(\mb{g,h,k,l}),
	\label{eqn:anomaly-beta}
\end{equation}
to get a consistent SET state if we choose $\beta(\mb{g,h,k,l})$ to be in the same cohomology class as $\chi_{\nu(\mb{k,l})}(\mb{g,h})$. The SET state then lives on the boundary of a nontrivial 3D SPT state characterized by $\beta$, which shows that the symmetry fractionalization class is anomalous. This result generalizes the obstruction-vanishing condition \eqref{eqn:obstruction}, which was obtained for onsite unitary symmetries, to anti-unitary symmetries for $D(N)$. We notice that \Ref{ChenASF2014} conjectured that the same formula of the obstruction class should apply to anti-unitary symmetries as well, and we give a strong evidence for the conjecture by proving it for $D(N)$. Furthermore, our approach establishes directly the bulk-boundary correspondence for this class of SET: the group cohomology class for the bulk SPT phase is given by $O(\mb{g,h,k,l})=\chi_{\nu(\mb{k,l})}(\mb{g,h})$.

\section{$\mathbb{Z}_2$ Toric Code with Unitary Onsite $\mathbb{Z}_2$ Symmetry}
\label{sec:z2tc-unitary}
\subsection{Classification}
\label{sec:z2tc-unitary-clas}
As a concrete example, let us consider the $\mathbb{Z}_2$ toric code with a global $\mathbb{Z}_2$ symmetry~\cite{SET1, lu2013}.  We shall carry out the classification explicitly and compare with the known results.

We need to determine $\mathbb{Z}_2$-extensions of the $\cal{C}=\text{Vec}_{\mathbb{Z}_2}=\{I, e\}$ category. To classify the extensions, we recall that the total quantum dimension of $\cal{C}_\mb{g}$ must be equal to that of $\cal{C}_\mb{1}$, which is $2$. So there are two scenarios: (1) There are two labels in $\cal{C}_\mb{g}$ both with dimension $1$, denoted by $\sigma^+$ and $\sigma^-=\sigma^+\times e$. Furthermore, depending on whether $\sigma^+\times\sigma^+=I$ or $e$, the fusion rules can be regarded as group multiplications of $\mathbb{Z}_2\times\mathbb{Z}_2$ or $\mathbb{Z}_4$. (2) There is a single label $\sigma$ with quantum dimension $\sqrt{2}$, and the fusion rule has to be $\sigma\times\sigma=1+e$. This is the famous Ising fusion rules. We now consider the three possibilities:
\begin{enumerate}
	\item $\cal{C}_{\mathbb{Z}_2}=\text{Vec}_{\mathbb{Z}_2\times\mathbb{Z}_2}$. The F symbols of $\text{Vec}_{\mathbb{Z}_2\times\mathbb{Z}_2}$ are classified by $\cal{H}^3[\mathbb{Z}_2\times\mathbb{Z}_2, \mathrm{U}(1)]=\mathbb{Z}_2^3$. One of the $\mathbb{Z}_2$ factor corresponds to $\omega(e,e,e)=-1$, i.e., a double semion topological order. The other two $\mathbb{Z}_2$ factors correspond to $\kappa_{\sigma^+}=\omega(\sigma^+, \sigma^+, \sigma^+)=[F^{\sigma^+\sigma^+\sigma^+}_{\sigma^+}]_{II}=\pm 1$ and $\kappa_{\sigma^-}=\omega(\sigma^-, \sigma^-, \sigma^-)=\pm 1$. Notice that the labeling in the $\cal{C}_\mb{g}$ sector is arbitrary; one is free to relabel $\sigma^\pm\rightarrow \sigma^\mp$. So there are only two nontrivial distinct extensions. One of them is $(\kappa_{\sigma^+},\kappa_{\sigma^-})=(1,-1)$, and physically it can be thought as a ``trivial'' SET with an additional layer of a $\mathbb{Z}_2$ SPT phase. The other one $(\kappa_{\sigma^+},\kappa_{\sigma^-})=(-1,-1)$ describes a SET where the $e$ (or $m$) particle carries a ``half'' charge under the $\mathbb{Z}_2$ symmetry. We notice that such SETs have been previously constructed in commuting-projector models~\cite{SnakeMonster, Ware_unpub}.
	\item $\cal{C}_{\mathbb{Z}_2}=\text{Vec}_{\mathbb{Z}_4}$. F symbols are classified by $\cal{H}^3[\mathbb{Z}_4, \mathrm{U}(1)]=\mathbb{Z}_4$. Representative $3$-cocycles are $\omega(a,b,c)=e^{\frac{\pi in}{8}a(b+c-[b+c]_4)}$~\cite{Propitius1995}, where we denote $[0]=1, [1]=\sigma^+, [2]=e, [3]=\sigma^-$, and $[a+b]_4$ means $(a+b)\,\text{mod }4$. Requiring $\omega([2],[2],[2])=1$ we find $n=0$ or $2$. The $n=0$ extension also yields a SET where $e/m$ carries a half $\mathbb{Z}_2$ charge, while in the $n=2$ SET both $e$ and $m$ carry half $\mathbb{Z}_2$ charges, corresponding to $\coho{w}(\mb{g,g})=\psi$.
	\item $\cal{C}_{\mathbb{Z}_2}=\text{Ising}$. As we shall see explicitly below, this extension corresponds to a SET where the $\mathbb{Z}_2$ symmetry permutes $e$ and $m$. This is known as the electro-magnetic duality symmetry. There are two gauge-inequivalent F symbols, distinguished by $[F^{\sigma\sigma\sigma}_{\sigma}]_{II}=\pm 1$.
\end{enumerate}
The resulting classification agrees completely with the one obtained in \Ref{SET1} based on $G$-crossed braided tensor category, as well as the Chern-Simons field theory analysis in \Ref{lu2013}.

\subsection{$\mathbb{Z}_2$ symmetry fractionalization}
First let us consider the example of $\mathcal{C}_G=\text{Vec}_{\mathbb{Z}_4}$ and a trivial $F$ symbol. The gauged model is simply the $\mathbb{Z}_4$ toric code, so we will formulate the un-gauged model on a square lattice.
There is a $\mathbb{Z}_4$ spin on each edge of the lattice, and a $\mathbb{Z}_2$ spin in each plaquette. We define $U_e\ket{n}_e=i^n\ket{n}_e, V_e\ket{n}_e=\ket{n+1}_e$, where $n\in\mathbb{Z}_4$ and $\ket{n}_e$ represents the basis states on an edge $e$.

 The following edge projectors are added to the Hamiltonian:
\begin{equation}
	Q_e=\frac{1+{U_e^2}}{2} \frac{1+\tau_p^z\tau_q^z}{2} + \frac{1-{U_e^2}}{2} \frac{1-\tau_p^z\tau_q^z}{2},
	\label{eqn:Z2SET1}
\end{equation}
where $p$ and $q$ denote plaquettes adjacent to the edge $e$. The projector imposes the $\mathbb{Z}_2$-grading on the edges.

We also have the vertex and plaquette terms:
\begin{equation}
	H=-\sum_v (A_\vr + A_\vr^\dag)-   \sum_\vr \tau_\vr^x(B_\vr + B_\vr^\dag)-\sum_eQ_e.
	\label{eqn:Z4}
\end{equation}
Here the vertex operator $A_\vr$ is given by $A_\vr=  U_{\vr,x} U_{\vr, y} U_{\vr-\hat{x}, x}^\dag U_{\vr-\hat{y},y}^\dag $, and the plaquette operator $B_\vr = V_{\vr,x}V_{\vr+\hat{x},y}V_{\vr+\hat{y},x}^\dag V_{\vr,y}^\dag $. We label the edges as $\vr, \mathbf{e}$, i.e., the edge connecting $\vr$ and $\vr+\mathbf{e}$, where $\mathbf{e}=\hat{x},\hat{y}$ are the two basis vectors of the square lattice. Correspondingly, we assign a direction to the edge pointing from $\vr$ to $\vr+\mathbf{e}$.
The global $\mathbb{Z}_2$ symmetry in the model is defined as $X=\prod_p \tau^x_p$. For comparison, the Hamiltonian of the original $\mathbb{Z}_4$ toric code, which can be thought as the gauged SET, reads
\begin{equation}
	H_{\mathbb{Z}_4}=-\sum_\vr (A_\vr + A_\vr^\dag)-   \sum_\vr (B_\vr + B_\vr^\dag).
	\label{}
\end{equation}

We first prove that if we break the $\mathbb{Z}_2$ symmetry the Hamiltonian Eq. \eqref{eqn:Z4} is adiabatically connected to a $\mathbb{Z}_2$ toric code. We add to the Hamiltonian a ``Zeeman'' term:
\begin{equation}
	H'= H - J_z\sum_\vr \tau_\vr^z.
	\label{}
\end{equation}
Imagine $J_z$ is turned on adiabatically. The $\tau_\vr^z$ term commutes with the vertex and edge terms in Eq. \eqref{eqn:Z4}, as well as plaquette operators except the one at $\vr$. So to study the spectrum of the model, we can fix a single plaquette, and define:
\begin{equation}
	h_\vr = -\tau_\vr^x(B_\vr + B_\vr^\dag) - J_z \tau_\vr^z.
	\label{}
\end{equation}
To solve for the spectrum of $h_\vr$, we notice that
\begin{equation}
	h_\vr^2= B_\vr^2+\big(B_\vr^\dag\big)^2 + 2 + J_z^2.
	\label{eqn:hsq}
\end{equation}
$B_\vr^2$ commutes with all terms in the Hamiltonian $H'$, so it is a conserved quantity. Since when $J_z=0$ we have $B_\vr^2=1$, we can set the value of $B_\vr^2$ to $1$ in Eq. \eqref{eqn:hsq}. Therefore the eigenvalues of $h_\vr$  are $\pm \sqrt{J_z^2+4}$. In particular, the gap between the ground state and the excited state of $h_\vr$ never closes regardless of the value of $J_z$. Therefore, we have constructed an adiabatic path between the $J_z=0$ and $J_z\rightarrow \infty$ states. When $J_z\rightarrow \infty$, all spins are polarized $\tau_\vr^z=1$, thus only labels from the identity sector $\mathcal{C}_\mb{1}$ are allowed on the lattice. In this limit our construction apparently reduces to the usual $\mathbb{Z}_2$ toric code. This calculation shows that once we break the symmetry, the model Eq. \eqref{eqn:Z4} is adiabatically connected to a $\mathbb{Z}_2$ toric code.

We now describe the quasiparticles in the SET phase. It is actually quite instructive to start from the quasiparticle string operators in the $\mathbb{Z}_4$ toric code, and see how they are modified in the SET phase. In the $\mathbb{Z}_4$ toric code, there are two elementary types of string operators: ``electric'' strings can be written for a path $P$ on the lattice:
\begin{equation}
	W_{\tilde{e}}(P)=\prod_{e\in P} V_e^{s_e},
	\label{eqn:e-string}
\end{equation}
Here $s_e=+1$ ($-1$) if the direction of the string is parallel(anti-parallel) to the direction of the edge.

The ``magnetic'' strings are defined on a path $P^*$ in the dual lattice. To illustrate, let us consider a path $P^*$ parallel to the $x$ direction:
\begin{equation}
	W_{\tilde{m}}(P^*)=\prod_{e\in P^*} U_e.
	\label{}
\end{equation}

In the SET phase, we first notice that there do not exist any open $W_{\tilde{e}}(P)$ strings. The reason is that $V$'s have to be accompanied with spin flips to stay within the low-energy subspace defined by $H_1$, so the path $P$ in the definition Eq. \eqref{eqn:e-string} must be aligned with domain walls of the $\mathbb{Z}_2$ spins, which are always closed. However,
\begin{equation}
	W_{\tilde{e}^2}\equiv W_{\tilde{e}}^2
	\label{}
\end{equation}
 remains as a deconfined string since $V^2$ do not change the $G$-graded sectors of the edge labels, and should be identified with the $e$ particle in the SET: $\tilde{e}^2\sim e$.

On the other hand, we observe that $W_{\tilde{m}}^2$ become a ``trivial'' string. This is because the edge projectors $Q_e$ identify $U_e^2$ with $\tau_p^z\tau_q^z$, where $p$ and $q$ denote the two plaquettes adjacent to the edge $e$, and as a result an open $W_{\tilde{m}}^2(P^*)$ string acting on the ground state is identical to the product of the two $\tau^z$ at the two ends of the string $P^*$. Thus $\tilde{m}^2$ is now a local excitation. More precisely, $\tilde{m}^2$ becomes the charge of the global $\mathbb{Z}_2$ symmetry, so that moving $\tilde{m}^2$ detects the $\mathbb{Z}_2$ symmetry domain walls along the way. $W_{\tilde{m}}\sim W_{\tilde{m}}^3$ is still a nontrivial string, but now with $\mathbb{Z}_2$ fusion rules. Therefore in a very precise sense the $\tilde{m}^2$ particles are condensed, which has the effect of confining $\tilde{e}$ and $\tilde{e}^3$ while identifying $\tilde{m}$ with $\tilde{m}^3$.

To summarize, we have found the following relations between the string operators in the SET model and those of the gauged model:
\begin{equation}
	W_{\mathds{1}}\sim W_{\tilde{m}^2}, W_{e}\sim W_{\tilde{e}^2}, W_m\sim W_{\tilde{m}}.
	\label{}
\end{equation}
Of course, this is what we expect from anyon condensation: when $\tilde{m}^2$ condenses, the remaining deconfined anyons all have the form $\tilde{e}^{2a}\tilde{m}^b$ where $a, b=0,1$. All the other particles, such as $\tilde{e}$, are confined because they have nontrivial braiding statistics with $\tilde{m}^2$. In our picture, it is simply because $\tilde{e}$ string has to be accompanied by spin flips and is therefore forced to align with the domain walls, as already explained above.

To extract the symmetry quantum numbers of anyons, we need to find the localized form of the symmetry transformation $X$~\cite{essin2013, SET1, Tarantino_arxiv}. Let us consider locally flipping a $\mathbb{Z}_2$ spin in the plaquette $\vr$. In order to stay in the restricted Hilbert space defined by the projector $Q_e=1$, one also needs to change the spins on the edges of the plaquette by $V$ or $V^\dag$. We choose the local symmetry action projected to the low-energy subspace to takes the form
\begin{equation}
	\cal{U}_{X}=\tau_{\vr}^x V_{\vr,x}V_{\vr+\hat{x},y}V_{\vr+\hat{y},x}^\dag V_{\vr,y}^\dag,
	\label{}
\end{equation}
which is simply a single plaquette operator $\tau_x B_\vr$ of the SET model. The reason to choose this particular combination of $V$'s is that they commute with the vertex terms. However, $\cal{U}_{X}$ fails to be an exact $\mathbb{Z}_2$ operator, since $\cal{U}_{X}^2=(V_{\vr,x}V_{\vr+\hat{x},y}V_{\vr+\hat{y},x}^\dag V_{\vr,y}^\dag)^2$ is nothing but a $\tilde{e}^2=e$ string around the plaquette [notice $V^2=(V^\dag)^2$],  which implies $\big(\cal{U}_{X}^{(e)}\big)^2=1$, $\big(\cal{U}_{X}^{(m)}\big)^2=-1$, here $\cal{U}_X^{(a)}$ refers to $\cal{U}_X$ acting on a region containing a quasiparticle of type $a$. Therefore, the fractionalization class is $\coho{w}(\mb{g,g})=e$.

The example illustrates some general features of the construction for a unitary symmetry group. We see that one can build up the quasiparticles of the SET model from those of the ``parent'' gauged model, corresponding to the condensation of $G$-charges in the gauged model, as we discussed in Sec. \ref{sec:parent_ham}. Furthermore, the local symmetry actions can be found exactly due to the fixed-point nature of the wavefunction, which are basically the plaquette operators in the parent Hamiltonian. Physically, this is because for unitary symmetries the localized symmetry transformation on a region can be implemented by transporting a symmetry defect around the region, which is precisely the plaquette operator fusing a string of ``gauge flux'' to the edges of a plaquette in this model.

As we mentioned in Sec. \ref{sec:z2tc-unitary-clas}, for the $\mathbb{Z}_4$ fusion rule there is another extension $\mathrm{U}(1)_4$. We will not go into details into the construction, but a similar analysis can be done to confirm that the extension realizes a $\mathbb{Z}_2$ toric code where both $e$ and $m$ carry half $\mathbb{Z}_2$ charge, which has eluded previous constructions.

\subsection{Electro-magnetic duality symmetry}
\label{sec:emd}
We now turn to the Ising extension. In the following, we draw the three types of strings:
\begin{equation}
	\begin{split}
		I:\:
    \begin{pspicture}[shift=-0.2](0,0.0)(0,0.6)
        \psid(0,0)(0,0.6)
    \end{pspicture}\, \\
     \sigma: \:
    \begin{pspicture}[shift=-0.2](0,0.0)(0,0.6)
        \pssigma(0,0)(0,0.6)
    \end{pspicture}\, \\
    \psi: \:
    \begin{pspicture}[shift=-0.2](0,0)(0,0.6)
        \pspsi(0,0)(0,0.6)
    \end{pspicture}\, \\
\end{split}
	\label{}
\end{equation}
 Notice that we rename the label $e$ as $\psi$, to be consistent with the usual labeling of the Ising category. The nontrivial $F$ symbols are given by
	\begin{equation}
		\begin{gathered}
			{}[F^{\sigma\psi\sigma}_{\psi}]_{\sigma\sigma}=[F^{\psi\sigma\psi}_\sigma]_{\sigma\sigma}=-1,\\
		[F^{\sigma\sigma\sigma}_\sigma]_{ab}=\frac{\kappa_\sigma}{\sqrt{2}}\begin{pmatrix}
			1 & 1 \\
			1 & -1
		\end{pmatrix}.
		\end{gathered}
				\label{}
	\end{equation}
	All other F symbols are $1$ as long as the fusions involved are allowed.

	There are two gauge-inequivalent $F$ symbols, distinguished by $\kappa_\sigma=\pm 1$. The UFC corresponding to $\kappa_\sigma=-1$ is also known as the $\mathrm{SU}(2)_2$ category.  We will focus on the $\kappa_\sigma=1$ case in this section. The $F$ symbols of the Ising category have tetrahedral symmetry, so in drawing the pictures one can freely bend lines or rotate vertices.

Before we write the SET model, it is convenient to first have the Hamiltonian of the Levin-Wen model for $\mathcal{Z}(\text{Ising})$. On a trivalent lattice, we associate each edge with three types of strings labeled as $I,\sigma, \psi$, and the Hamiltonian consists of the vertex and plaquette terms following the standard construction:
\begin{equation}
	H_\text{gauged}=-\sum_v Q_v-\sum_p B_p.
	\label{}
\end{equation}
Here $B_p=\frac{1}{4}\sum_p(\mathds{1}+B_p^\psi + \sqrt{2}B_p^\sigma)$.

In the SET phase we also have $\mathbb{Z}_2$ spins in the plaquettes. Again, we have edge projectors to enforce $\mathbb{Z}_2$-grading:
\begin{equation}
	Q_e=\ket{\sigma}_e\bra{\sigma}_e\frac{1-\tau_{p_e}^z\tau_{q_e}^z}{2} + (\mathds{1}-\ket{\sigma}_e\bra{\sigma}_e)\frac{1+\tau_{p_e}^z\tau_{q_e}^z}{2}.
	\label{}
\end{equation}
So, the $\sigma$ strings comfort to the domain walls of the spins.

We also need to modify the plaquette terms accordingly, and the final Hamiltonian becomes
\begin{equation}
	H=-\sum_v Q_v-\frac{1}{4}\sum_p(\mathds{1}+B_p^\psi+\sqrt{2}\tau_p^x B_p^\sigma)-\sum_e Q_e.
	\label{eqn:Hem}
\end{equation}
It is straightforward to check that all the terms in Eq. \eqref{eqn:Hem} commute with each other.
The Hamiltonian has a global Ising symmetry: $X = \prod_{p}\tau_p^x$.

Simiar to the analysis of Eq. \eqref{eqn:Z4}, one can show that the Hamiltonian is adiabatically connected to a $\mathbb{Z}_2$ toric code if the $\mathbb{Z}_2$ symmetry is broken. Again, we add a Zeeman term $-J_z\sum_p\tau_p^z$ to the Hamiltonian, and focus on one plaquette:
\begin{equation}
	h_p = -\frac{B_p^\sigma}{2\sqrt{2}}\tau_p^x - J_z\tau_p^z.
	\label{}
\end{equation}
Using $(B_p^\sigma)^2=\mathds{1}+B_p^\psi$, we have
\begin{equation}
	h_p^2= \frac{\mathds{1}+B_p^\psi}{8}+J_z^2.
	\label{}
\end{equation}
Because $B_p^\psi$ commutes with every other term in the Hamiltonian, the value is fixed. Further, because $(B_p^\psi)^2=\mathds{1}$, it can only take $\pm 1$. We see that the spectrum of $h_p$ remains gapped when $J_z$ is increased. So, the ground state of the SET Hamiltonian \eqref{eqn:Hem} and the ground state of a plain Levin-Wen Hamiltonian is adiabatically connected once the symmetry is broken.

\subsubsection{Symmetry action on quasiparticles}
\label{sec:string}
We now analyze the SET order in the model \eqref{eqn:Hem}. Since we obtain this model by ``un-gauging'' the parent $\cal{Z}(\text{Ising})$ Hamiltonian, one can expect that if we gauge the $\mathbb{Z}_2$ symmetry we will get back the parent state. This is consistent with $\mathbb{Z}_2$ symmetry permuting $e$ and $m$ anyons~\cite{SET1, SET2}. Below, we will explicitly construct the quasiparticle states in the model and determine the symmetry action directly.

Quasiparticles in Levin-Wen models are associated with string operators. For example, an open string operator acting on the ground state creates a particle-anti particle pair at the end of the string. We will briefly review the definition of string operators~\cite{Levin05a}. A string operator $W_{\bm{a}}$ is represented by a directed string acting along an open or closed path on the lattice (in fact, on the \emph{fattened} lattice), as shown in Fig. \ref{fig:string-op}. Graphically, we draw a string lying on top of the graph state to represent the string operator. Its action on a given basis state is defined using the following rule to resolve each overcrossing:
\begin{equation}
	\begin{gathered}
		\begin{pspicture}[shift=-0.3](-0.2,-0.4)(1.2,0.4)
            \scriptsize
            \psframe*[linecolor=lightgray, framearc=0.3](0,-0.45)(1,0.45)
            \psstring(0,0)(1,0)\rput(0.2,0.15){$i$}
            \psline{->}(0.8,0)(0.9,0)
            \psstring[border=1.5pt, bordercolor=lightgray](0,-0.2)(0.5,-0.2)(0.5,0.2)(1,0.2)\rput(0.2,-0.35){$\bm{a}$}
            \psline{->}(0.8,0.2)(0.9,0.2)
        \end{pspicture}
        =\sum_{jst} \Omega^j_{\bm{a},ist}
        \begin{pspicture}[shift=-0.3](-0.2,-0.4)(1.2,0.4)
            \scriptsize
            \psframe*[linecolor=lightgray, framearc=0.3](0,-0.45)(1,0.45)
            \psstring[ArrowInside=->](0.7,0)(1.0,0)\rput(0.9,-0.15){$i$}
            \psstring[ArrowInside=->](0.3,0)(0.7,0)\rput(0.5,-0.15){$j$}
            \psstring[ArrowInside=->](-0.0,0)(0.3,0)\rput(0.1,0.20){$i$}
            \psstring(0.3,0)(0.3,-0.2)(0,-0.2)\rput(0.1,-0.35){$s$}
            \psline{->}(0.1,-0.2)(0.2,-0.2)
			\psstring(1.0,0.2)(0.7,0.2)(0.7,0)\rput(0.9,0.35){$t$}
            \psline{->}(0.8,0.2)(0.9,0.2)
        \end{pspicture},\\
		        \begin{pspicture}[shift=-0.3](-0.2,-0.4)(1.2,0.4)
            \scriptsize
            \psframe*[linecolor=lightgray, framearc=0.3](0,-0.45)(1,0.45)
            \psstring(1,0)(0,0)\rput(0.2,-0.15){$i$}
            \psline{->}(0.8,0)(0.9,0)
            \psstring[border=1.5pt, bordercolor=lightgray](0,0.2)(0.5,0.2)(0.5,-0.2)(1,-0.2)\rput(0.2,0.3){$\bm{a}$}
            \psline{->}(0.8,-0.2)(0.9,-0.2)
        \end{pspicture}
        =\sum_{jst} \bar{\Omega}^j_{\bm{a},ist}
        \begin{pspicture}[shift=-0.3](-0.2,-0.4)(1.2,0.4)
            \scriptsize
            \psframe*[linecolor=lightgray, framearc=0.3](0,-0.45)(1,0.45)
            \psstring[ArrowInside=->](0.7,0)(1,0)\rput(0.9,0.15){$i$}
            \psstring[ArrowInside=->](0.3,0)(0.7,0)\rput(0.5,-0.15){$j$}
            \psstring[ArrowInside=->](0,0)(0.3,0)\rput(0.1,-0.15){$i$}
            \psstring(0.3,0)(0.3,0.2)(0,0.2)\rput(0.1,0.35){$s$}
            \psline{->}(0.1,0.2)(0.2,0.2)
            \psstring(1,-0.2)(0.7,-0.2)(0.7,0)\rput(0.9,-0.35){$t$}
            \psline{->}(0.8,-0.2)(0.9,-0.2)
        \end{pspicture}.
\end{gathered}
\end{equation}
We then join the string tails between adjacent overcrossings, discarding the diagrams where these joined strings do not match, and finally using the local moves to reduce the diagrams to the lattice state. The coefficients $\Omega_{\bm{a},sti}^j$ need to satisfy a set of consistency equations (basically to make sure that the string operator is path-independent)~\cite{Levin05a}, so that $W_{\bm{a}}$ commutes with the Levin-Wen Hamiltonian away from the ends of the string.

We can easily generalize the construction to the SET model, with some important differences:
\begin{enumerate}
	\item The coefficients are allowed to also depend on the group elements on the two sides of the $i$ string:
		\begin{equation}
			\begin{split}
		\begin{pspicture}[shift=-0.3](-0.2,-0.4)(1.2,0.4)
            \scriptsize
            \psframe*[linecolor=lightgray, framearc=0.3](0,-0.45)(1,0.45)
            \psstring(0,0)(1,0)\rput(0.2,0.15){$i$}
            \psline{->}(0.8,0)(0.9,0)
            \psstring[border=1.5pt, bordercolor=lightgray](0,-0.2)(0.5,-0.2)(0.5,0.2)(1,0.2)\rput(0.2,-0.35){$\bm{a}$}
            \psline{->}(0.8,0.2)(0.9,0.2)
			\rput(0.8, -0.3){$\mathbf{g}$}
        \end{pspicture}
		&=\sum_{jst} {}^{\mathbf{g}}\Omega^j_{\bm{a},ist}
        \begin{pspicture}[shift=-0.3](-0.2,-0.4)(1.2,0.4)
            \scriptsize
            \psframe*[linecolor=lightgray, framearc=0.3](0,-0.45)(1,0.45)
            \psstring[ArrowInside=->](0.7,0)(1.0,0)\rput(0.9,-0.15){$i$}
            \psstring[ArrowInside=->](0.3,0)(0.7,0)\rput(0.5,-0.15){$j$}
            \psstring[ArrowInside=->](-0.0,0)(0.3,0)\rput(0.1,0.20){$i$}
            \psstring(0.3,0)(0.3,-0.2)(0,-0.2)\rput(0.1,-0.35){$s$}
            \psline{->}(0.1,-0.2)(0.2,-0.2)
			\psstring(1.0,0.2)(0.7,0.2)(0.7,0)\rput(0.9,0.35){$t$}
            \psline{->}(0.8,0.2)(0.9,0.2)
        \end{pspicture},\\
		\begin{pspicture}[shift=-0.3](-0.2,-0.4)(1.2,0.4)
            \scriptsize
            \psframe*[linecolor=lightgray, framearc=0.3](0,-0.45)(1,0.45)
            \psstring(1,0)(0,0)\rput(0.2,-0.15){$i$}
            \psline{->}(0.8,0)(0.9,0)
            \psstring[border=1.5pt, bordercolor=lightgray](0,0.2)(0.5,0.2)(0.5,-0.2)(1,-0.2)\rput(0.2,0.33){$\bm{a}$}
            \psline{->}(0.8,-0.2)(0.9,-0.2)
			\rput(0.8, 0.2){$\mathbf{g}$}
        \end{pspicture}
		&=\sum_{jst} {}^{\mb{g}}\bar{\Omega}^j_{\bm{a},ist}
        \begin{pspicture}[shift=-0.3](-0.2,-0.4)(1.2,0.4)
            \scriptsize
            \psframe*[linecolor=lightgray, framearc=0.3](0,-0.45)(1,0.45)
            \psstring[ArrowInside=->](0.7,0)(1,0)\rput(0.9,0.15){$i$}
            \psstring[ArrowInside=->](0.3,0)(0.7,0)\rput(0.5,-0.15){$j$}
            \psstring[ArrowInside=->](0,0)(0.3,0)\rput(0.1,-0.15){$i$}
            \psstring(0.3,0)(0.3,0.2)(0,0.2)\rput(0.1,0.35){$s$}
            \psline{->}(0.1,0.2)(0.2,0.2)
            \psstring(1,-0.2)(0.7,-0.2)(0.7,0)\rput(0.9,-0.35){$t$}
            \psline{->}(0.8,-0.2)(0.9,-0.2)
        \end{pspicture}
		.
	\end{split}
\end{equation}

	\item $j$ and $i$ must have the same grading (so $s, t\in \mathcal{C}_{\mb{0}}$). Otherwise, the string has to be accompanied by a spin flip in the adjacent plaquettes to change the grading, and therefore the path of the string is forced to align with a domain wall of the spin configurations, which forbids open strings.
	\item Certain seemingly nontrivial string operators actually represent local excitations which transform nontrivially under the global symmetry group. Therefore, one needs to consider equivalence classes of string operators moding out those local symmetry charges.
\end{enumerate}

Once the (irreducible) string operators are obtained, one can compute the braiding and exchange statistics of the corresponding quasiparticle excitations. The topological twist is given by the following formula, generalizing the results of \Ref{Levin05a}:
\begin{equation}
	\theta_{\bm{a}}=\frac{\sum_{i\in \mathcal{C}_1} {}^{\mb{1}}\Omega_{\bm{a},i\bar{i}\bar{i}}^0 d_i}{\sum_{i\in\mathcal{C}_1} {}^{\mb{1}}\Omega_{\bm{a},0ii}^i d_i}.
	\label{eqn:string-twist}
\end{equation}
Notice that $\Omega_{\bm{a},0ii}^i=0,1$.
The topological $S$ matrix can be evaluated using the following formula:
\begin{equation}
	S_{\bm{a},\bm{b}}=\frac{1}{\mathcal{D}}\sum_{ijk}{}^{\mb{1}}\Omega_{\bm{a}, ijj}^k {}^{\mb{1}}\Omega_{\bm{b}, jii}^k d_id_j
	\label{eqn:string-Smatrix}
\end{equation}

We can directly solve for the string operators in the SET model. For pedagogical purposes, we will do it in a slightly different way.
There are nine string operators in the Ising Levin-Wen model~\cite{Kaushal}. Since $\mathcal{Z}(\text{Ising})=\text{Ising}\times \overline{\text{Ising}}$, the string operators can be labeled as $(a_1,a_2)$ where $a_1\in \{I,\sigma,\psi\}$. Using these string operators in the $\mathcal{Z}(\text{Ising})$ phase, we will see how they can be ``ungauged'' to give string operators in the SET model.

First, we consider the $(\psi, \psi)$ string:
\begin{equation}
	\begin{split}
	\begin{pspicture}[shift=-0.25](-0.1,-0.1)(0.6,0.6)
            \psid(0,0.5)(0.5,0)\psline[border=1.5pt](0,0)(0.5,0.5)
        \end{pspicture}
        &=
        \begin{pspicture}[shift=-0.25](-0.1,-0.1)(0.6,0.6)
            \psid(0,0.5)(0.2,0.3)\psid(0.3,0.2)(0.5,0)\psid(0,0.1)(0.2,0.3)(0.3,0.2)(0.5,0.4)
        \end{pspicture}\\
            \begin{pspicture}[shift=-0.25](-0.1,-0.1)(0.6,0.6)
                \pssigma(0,0.5)(0.5,0)\psline[border=1.5pt](0,0)(0.5,0.5)
            \end{pspicture}
            &=-
            \begin{pspicture}[shift=-0.25](-0.1,-0.1)(0.6,0.6)
                \pssigma(0,0.5)(0.5,0)\psid(0,0.1)(0.2,0.3)\psid(0.3,0.2)(0.5,0.4)
            \end{pspicture}\\
            \begin{pspicture}[shift=-0.25](-0.1,-0.1)(0.6,0.6)
                \pspsi(0,0.5)(0.5,0)\psline[border=1.5pt](0,0)(0.5,0.5)
            \end{pspicture}
            &=
			\begin{pspicture}[shift=-0.25](-0.1,-0.1)(0.6,0.6)
                \pspsi(0,0.5)(0.5,0)\psid(0,0.1)(0.2,0.3)\psid(0.3,0.2)(0.5,0.4)
            \end{pspicture}
		\end{split}
	\label{}
\end{equation}
Basically, whenever the string operator crosses a $\sigma$ string we pick up a $-1$ phase. One can check that this remains a string operator in the SET model: ${}^{\uparrow}\Omega_{ist}^j={}^{\downarrow}\Omega_{ist}^j=\Omega_{(\psi,\psi), ist}^j$. However, due to the edge projectors this is equal to $\tau_p^z\tau_q^z$,
which implies that the $(\psi,\psi)$ quasiparticles become local $\mathbb{Z}_2$ charges, i.e., $(\psi,\psi)$ is ``condensed''. As a result, $(\sigma, I), (\sigma,\psi), (I,\sigma)$ and $(\psi,\sigma)$ are all ``confined''. Looking at the string operators, all these four need to fuse a $\sigma$ string to the edges, which are not allowed in the SET phase.

The $(\sigma,\sigma)$ string in $\mathcal{Z}(\text{Ising})$ is defined by the following rules:
\begin{equation}
	\begin{split}
	\begin{pspicture}[shift=-0.25](-0.1,-0.1)(0.6,0.6)
            \psid(0,0.5)(0.5,0)\psline[border=1.5pt](0,0)(0.5,0.5)
        \end{pspicture}
        &=
        \begin{pspicture}[shift=-0.25](-0.1,-0.1)(0.6,0.6)
            \psid(0,0.5)(0.5,0)\psid(0,0.1)(0.2,0.3)\psid(0.3,0.2)(0.5,0.4)
        \end{pspicture}
        +
        \begin{pspicture}[shift=-0.25](-0.1,-0.1)(0.6,0.6)
            \psid(0,0.5)(0.2,0.3)\psid(0.3,0.2)(0.5,0)\pspsi(0,0.1)(0.2,0.3)(0.3,0.2)(0.5,0.4)
        \end{pspicture}\\
            \begin{pspicture}[shift=-0.25](-0.1,-0.1)(0.6,0.6)
                \pssigma(0,0.5)(0.5,0)\psline[border=1.5pt](0,0)(0.5,0.5)
            \end{pspicture}
            &=e^{\frac{\pi i}{4}}
            \begin{pspicture}[shift=-0.25](-0.1,-0.1)(0.6,0.6)
                \pssigma(0,0.5)(0.5,0)\psid(0,0.1)(0.2,0.3)\pspsi(0.3,0.2)(0.5,0.4)
            \end{pspicture}
            +e^{-\frac{\pi i}{4}}
            \begin{pspicture}[shift=-0.25](-0.1,-0.1)(0.6,0.6)
            \pssigma(0,0.5)(0.5,0)\pspsi(0,0.1)(0.2,0.3)\psid(0.3,0.2)(0.5,0.4)
                \end{pspicture}\\
            \begin{pspicture}[shift=-0.25](-0.1,-0.1)(0.6,0.6)
            \pspsi(0,0.5)(0.5,0)\psline[border=1.5pt](0,0)(0.5,0.5)
            \end{pspicture}
            &=-
            \begin{pspicture}[shift=-0.25](-0.1,-0.1)(0.6,0.6)
            \pspsi(0,0.5)(0.5,0)\psid(0,0.1)(0.2,0.3)\psid(0.3,0.2)(0.5,0.4)
            \end{pspicture}
            +
            \begin{pspicture}[shift=-0.25](-0.1,-0.1)(0.6,0.6)
            \pspsi(0,0.5)(0.2,0.3)(0,0.1)\pspsi(0.5,0)(0.3,0.2)(0.5,0.4)\psid(0.2,0.3)(0.3,0.2)
            \end{pspicture}
	\end{split}
\label{eqn:sigma-sigma-string}
\end{equation}
A noticeable feature of the rules is that if we ignore the diagram for crossing on $\sigma$ edge, the rules to resolve crossings essentially decompose into two sets: one is that we only use the first diagram on the right hand side, which will be refereed to as the $m$-type diagram, and the other is to use the second diagram, referred to as the $e$-type diagram. The types of diagrams are interchanged whenever there is a $\sigma$ string. In the theory of anyon condensation, the $(\sigma,\sigma)$ particle has to split into two after condensation~\cite{bais2009, burnell2012}.

Motivated by this observation, we find two nontrivial string operators $W^{\uparrow/\downarrow}$ in the SET model, and the corresponding quasiparticles will be denoted by $v_{\uparrow/\downarrow}$. The non-vanishing $\Omega$ symbols for $W^\uparrow$ are given by:
\begin{equation}
	\begin{gathered}
	{}^{\uparrow}\Omega_{v_\uparrow, I\psi\psi}^\psi={}^{\uparrow}\Omega_{v_\uparrow,\psi\psi\psi}^I={}^{\uparrow}\Omega_{v_\uparrow,\sigma\psi I}^\sigma=1,\\
	{}^{\downarrow}\Omega_{v_\uparrow,III}^I=1, {}^{\downarrow}\Omega_{v_\uparrow,\psi II}^\psi=-1,
	{}^{\downarrow}\Omega_{v_\uparrow,\sigma I\psi}^\sigma=1
	\end{gathered}
	\label{eqn:Wup}
\end{equation}
Similarly we can define $W^\downarrow$ by interchanging $\uparrow$ and $\downarrow$ in Eq. \eqref{eqn:Wup}.

To have an intuitive understanding of the solutions, let us imagine freezing all the Ising spins and the bulk breaks into domains of Ising spins. In each domain, the wavefunction is locally the same as the toric code. Within a domain of $\uparrow$/$\downarrow$ spins, the action of $W^\uparrow$ is defined using the $e$-type /$m$-type diagram in Eq. \eqref{eqn:sigma-sigma-string}. Therefore, when the string crosses a domain wall, the string changes its type in accordance with Eq. \eqref{eqn:sigma-sigma-string} (see Fig. \ref{fig:string-op} for an illustration).

\begin{figure}
	\includegraphics{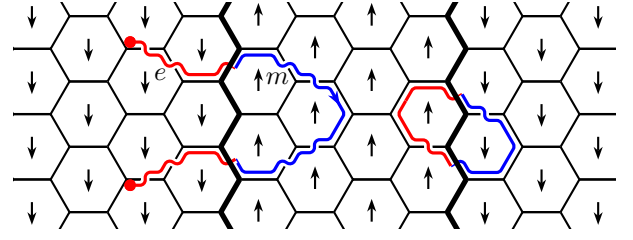}
  \caption{Illustration of an open string operator ${W}^\downarrow$ and closed string operator ${W}^\uparrow$.}
	\label{fig:string-op}
\end{figure}

Finally, the $(\psi, I)$ /$(I,\psi)$ quasiparticles are created by the following string operators:
\begin{equation}
	\begin{split}
	\begin{pspicture}[shift=-0.25](-0.1,-0.1)(0.6,0.6)
            \psid(0,0.5)(0.5,0)\psline[border=1.5pt](0,0)(0.5,0.5)
        \end{pspicture}
        &=
        \begin{pspicture}[shift=-0.25](-0.1,-0.1)(0.6,0.6)
            \psid(0,0.5)(0.2,0.3)\psid(0.3,0.2)(0.5,0)\pspsi(0,0.1)(0.2,0.3)(0.3,0.2)(0.5,0.4)
        \end{pspicture}\\
            \begin{pspicture}[shift=-0.25](-0.1,-0.1)(0.6,0.6)
                \pssigma(0,0.5)(0.5,0)\psline[border=1.5pt](0,0)(0.5,0.5)
            \end{pspicture}
            &=\pm i
            \begin{pspicture}[shift=-0.25](-0.1,-0.1)(0.6,0.6)
                \pssigma(0,0.5)(0.5,0)\pspsi(0,0.1)(0.2,0.3)\pspsi(0.3,0.2)(0.5,0.4)
            \end{pspicture}\\
            \begin{pspicture}[shift=-0.25](-0.1,-0.1)(0.6,0.6)
                \pspsi(0,0.5)(0.5,0)\psline[border=1.5pt](0,0)(0.5,0.5)
            \end{pspicture}
            &=-
            \begin{pspicture}[shift=-0.25](-0.1,-0.1)(0.6,0.6)
                \pspsi(0,0.5)(0.2,0.3)(0,0.1)\pspsi(0.5,0)(0.3,0.2)(0.5,0.4)\psid(0.2,0.3)(0.3,0.2)
            \end{pspicture}
		\end{split}
	\label{eqn:psi-string}
\end{equation}
They also persist in the SET phase. However, the difference between them is a $(\psi, \psi)$ string, which as we described earlier becomes local excitations in the SET phase. So, $(\psi, I)$ and $(I, \psi)$ belong to the same type of quasiparticle, which intuitively should be a fermion $\psi$.

With all the solutions for string operators, we can compute the braiding statistics of the quasiparticles using Eqs. \eqref{eqn:string-twist} and \eqref{eqn:string-Smatrix}. We find $\theta_{v_\uparrow}=\theta_{v_\downarrow}=1, \theta_\psi=-1$, as expected. The $S$ matrix is also identical to the one of the toric code:
\begin{equation}
	S=\frac{1}{2}
	\begin{pmatrix}
		1 & 1 & 1 & 1\\
		1 & 1 & -1 & -1\\
		1 & -1 & 1 & -1\\
		1 & -1 & -1 & -1
	\end{pmatrix},
	\label{}
\end{equation}
where the row/column is ordered as $1,v_\uparrow, v_\downarrow, \psi$. This justifies the identification of $v_\uparrow$ and $v_\downarrow$ with the $e$ and $m$ anyons in the SET phase.

From the definition of $W^{\uparrow/\downarrow}$, it is obvious that under the global $\mathbb{Z}_2$ symmetry transformation, $XW^\uparrow X^{-1}=W^\downarrow$, $XW^{\downarrow}X^{-1}=W^\uparrow$. Therefore, we have shown explicitly using string operators that the $\mathbb{Z}_2$ symmetry indeed permutes $e$ and $m$.

\subsubsection{More examples of EMD symmetries}
We will discuss a couple of other examples where the electro-magnetic duality symmetry in a discrete gauge theory is realized by an onsite unitary $\mathbb{Z}_2$ symmetry.

\begin{description}
	\item[$\mathbb{Z}_N$]
		Consider odd $N$ for simplicity. As shown in \Refs{SET1,SET2}, gauging the $e\leftrightarrow m$ symmetry yields the theory $\mathcal{Z}(\mathrm{TY}_N)$. Here $\mathrm{TY}_N$ is the Tambara-Yamagami category for $\mathbb{Z}_N$~\cite{Tambara1998}, with $N+1$ labels $[0],[1],\dots, [N-1], \sigma$. The first $N$ labels have $\mathbb{Z}_N$ fusion rules. The last label $\sigma$ represents the symmetry defect:
\begin{equation}
	\begin{gathered}
		{}[a]\times\sigma=\sigma,\\
	\sigma\times\sigma=[0]+[1]+\cdots + [N-1].
	\end{gathered}
	\label{}
\end{equation} The $F$ symbols are given by:
\begin{equation}
	\begin{gathered}
		{}[F^{a\sigma b}_\sigma]_{\sigma\sigma}=[F^{\sigma a\sigma}_b]_{\sigma\sigma}=e^{\frac{2\pi i ab}{N}},\\
	[F^{\sigma\sigma\sigma}]_{ab}=\frac{1}{\sqrt{N}}e^{-\frac{2\pi i ab}{N}}.
	\end{gathered}
	\label{}
\end{equation}
The $\mathrm{TY}_N$ category is $\mathbb{Z}_2$-graded: $\mathcal{C}_\mb{1}=\{[0],[1],\dots,[N-1]\}, \mathcal{C}_\mb{g}=\{\sigma\}$.
Therefore, we can construct a $\mathbb{Z}_N$ SET with a $\mathbb{Z}_2$ EMD symmetry using the $\mathrm{TY}_N$ category.

\item[$S_3$]
	There is also an EMD symmetry in the $S_3$ gauge theory~\cite{Beigi2010}, between the pure gauge charge $C$ corresponding to the two-dimensional irreducible representation of $S_3$, and the pure fluxon $F$ corresponding to the $2$-dimensional conjugacy class of $S_3$ (see e.g. \Ref{Cui2014} for a complete description of the $S_3$ gauge theory). The appropriate extension is the fusion category $\mathrm{SU}(2)_4$, whose integer-spin subcategory coincides, as a fusion category, with the category of irreducible linear representations of $S_3$. As shown in \Ref{SET1}, gauging the $\mathbb{Z}_2$ EMD symmetry one obtains $\mathcal{Z}(\mathrm{SU}(2)_4)$, so one can construct a $S_3$ gauge theory with a $\mathbb{Z}_2$ EMD symmetry using the $\mathrm{SU}(2)_4$ extension.

\end{description}


\section{$\mathbb{Z}_2$ Toric Code with Anti-unitary Onsite $\mathbb{Z}_2^T$ Symmetry}

\subsection{Classification}
\label{sec:z2tc-au-classification}
In this section, we will study $\mathbb{Z}_2$ toric code enriched by the time-reversal symmetry $\mathbb{Z}_2^T$.
Let us classify $\mathbb{Z}_2^T$ extensions of $\text{Vec}_{\mathbb{Z}_2}=\{1,e\}$. The classification of possible fusion rules is the same as the case of the unitary $\mathbb{Z}_2$ extension, and we get three types of fusion rules: $\mathbb{Z}_2\times\mathbb{Z}_2^T, \mathbb{Z}_4^T$ and $\text{Ising}$. For the first two cases, the solutions of twisted pentagon equations are classified by the corresponding twisted group cohomology.

\begin{description}
	\item[$\mathbb{Z}_2\times \mathbb{Z}_2^T$] It is known that $\cal{H}^3[\mathbb{Z}_2\times\mathbb{Z}_2^T, \mathrm{U}(1)]=\mathbb{Z}_2^2$, and as before one of the $\mathbb{Z}_2$ factors corresponds to double-semion topological order. The other $\mathbb{Z}_2$ factor can be understood in the SPT picture: it describes a $\mathbb{Z}_2\times\mathbb{Z}_2^T$ SPT where the $\mathbb{Z}_2$ flux is a Kramers doublet. After gauging the $\mathbb{Z}_2$ symmetry, we obtain a toric code where the magnetic flux $m$ is a Kramers doublet.
	\item[$\mathbb{Z}_4^T$] One can find that $\cal{H}^3[\mathbb{Z}_4^T, \mathrm{U}(1)]=\mathbb{Z}_1$, i.e., there is only the trivial $3$-cocycle. As we will see in the following, the physics of this SET is also that $e$ or $m$ is a Kramers doublet.
	\item[Ising] Similar to the Ising extension for the unitary $\mathbb{Z}_2$ symmetry, this represents a SET with time-reversal transformation permuting $e$ and $m$. The $F$ symbols are given by the familiar ones of the Ising category.
\end{description}

We will now examine the $\mathbb{Z}_4^T$ and Ising extensions more carefully.

\subsection{$\cal{T}^2=-1$ fractionalization}
We have two extensions that correspond to $\cal{T}^2=-1$ fractionalization, and we will only consider the $\mathbb{Z}_4^T$ extension in this section.

The Hamiltonian is essentially the same one as in Eq. \eqref{eqn:Z4}, and we define $\cal{T}=\prod_p \tau^x_p K$. Notice that in the chosen basis $KU_eK^{-1}=U_e^*=U_e^\dag, KV_eK^{-1}=V_e$, so the Hamiltonian is indeed invariant under $T$. We will not repeat the analysis of quasiparticle string operators. The $\tilde{e}$ particle is easily seen to transform trivially under $\cal{T}$. The $\tilde{m}$ string, however involves $\pm i$ factors whenever the string crosses a domain wall. Therefore, if we have an open $\tilde{m}$ string connecting plaquettes $p$ and $q$, then under complex conjugation we have $KW_{m}K^{-1}=W_{m^3}=W_m\cdot W_{m^2}=\tau_p^z\tau_q^z W_m$. So the time-reversal transformation acting on a state with two $m$'s becomes:
\begin{equation}
	\begin{split}
	\cal{T}W_m\ket{0}&=\Big(\prod_r \tau_r^x\Big) \tau_p^z\tau_q^z W_m\ket{0}\\
	&=\tau_p^y\tau_q^y \Big( \prod_{r\neq p,q}\tau_r^x \Big)W_m\ket{0}.
	\end{split}
	\label{}
\end{equation}
Therefore, the local time-reversal action on the $m$ located at the plaquette $p$ can be chosen as $\cal{U}_\cal{T}^{(m)}=\tau_p^y$, which satisfies $\cal{U}_\cal{T}\cal{U}_\cal{T}^*=-1$.

\subsection{$\cal{T}$ as the EM duality}
We consider the Ising extension, where the time-reversal symmetry permutes $e$ and $m$. Because the usual $F$ symbols of Ising category are real, they are automatically solutions of the twisted Pentagon equations. One caveat here is that the (untwisted) Pentagon equations have two gauge-inequivalent solutions for Ising fusion rules, distinguished by the Frobenius-Schur indicator $\kappa_\sigma=\frac{[F^{\sigma\sigma\sigma}_{\sigma}]_{II}}{|[F^{\sigma\sigma\sigma}_{\sigma}]_{II}|}=\pm 1$. However, with the twisted Pentagon equations and the gauge transformations, these two solutions become identical under the gauge transformation $u^{\sigma\sigma}_I=u^{\sigma\sigma}_\psi=i$.

As a result, we can just take the same Hamiltonian Eq. \eqref{eqn:Hem} in the unitary $\mathbb{Z}_2$ case, but now define $\cal{T}=\prod_p\tau_p^x K$ where $K$ is the complex conjugation. Notice that microscopically $\mathcal{T}^2=1$. It follows immediately that $\cal{T}$ exchanges $e$ and $m$.

It is known from general consideration~\cite{SET1} that if $e$ and $m$ are interchanged under $\mathcal{T}$, their fermionic bound state $\psi$ should be a Kramers doublet with $\cal{T}^2=-1$. We now explicitly verify this result in our model.
We have explained the construction of quasiparticle string operators in Sec. \ref{sec:string}. In particular, the $\psi$ particles are created by the string operator defined in Eq. \eqref{eqn:psi-string}.
Denote the one with $\pm i$ as $W_\psi^\pm(P)$, where $P$ is the path of the string with end points in the dual lattice $p$ and $q$. Under complex conjugation, we have $KW_\psi^\pm(P)K^{-1}=W_\psi^\mp(P)=\tau_p^z\tau_q^z W_\psi^\pm(P)$. Now, consider the time-reversal transformation acting on a state with two $\psi$ quasiparticles created by $W_\psi^+(P)$:
\begin{equation}
	\begin{split}
	\mathcal{T}W_{\psi}^+(P)\ket{\Psi}&=\tau^x_p\tau^x_p \Big(\prod_{r\neq p,q}\tau_r^x\Big)\tau_p^z\tau^z_q  W_\psi^+(P)\ket{\Psi}\\
	&=\tau^y_p\tau^y_q\Big(\prod_{r\neq p,q}\tau_r^x\Big) W_\psi^+(P)\ket{\Psi}.
	\end{split}
	\label{}
\end{equation}
Therefore we identify the local $\mathcal{T}$ action as being given by $\cal{U}_\cal{T}=\tau^y$. So the local $\cal{T}^2$ value is $\cal{U}_\cal{T}\cal{U}_\cal{T}^*=-1$.

\subsection{$eTmT$ state}
\label{sec:etmt}

According to the group cohomology classification of SPT phases~\cite{chen2013}, there exists a nontrivial SPT phase protected by $\mathbb{Z}_2^T$ symmetry in $3+1$ dimensions, since $\mathcal{H}^4[\mathbb{Z}_2^T, \mathrm{U}(1)]=\mathbb{Z}_2$. Therefore, it should be possible to construct an anomalous SET living on the surface of this bosonic SPT phase, using the obstructed pentagon equation given in Eq.~\eqref{eq:pentagon-obs}, with the nontrivial 4-cocyle $\beta\in\cal H^4[\mathbb Z_2^T, U(1)]$. One choice of $\beta$ is
\begin{equation}
  \label{eq:beta-TSC}
  \beta(\mb{g,h,k,l})=
  \begin{cases}
	  -1 & \mb{g}=\mb{h}=\mb{k}=\mb{l}=\cal{T}\\
	  1 & \text{otherwise}
  \end{cases}.
\end{equation}
We can now solve for anti-unitary extensions for the three types of fusion rules listed in Sec. \ref{sec:z2tc-au-classification}. We find that only the $\mathbb{Z}_4^T$ fusion rules allow solutions, while $\mathbb{Z}_2\times\mathbb{Z}_2^T$ and $\text{Ising}$ do not. In fact, the solution can be parametrized using Eq. \eqref{eq:w=ca} with $\chi_e(\cal{T},\cal{T})=-1$. According to the argument in Sec. \ref{sec:qn-sf}, both the $e$ and $m$ particles in the resulting SET have $\mathcal{T}^2=-1$. This SET is referred to ``$eTmT$'' in literature, and has been known to exist on the surface of 3D $\mathbb{Z}_2^T$ SPT state via very different arguments~\cite{VishwanathPRX2013, wang2013}. We notice that another exactly solvable model for this anomalous SET was found in \cite{Song_thesis}.

\section{Conclusions}

In this work, we construct exactly solvable models to realize SET phases. Starting from the topological phase of the quantum double $\cal{Z}(\cal{C})$ of a UFC $\cal C$ and a symmetry group $G$, such SET models are described by a generalized form of the $G$-extension of $\cal C$, denoted by $\cal C_G$, where the $F$ symbols satisfy the generalized pentagon equation in Eq.~\eqref{eq:pentagon-t}, with a nontrivial symmetry action. When $G$ is onsite and unitary, the symmetry action is trivial, and our models can be considered as ``ungauging'' the quantum double of $G$-extensions of $\cal C$, which fully classify the $G$-enriched phases of $\cal{Z}(\cal C)$. When $G$ contains anti-unitary and/or mirror-reflection symmetry operations, $\cal C_G$ obeys a twisted pentagon equation, where the anti-unitary and mirror-reflection symmetries act on one of the $F$ symbols by complex conjugation. Finally, our models can also describe anomalous SET states realized on the surface of a 3D nontrivial SPT bulk, using solutions of the ``obstructed'' pentagon equation in Eq.~\eqref{eq:pentagon-obs}.

When $\cal C$ describes an untwisted Abelian gauge theory, we explicitly construct solutions of the twisted pentagon equation describing all possible patterns of symmetry fractionalization, when symmetries do not permute anyons. We also demonstrate the bulk-boundary correspondence between the surface symmetry fractionalization and the bulk SPT state directly in this construction. The results can be straightforwardly generalized to twisted gauge theories.

As a concrete example of our general framework, we explicitly construct all SET phases of the $\mathbb{Z}_2$ toric code topological order, enriched by either an onsite unitary $\mathbb Z_2$ symmetry, or the time-reversal symmetry $\mathbb Z_2^T$. To the best of our knowledge, our construction yields the first onsite realization of the EMD symmetry using commuting-projector Hamiltonians.

There are several potential directions for further investigation.
In this work we have focused on finite symmetry groups. It would be interesting to generalize the construction to continuous symmetries. For spatial symmetries, we only consider mirror symmetries, and it is certainly desirable to have a more systematical treatment of space-group symmetries.
Finally, it is also an interesting direction to use the concept of equivalent classes of symmetric local unitary transformations to classify fermionic SET phases and three-dimensional SETs~\cite{Cheng_unpub_3D}.

\section{Acknowledgement}
M.C. would like to thank Kaushal Patel for sharing illustrations on string-net models. We gratefully acknowledge Maissam Barkeshli, Parsa Bonderson, Ying Ran, Kevin Walker, Zhenghan Wang, Chenjie Wang, Dominic Williamson, and Yong-Shi Wu for helpful discussions. Z. C. G acknowledges start-up support from Department of Physics, The Chinese University of Hong Kong, Direct Grant No. 3132745 from The Chinese University of Hong Kong and the funding from RGC/ECS(No.2191110). S. J. thanks Prof. Ying Ran for support during this work under National Science Foundation under Grant No. DMR-1151440. M.C. acknowledges Aspen Center for Physics supported by NSF grant PHY-1066293 for hospitality while the manuscript was finalized. This research was supported in part by Perimeter Institute for Theoretical Physics. Research at Perimeter Institute is supported by the Government of Canada through the Department of Innovation, Science and Economic Development Canada and by the Province of Ontario through the Ministry of Research, Innovation and Science.

\textbf{Note added}. During the preparation of the manuscript, we became aware of a related work~\cite{Heinrich2016} which also constructed commuting-projector Hamiltonians for SET phases using the idea of unitary $G$-extensions of UFCs.

\appendix
\section{Review of group cohomology}
\label{sec:review-coho}

This appendix provides a brief review of group cohomology, which is used throughout the main text of the paper. Here, we only discuss properties of group cohomology that are relevant to our paper, and for more details, we refer the readers to \Ref{chen2013}.

Given a finite group $G$, the group cohomology $\cal H^n_\rho[G, M]$ is defined for a $G$-module $(M, \rho)$, which is an Abelian group $M$ (sometimes called the coefficient of the group cohomology) equipped with a $G$-action $\rho$. The action $\rho:G\times M\rightarrow M$ specifies how the group $G$ acts on $M$. In particular, an group element $\mb g\in G$ maps $m\in M$ to $\rho_{\mb g}(m)$. For example, when computing the classification of SPT states, we choose $M$ to be the U(1) group and $G$ to be the symmetry group, and antiunitary symmetry operations have a nontrivial action on $M$: if $\mb g\in G$ is antiunitary, then $\rho_{\mb g}(\phi) = \phi^\ast$, for any $\phi\in\mathrm U(1)$. Mathematically, the definition of a $G$-module requires that the action $\rho$ is compatible with group multiplications,
\begin{equation}
  \label{eq:gaction-compatibility}
  \rho_\mb{g}(\rho_\mb{h}(a))=\rho_\mb{gh}(a),\quad
  \rho_\mb{g}(a)\rho_\mb{g}(b)=\rho_\mb{g}(ab).
\end{equation}

Given a finite group $G$ and a $G$-module, the group cohomology can be defined and computed using cochains and the coboundary mappings. In this appendix, we construct the group cohomology using the so-called inhomogeneous cochains. An $n$-cochain can be viewed as a function $\omega:G^n\rightarrow M$. In other words, for any $n$ group elements $\mb g_1,\ldots\mb g_n$, $\omega(\mb g_1,\ldots \mb g_n)$ is an element of $M$. We denote the collection of all $n$-cochains by $C^n_\rho[G, M]$. $C^n_\rho[G, M]$ naturally forms a group, using the multiplication of $M$.

Next, we define the coboundary mapping $\di: C^n_\rho[G, M]\rightarrow C^{n+1}_\rho[G, M]$,
\begin{equation}
  \label{eq:def:coboundary}
  \begin{split}
  \di\omega(\mb g_1,\ldots \mb g_{n+1})
  =&\rho(\mb g_1)[\omega(\mb g_2,\ldots\mb g_{n+1})]\\
    &\times\prod_{i=1}^n
	\omega^{(-1)^i}(\mb g_1,\ldots,\mb g_i\mb g_{i+1},\ldots,\mb g_{n+1})\\
	&\times\omega^{(-1)^{n+1}}(\mb g_1,\ldots\mb g_{n}).
  \end{split}
\end{equation}

One can directly verify that $\mathrm{d} \mathrm{d}\omega=1$ for any $\omega \in C^n(G, M)$, where $1$ is the trivial cochain in $C^{n+2}(G, M)$. This is why $\mathrm{d}$ is considered a ``boundary operator.''

With the coboundary map, we next define $\omega\in C^n(G, M)$ to be an $n$-cocycle if it satisfies the condition $\mathrm{d}\omega=1$. We denote the set of all $n$-cocycles by
\begin{equation}
\begin{split}
Z^n_{\rho}(G, M) &= \text{ker}[\mathrm{d}: C^n(G, M) \rightarrow C^{n+1}(G, M)]  \\
& = \{ \, \omega\in C^n(G, M) \,\, | \,\, \mathrm{d}\omega=1 \, \}.
\end{split}
\label{}
\end{equation}
We also define $\omega\in C^n(G, M)$ to be an $n$-coboundary if it satisfies the condition $\omega= \mathrm{d} \mu $ for some $(n-1)$-cochain $\mu \in C^{n-1}(G, M)$. We denote the set of all $n$-coboundaries by
\begin{equation}
\begin{split}
& B^n_{\rho}(G, M)  = \text{im}[ \mathrm{d}: C^{n-1}(G, M) \rightarrow C^{n}(G, M) ] \\
& =\{ \, \omega\in C^n(G, M) \,\, | \,\, \exists \mu \in C^{n-1}(G, M) : \omega = \mathrm{d}\mu \, \}
.
\end{split}
\label{}
\end{equation}

Clearly, $B^n_{\rho}(G, M) \subset Z^n_{\rho}(G, M) \subset C^n(G, M)$. In fact, $C^n$, $Z^n$, and $B^n$ are all groups and the co-boundary maps are homomorphisms. It is easy to see that $B^n_{\rho}(G, M)$ is a normal subgroup of $Z^n_{\rho}(G, M)$. Since d is a boundary map, we think of the $n$-coboundaries as being trivial $n$-cocycles, and it is natural to consider the quotient group
\begin{equation}
	\mathcal{H}^n_{\rho}(G, M)=\frac{Z^n_{\rho}(G, M)}{B^n_{\rho}(G, M)}
,
\label{}
\end{equation}
which is called the $n$-th cohomology group. In other words, $\mathcal{H}^n_{\rho}(G, M)$ collects the equivalence classes of $n$-cocycles that only differ by $n$-coboundaries.

It is instructive to look at the lowest several cohomology groups. Let us first consider $\mathcal{H}^1_{\rho}(G, M)$:
\begin{equation}
	\begin{split}
Z^1_{\rho}(G, M) &= \{\, \omega \, \,| \, \, \omega(\mathbf{g}_1)\rho_\mathbf{g}[\omega(\mathbf{g}_2)]=\omega(\mathbf{g}_1\mathbf{g}_2) \,\} \\
B^1_{\rho}(G, M) &= \{\, \omega \,\, | \,\, \omega(\mathbf{g})=\rho_\mathbf{g}(\mu)\mu^{-1} \, \}
.
\end{split}
\label{}
\end{equation}
If the $G$-action on $M$ is trivial, then $B^1_{\rho}(G, M) = \{ 1 \}$ and $Z^1_{\rho}(G, M)$ is the group homomorphisms from $G$ to $M$. In general, $\mathcal{H}^1_{\rho}(G, M)$ classifies ``crossed group homomorphisms'' from $G$ to $M$.

For the second cohomology, we have
\begin{equation}
	\begin{split}
		Z^2_{\rho}(G, M)&=\{ \, \omega \,\, | \,\, \frac{\rho_{\mathbf{g}_1}[\omega(\mathbf{g}_2,\mathbf{g}_3)]\omega(\mathbf{g}_1,\mathbf{g}_2\mathbf{g}_3) }{ \omega(\mathbf{g}_1,\mathbf{g}_2)\omega(\mathbf{g_1}\mathbf{g}_2,\mathbf{g}_3)}=1 \,\}\\
		B^2_{\rho}(G, M)&=\{\, \omega \,\, | \,\, \omega(\mathbf{g}_1,\mathbf{g}_2)=\frac{\varepsilon(\mathbf{g}_1)\rho_{\mathbf{g}_1}[\varepsilon(\mathbf{g}_2)]}{\varepsilon(\mathbf{g}_1\mathbf{g}_2)}\, \}
.
\end{split}
\label{}
\end{equation}
If $M=\mathrm{U}(1)$, it is well-known that $Z^2(G, \mathrm{U}(1))$ is exactly the factor sets (also known as the Schur multipliers) of projective representations of $G$, with the cocycle condition coming from the requirement of associativity. $\mathcal{H}^2(G, \mathrm{U}(1))$ classifies all inequivalent projective representations of $G$.

For the third cohomology, we have
\begin{equation}
\begin{split}
Z^3_{\rho}(G, M)=\{ \, \omega \, \, |\,\, \omega(\mathbf{g}_1\mathbf{g}_2, \mathbf{g}_3, \mathbf{g}_4)\omega(\mathbf{g}_1, \mathbf{g}_2, \mathbf{g}_3\mathbf{g}_4)\\
    	=\rho_{\mathbf{g}_1}[\omega(\mathbf{g}_2, \mathbf{g}_3, \mathbf{g}_4)]\omega(\mathbf{g}_1, \mathbf{g}_2\mathbf{g}_3, \mathbf{g}_4)\omega(\mathbf{g}_1, \mathbf{g}_2, \mathbf{g}_3) \, \}
\end{split}
\label{}
\end{equation}
For $M=\mathrm{U}(1)$ and trivial $G$ action, $Z^3(G, \mathrm{U}(1))$ is the set of $F$-symbols for the fusion category $\text{Vec}_G$, with the $3$-cocycle condition being the Pentagon identity. $B^3(G, \mathrm{U}(1))$ is identified with all the $F$-symbols that are gauge-equivalent to the trivial one. $\mathcal{H}^3(G, \mathrm{U}(1))$ then classifies the gauge-equivalent classes of $F$-symbols on $\text{Vec}_G$.

Throughout the paper, we adopt a canonical gauge fixing for the cocycles: for a $n$-cocycle $\omega(\mb{g}_1, \mb{g}_2, \dots, \mb{g}_n)$, as long as any of the $\mb{g}_i$ with $1\leq i \leq n$ is the identity element $1$, the cocycle is set to $1$.

\newcommand{\FmL}{\coordinate (o1) at (.3, -.3);
\coordinate (o3) at (.6, -.6);
\draw (0, 0) node [above] {\small $\ag ag$} -- (o1) [string];
\draw (.6, 0) node [above] {\small $\ag bh$} -- (o1) [string];
\draw (1.2, 0) node [above] {\small $\ag ck$} -- (o3) [string];
\draw (o1) -- (o3) [string] node [midway, left] {\small $e$};
\draw (o3) -- ++(.3, -.3) node [below] {\small $d$} [string];

\node at (.1, -.9) [group] {\small $\mb g_0$};}
\newcommand{\FmR}{\coordinate (o2) at (.9, -.3);
    \coordinate (o3) at (.6, -.6);
    \draw (0, 0) node [above] {\small $\ag ag$} -- (o3) [string];
    \draw (.6, 0) node [above] {\small $\ag bh$} -- (o2) [string];
    \draw (1.2, 0) node [above] {\small $\ag ck$} -- (o2) [string];
    \draw (o2) -- (o3) [string] node [midway, right] {\small $f$};
    \draw (o3) -- ++(.3, -.3) node [below] {\small $d$} [string];

    \node at (.1, -.9) [group] {\small $\mb g_0$};}
\newcommand{\OmS}{
    \coordinate (o1) at (0, 0);
    \coordinate (o2) at (0, -.6);
    \draw (0, .3) node [above] {\small $\ag c{gh}$} -- (o1) [string];
    \draw (o2) -- ++(0, -.3) node [below] {\small $\ag c{gh}$} [string];
    \draw (o1) ..controls (-.3, -.1).. (-.3,-.2) -- (-.3, -.4)
      ..controls (-.3, -.5).. (o2) [string]
      node [midway, left] {\small $\ag ag$};
    \draw (o1) ..controls (.3, -.1).. (.3,-.2) -- (.3, -.4)
      ..controls (.3, -.5).. (o2) [string]
      node [midway, right] {\small $\ag bh$};
    \node at (-.5, .2) [group] {\small $\mb g_0$};
  }
\newcommand{\OmSb}{
    \coordinate (o1) at (0, 0);
    \coordinate (o2) at (0, -.6);
    \draw (0, .3) node [above] {\small $\ag c{gh}$} -- (o1) [string];
    \draw (o2) -- ++(0, -.3) node [below]
    {\small $c_\mb{gh}^\prime$} [string];
    \draw (o1) ..controls (-.3, -.1).. (-.3,-.2) -- (-.3, -.4)
      node [midway, left] {\small $\ag ag$}
      ..controls (-.3, -.5).. (o2) [string];
    \draw (o1) ..controls (.3, -.1).. (.3,-.2) -- (.3, -.4)
      node [midway, right] {\small $\ag bh$}
      ..controls (.3, -.5).. (o2) [string];
    \node at (-.5, .2) [group] {\small $\mb g_0$};
  }
\newcommand{\OmD}{
    \coordinate (o1) at (0, 0);
    \coordinate (o2) at (0, -.6);
    \coordinate (o3) at (0, -.9);
    \coordinate (o4) at (0, -1.5);
    \draw (0, .3) node [above] {\small $\ag c{gh}$} -- (o1) [string];
    \draw (o1) ..controls (-.3, -.1).. (-.3,-.2) -- (-.3, -.4)
      node [midway, left] {\small $\ag ag$}
      ..controls (-.3, -.5).. (o2) [string];
    \draw (o1) ..controls (.3, -.1).. (.3,-.2) -- (.3, -.4)
      node [midway, right] {\small $\ag bh$}
      ..controls (.3, -.5).. (o2) [string];

    \draw (o2) -- (o3) [string] node [midway, right] {\small $\ag c{gh}$};
    \draw (o4) -- ++(0, -.3) node [below] {\small $\ag c{gh}$} [string];
    \draw (o3) ..controls (-.3, -1).. (-.3, -1.1) -- (-.3, -1.3)
      node [midway, left] {\small $\ag ag$}
      ..controls (-.3, -1.4).. (o4) [string];
    \draw (o3) ..controls (.3, -1).. (.3, -1.1) -- (.3, -1.3)
      node [midway, right] {\small $\ag bh$}
      ..controls (.3, -1.4).. (o4) [string];

    \node at (-.7, -.75) [group] {\small $\mb g_0$};
  }
\newcommand{\Line}{
    \draw (0, 0) node [above] {\small $\ag c{gh}$} -- (0, -.6) [string];
    \node at (-.3, -.3) [group] {\small $\mb g_0$};
  }
\newcommand{\LineD}{
    \draw (0, 0) node [above] {\small $\ag d{ghk}$} -- (0, -.6) [string];
    \node at (-.3, -.3) [group] {\small $\mb g_0$};
  }

\newcommand{\PentagonA}{
    \coordinate (o1) at (.3, -.3);
    \coordinate (o3) at (.6, -.6);
    \coordinate (o6) at (.9, -.9);

    \draw (0, 0) node [above] {\small $\ag ag$} -- (o1) [string];
    \draw (.6, 0) node [above] {\small $\ag bh$} -- (o1) [string];
    \draw (1.2, 0) node [above] {\small $\ag ck$} -- (o3) [string];
    \draw (1.8, 0) node [above] {\small $\ag dl$} -- (o6) [string];
    \draw (o1) -- (o3) [string] node [midway, left] {\small $f$};
    \draw (o3) -- (o6) [string] node [midway, left] {\small $m$};
    \draw (o6) -- ++(.3, -.3) [string] node [below] {\small $e$};

    \node at (.15, -1.1) [group] {\small $\mb g_0$};
  }

\newcommand{\PentagonB}{
    \coordinate (o2) at (.9, -.3);
    \coordinate (o3) at (.6, -.6);
    \coordinate (o6) at (.9, -.9);

    \draw (0, 0) node [above] {\small $\ag ag$} -- (o3) [string];
    \draw (.6, 0) node [above] {\small $\ag bh$} -- (o2) [string];
    \draw (1.2, 0) node [above] {\small $\ag ck$} -- (o2) [string];
    \draw (1.8, 0) node [above] {\small $\ag dl$} -- (o6) [string];
    \draw (o2) -- (o3) [string] node [midway, right] {\small $n$};
    \draw (o3) -- (o6) [string] node [midway, left] {\small $m$};
    \draw (o6) -- ++(.3, -.3) [string] node [below] {\small $e$};

    \node at (.15, -1.1) [group] {\small $\mb g_0$};
  }
\newcommand{\PentagonC}{
    \coordinate (o2) at (.9, -.3);
    \coordinate (o5) at (1.2, -.6);
    \coordinate (o6) at (.9, -.9);

    \draw (0, 0) node [above] {\small $\ag ag$} -- (o6) [string];
    \draw (.6, 0) node [above] {\small $\ag bh$} -- (o2) [string];
    \draw (1.2, 0) node [above] {\small $\ag ck$} -- (o2) [string];
    \draw (1.8, 0) node [above] {\small $\ag dl$} -- (o5) [string];
    \draw (o2) -- (o5) [string] node [midway, left] {\small $n$};
    \draw (o5) -- (o6) [string] node [midway, right] {\small $p$};
    \draw (o6) -- ++(.3, -.3) [string] node [below] {\small $e$};

    \node at (.15, -1.1) [group] {\small $\mb g_0$};
  }
\newcommand{\PentagonD}{
    \coordinate (o4) at (1.5, -.3);
    \coordinate (o5) at (1.2, -.6);
    \coordinate (o6) at (.9, -.9);

    \draw (0, 0) node [above] {\small $\ag ag$} -- (o6) [string];
    \draw (.6, 0) node [above] {\small $\ag bh$} -- (o5) [string];
    \draw (1.2, 0) node [above] {\small $\ag ck$} -- (o4) [string];
    \draw (1.8, 0) node [above] {\small $\ag dl$} -- (o4) [string];
    \draw (o4) -- (o5) [string] node [midway, right] {\small $q$};
    \draw (o5) -- (o6) [string] node [midway, right] {\small $p$};
    \draw (o6) -- ++(.3, -.3) [string] node [below] {\small $e$};

    \node at (.15, -1.1) [group] {\small $\mb g_0$};
  }

\newcommand{\PentagonE}{
    \coordinate (o1) at (.3, -.3);
    \coordinate (o4) at (1.5, -.3);
    \coordinate (o6) at (.9, -.9);

    \draw (0, 0) node [above] {\small $\ag ag$} -- (o1) [string];
    \draw (.6, 0) node [above] {\small $\ag bh$} -- (o1) [string];
    \draw (1.2, 0) node [above] {\small $\ag ck$} -- (o4) [string];
    \draw (1.8, 0) node [above] {\small $\ag dl$} -- (o4) [string];
    \draw (o1) -- (o6) [string] node [midway, left] {\small $f$};
    \draw (o4) -- (o6) [string] node [midway, right] {\small $q$};
    \draw (o6) -- ++(.3, -.3) [string] node [below] {\small $e$};

    \node at (.15, -1.1) [group] {\small $\mb g_0$};
  }
\newcommand{\HmL}{
    \coordinate (lu) at (0, 0);
    \coordinate (ld) at (0, -.6);
    \coordinate (ru) at (.6, 0);
    \coordinate (rd) at (.6, -.6);
    \draw (0, .3) node [above] {\small $\ag a{gh}$} -- (lu) [string];
    \draw (0.6, .3) node [above] {\small $\ag bk$} -- (rd) [string];
    \draw (lu) -- (ld) -- ++(0, -.3) node [below] {\small $\ag cg$} [string];
    \draw (rd) -- ++(0, -.3) node [below] {\small $\ag d{hk}$} [string];
    \draw (lu) -- (rd) [string] node [midway, below] {\small $\ag eh$};
    \node at (-.3, -.3) [group] {\small $\mb g_0$};
  }
\newcommand{\HmR}{
    \coordinate (o1) at (0, 0);
    \coordinate (o2) at (0, -.6);
    \draw (-.3, .3) node [above] {\small $\ag a{gh}$} -- (o1) [string];
    \draw (.3, .3) node [above] {\small $\ag bk$} -- (o1) [string];
    \draw (o1) -- (o2) [string] node [midway, right] {\small $\ag f{ghk}$};
    \draw (o2) -- ++(-.3, -.3) node [below] {\small $\ag cg$} [string];
    \draw (o2) -- ++(.3, -.3) node [below] {\small $\ag d{hk}$} [string];
    \node at (-.3, -.3) [group] {\small $\mb g_0$};
  }
\newcommand{\Hc}{
    \coordinate (ul) at (-.6, .6);
    \coordinate (ur) at (.6, .6);
    \coordinate (uul) at (-.6, 1.2);
    \coordinate (o1) at (0, 0);
    \coordinate (o2) at (0, -.6);
    \draw (-.6, 1.5) node [above] {\small $\ag a{gh}$} -- (uul) [string];
    \draw (.6, 1.5) node [above] {\small $\ag bk$} -- (ur) [string];
    \draw (uul) -- (ul) [thick];
    \draw (uul) -- (ur) [string] node [midway, above] {\small $\ag eh$};
    \draw (ul) -- (o1) [string] node [midway, left]
    {\small $\ag cg$};
    \draw (ur) -- (o1) [string] node [midway, right]
    {\small $\ag d{hk}$};
    \draw (o1) -- (o2) [string] node [midway, right] {\small $\ag f{ghk}$};
    \draw (o2) -- ++(-.3, -.3) node [below] {\small $\ag cg$} [string];
    \draw (o2) -- ++(.3, -.3) node [below] {\small $\ag d{hk}$} [string];
    \node at (-.3, -.3) [group] {\small $\mb g_0$};
  }
\newcommand{\Hd}{
    \coordinate (ul) at (-.6, .6);
    \coordinate (ur) at (.6, .6);
    \coordinate (uul) at (-.6, 1.2);
    \coordinate (o1) at (0, 0);
    \coordinate (o2) at (0, -.6);
    \draw (-.6, 1.5) node [above] {\small $\ag a{gh}$} -- (uul) [string];
    \draw (.6, 1.5) node [above] {\small $\ag bk$} -- (ur) [string];
    \draw (ur) -- (o1) [thick];
    \draw (uul) -- (ul) [string] node [midway, left] {\small $\ag cg$};
    \draw (uul) ..controls (-.3, 1.1).. (-.3, 1)
    -- (-.3, .8) node [midway, right] {\small $\ag eh$}
    ..controls (-.3, .7).. (ul) [string];
    \draw (ul) -- (o1) [string] node [midway, left]
    {\small $\ag a{gh}$};
    \draw (o1) -- (o2) [string] node [midway, right] {\small $\ag f{ghk}$};
    \draw (o2) -- ++(-.3, -.3) node [below] {\small $\ag cg$} [string];
    \draw (o2) -- ++(.3, -.3) node [below] {\small $\ag d{hk}$} [string];
    \node at (-.3, -.3) [group] {\small $\mb g_0$};
  }

\onecolumngrid
\section{
	Fixed-point Wave Functions and Symmetric Local Unitary Transformations
}
\label{fixwv}

In this section, we provide more details of the derivation of fixed-point wave functions, based on the idea that
different SET orders are classified as the
equivalence classes of many-body wave functions under symmetric local unitary (SLU)
transformations. We use generalized symmetric local unitary
transformations to define a wave function renormalization
procedure~\cite{ChenPRB2010, GuPRB2015}. The wave function renormalization can remove the
non-universal short-range entanglement and make generic
complicated wave functions flow to a simple fixed-point wave functions.

\subsection{Quantum state on a graph}
The basic setup of the quantum states on a graph has been described in Sec.~\ref{sec:fpwf}. We start with a trivalent graph with a branching structure.
In each plaquette of the graph, we put an element of the symmetry group. Then on each edge of the graph, we have states labeled as $a_\mb{g}$. The label set has a $G$-graded structure.

Our fixed-point state is a superposition of the basis states
\begin{align}
\label{Gstate}
|\Psi\>=\sum_\text{all conf.}
\Psi\left( \bmm \includegraphics[scale=0.18]{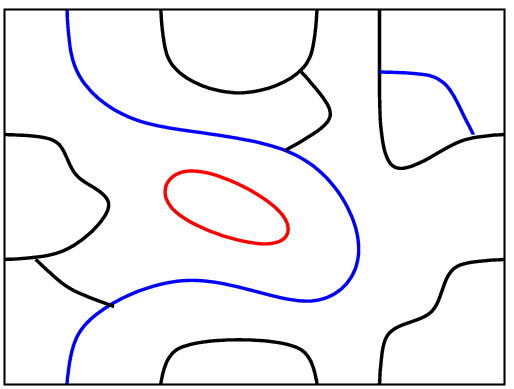}\emm \right)
\left
|\bmm \includegraphics[scale=0.2]{strnet}\emm\right \> .
\end{align}

In this appendix, for simplicity, we limit ourselves to the case where $N_{\ag ag \ag bh}^{c_\mb{gh} }$ is either 0 or 1. However, our results can be easily generalized to accommodate the more general case, where $N_{\ag ag \ag bh}^{c_\mb{gh} }$ can be greater than one. In that case,
each vertex can also have physical states, the number of which depends on the
fusion multiplicity $N_{\ag ag \ag bh}^{c_\mb{gh} }$.

In the rest of this appendix, we describe basic elements of the gSLU transformations (or moves) that relates wave function amplitudes on different configurations in Sec.~\ref{sec:appmoves:locrel}. We then list consistency conditions that constrains the forms of these moves, in Sec.~\ref{sec:appmoves:cons}.

\subsection{Generalized SLU transformations as local relations}
\label{sec:appmoves:locrel}

In this section, we list the basic types of the gSLU transformations, which are known as the ``moves'' in this paper. They are applied to a local patch of the wave function to remove unwanted short-range entanglement. We begin with the $F$ moves, which are introduced in Eq.~\eqref{eq:fmove} in the main text. Next, we introduce several variants of the $F$ moves. Finally we study the $O$ moves, which are introduced in Eq.~\eqref{eq:locrel2} in the main text.

\subsubsection{$F$ move}
The $F$ move is defined as
\begin{align}
\label{IHwave1}
\Psi\left(\tikzpicmidline{\FmL}\right)
=
\sum_{f_{\mb h\mb k}}
{^{\mb g_0}}[\cF^{a_{\mb g }b_{\mb h}c_{\mb k}}_{d_{\mb g\mb h \mb k}}]_{e_{\mb g \mb h} f_{\mb h \mb k}}
\Psi \left(\tikzpicmidline{\FmR}\right).
\end{align}
Here

  We require that
${^{\mb g_0}}[\cF^{a_{\mb g }b_{\mb h}c_{\mb k}}_{d_{\mb g\mb h \mb k}}]$ to be an
unitary matrix
\begin{equation}
\label{2FFstar}
\sum_{f_{\mb h \mb k}}
{^{\mb g_0}}[\cF^{a_{\mb g }b_{\mb h}c_{\mb k}}_{d_{\mb g\mb h \mb k}}]_{e_{\mb g \mb h}^\prime,f_{\mb h \mb k}}
{^{\mb g_0}}[\cF^{a_{\mb g }b_{\mb h}c_{\mb k}}_{d_{\mb g\mb h \mb k}}]^*_{e_{\mb g \mb h},f_{\mb h \mb k}}\nonumber
=\del_{e_{\mb g \mb h},e_{\mb g\mb h}^\prime},
\end{equation}

The $F$ move \eq{IHwave1} can be viewed as a relation between
wave functions on different v-graphs that only differ by a
local transformation.  Since we can locally transform one
v-graph to another v-graph through different paths, the $F$ move
\eq{IHwave1} must satisfy certain self-consistent
conditions.  For example, the v-graph $\tikzpicmidline{\PentagonA}$ can be transformed to $\tikzpicmidline{\PentagonD}$, through two
different paths; the first path is given by
\begin{align}
\label{FFFrelG}
\Psi\left(\tikzpicmidline{\PentagonA}\right)
&=\sum_{n_{\mb h \mb k }}
{^{\mb g_0}}[\cF^{a_{\mb g }b_{\mb h}c_{\mb k}}_{m_{\mb g \mb h \mb k}}]_{f_{\mb g \mb h},n_{\mb h \mb k}}
\Psi\left(\tikzpicmidline{\PentagonB}\right)\nonumber\\
&=\sum_{n_{ \mb h \mb k },p_{\mb h \mb k \mb l}}
{^{\mb g_0}}[\cF^{a_{\mb g }b_{\mb h}c_{\mb k}}_{m_{\mb g \mb h \mb k}}]_{f_{\mb g \mb h}, n_{\mb h \mb k}}
{^{\mb g_0}}[\cF^{a_{\mb g }n_{\mb h \mb k}d_{\mb l}}_{e_{\mb g \mb h \mb k \mb l}}]_{m_{\mb g \mb h \mb k }, p_{\mb h \mb k \mb l}}
\Psi\left(\tikzpicmidline{\PentagonC}\right)
\nonumber\\
&=\sum_{n_{\mb h \mb k},p_{ \mb h \mb k \mb l},q_{ \mb k \mb l}}
{^{\mb g_0}}[\cF^{a_{\mb g }b_{\mb h}c_{\mb k}}_{m_{\mb g \mb h \mb k }}]_{f_{\mb g \mb h },n_{\mb h \mb k }}
{^{\mb g_0}}[\cF^{a_{\mb g }n_{\mb h \mb k}d_{\mb l}}_{e_{\mb g \mb h \mb k \mb l}}]_{m_{\mb g \mb h \mb k }, p_{ \mb h \mb k \mb l}}
{^{\mb g_0 \mb g}}[\cF^{b_{\mb h }c_{\mb k}d_{\mb l}}_{p_{ \mb h \mb k \mb l}}]_{n_{\mb h \mb k }, q_{\mb k \mb l}}
\Psi\left(\tikzpicmidline{\PentagonD}\right) .
\end{align}
And the second path is
\begin{align}
\label{FFrelG}
\Psi\left(\tikzpicmidline{\PentagonA}\right)
&=\sum_{ q_{\mb k \mb l}}
{^{\mb g_0}}[\cF^{f_{\mb g \mb h }c_{ \mb k}d_{\mb l}}_{e_{\mb g \mb h \mb k \mb l}}]_{\bt\chi m_{\mb g \mb h \mb k }, q_{ \mb k \mb l}}
\Psi\left(\tikzpicmidline{\PentagonE}\right)
\nonumber\\ &=\sum_{ q_{\mb k \mb l},p_{ \mb h \mb k \mb l}}
{^{\mb g_0}}[\cF^{f_{\mb g \mb h }c_{ \mb k}d_{\mb l}}_{e_{\mb g \mb h \mb k \mb l}}]_{m_{\mb g \mb h \mb k }q_{\mb k \mb l}}
{^{\mb g_0}}[\cF^{a_{\mb g }b_{ \mb h}q_{\mb k \mb l}}_{e_{\mb g \mb h \mb k \mb l}}]_{f_{\mb g \mb h },p_{ \mb h \mb k \mb l}}
\Psi\left(\tikzpicmidline{\PentagonD}\right),
\end{align}
The consistence of the above two relations leads
the following condition on the generalized $F$-symbol.
\begin{equation}
\label{penidG1}
\begin{split}
\sum_{n_{\mb h \mb k}}
{^{\mb g_0}}[\cF^{a_{\mb g }b_{\mb h}c_{\mb k}}_{m_{\mb g \mb h \mb k }}]_{f_{\mb g \mb h },n_{\mb h \mb k }}
{^{\mb g_0}}[\cF^{a_{\mb g }n_{\mb h \mb k}d_{\mb l}}_{e_{\mb g \mb h \mb k \mb l}}]_{m_{\mb g \mb h \mb k }, p_{ \mb h \mb k \mb l}}
{^{\mb g_0 \mb g}}[\cF^{b_{\mb h }c_{\mb k}d_{\mb l}}_{p_{ \mb h \mb k \mb l}}]_{ n_{\mb h \mb k },q_{\mb k \mb l}}
 =
{^{\mb g_0}}[\cF^{f_{\mb g \mb h }c_{ \mb k}d_{\mb l}}_{e_{\mb g \mb h \mb k \mb l}}]_{ m_{\mb g \mb h \mb k }, q_{\mb k \mb l}}
{^{\mb g_0}}[\cF^{a_{\mb g }b_{ \mb h}q_{\mb k \mb l}}_{e_{\mb g \mb h \mb k \mb l}}]_{f_{\mb g \mb h },p_{ \mb h \mb k \mb l}}.
\end{split}
\end{equation}
which is the symmetry enriched form of the famous pentagon identity.

\subsubsection{ $Y$ move}
The following relation defines the $Y$ move:
\begin{align}
\label{PhiY}
\Psi\left(\tikzpicmidline{
  \draw (0, 0) node [above] {\small $\ag ag$} -- (0, -.6) [string];
  \draw (.6, 0) node [above] {\small $\ag bh$} -- (.6, -.6) [string];
  \node at (-.6, -.2) [group] {\small $\mb g_0$};
}\right)
=
\sum_{c_{\mb g \mb h}}
{^{\mb g_0}}[\cY^{a_{\mb g}b_{\mb h}}_{c_{\mb g \mb h}}]
\Psi\left(\tikzpicmidline{
  \coordinate (o1) at (.3, -.3);
  \coordinate (o2) at (.3, -.6);
  \draw (0, 0) node [above] {\small $\ag ag$} -- (o1) [string];
  \draw (.6, 0) node [above] {\small $\ag bh$} -- (o1) [string];
  \draw (o1) -- (o2) [string] node [midway, right] {\small $\ag c{gh}$};
  \draw (o2) -- ++(-.3, -.3) node [below] {\small $\ag ag$} [string];
  \draw (o2) -- ++(.3, -.3) node [below] {\small $\ag bh$} [string];
  \node at (0, -.45) [group] {\small $\mb g_0$};
}\right).
\end{align}

\subsubsection{$O$ move}
\label{sec:appmoves:O}

The $O$ move allows one to shrink a ``bubble'' in the diagram. First, we must have
\begin{align}
\Psi\left(\tikzpicmidline{\OmSb}\right)
=\del_{cc^\prime}
\Psi\left(\tikzpicmidline{\OmS}\right) .
\end{align}
We define the $O$ move as
\begin{align}
\label{PhiO}
\Psi\left(\tikzpicmidline{\OmS}\right)
= {}^{\mb g_0}[\cO_{c_{\mb g \mb h}}^{a_{\mb g}b_{\mb h}}]
\Psi\left(\tikzpicmidline{\Line}\right) .
\end{align}
The $O$ move should satisfy
\begin{align}
\label{Onorm}
\sum_{a_{\mb g},b_{\mb h}}
{^{\mb g_0}}[\cO_{c_{\mb g \mb h}}^{a_{\mb g}b_{\mb h}}](^{\mb g_0}[\cO_{c_{\mb g \mb h}}^{a_{\mb g}b_{\mb h}}])^*=1
\end{align}

We notice that despite the similarity in appearance, the $O$ moves defined here are \emph{different} from the moves defined in Eq.~\eqref{eq:locrel2}. As mentioned in the main text, the moves defined there ``eliminates'' bubbles by changing labels on some bonds to zero, but they neither changes the lattice structure, nor eliminates the degrees of the freedom on the bonds. On the contrary, the $O$ moves defined here eliminates the bubble by actually removing the bonds, and therefore changing the lattice structure. As a result, these two types of moves differ by a normalization factor. This will be further explained in Sec.~\ref{sec:appmoves:OY}, where the form of the $O$ moves will be fixed.

\subsection{Gauge freedom}

We note that the following transformation changes the wave function, but does not
change fixed-point property and the phase described by the wave function:
\begin{equation}
	\Psi\left(\begin{tikzpicture}[baseline={($ (current bounding box) - (0,0pt) $)}]
		\draw (0, .5) node [above] {$\ag ag$}--(.5, 0) [string];
		\draw (1, .5) node [above] {$\ag bh$}--(.5, 0) [string];
		\draw (.5, 0)--(.5, -.5) node [below] {$\ag c{gh}$} [string];
		\node [group] at (-.2, -.2) {$\mb g_0$};
	\end{tikzpicture}
	\right)\rightarrow
	{}^{\mb g_0}[v^{\ag ag\ag bh}_{\ag c{gh}}]
	\Psi\left(\begin{tikzpicture}[baseline={($ (current bounding box) - (0,0pt) $)}]
		\draw (0, .5) node [above] {$\ag ag$}--(.5, 0) [string];
		\draw (1, .5) node [above] {$\ag bh$}--(.5, 0) [string];
		\draw (.5, 0)--(.5, -.5) node [below] {$\ag c{gh}$} [string];
		\node [group] at (-.2, -.2) {$\mb g_0$};
	\end{tikzpicture}\right),
	\label{eqn:gauge-F-app}
\end{equation}
where ${}^{\mb{g}_0}[v^{a_{\mb g}b_{\mb h}}_{c_{\mb g \mb h}}]$ is a phase factor. This equation is identical to Eq.~\eqref{eqn:gauge-F} in the main text.

Similarly, we can have phase factors ${^{\mb g_0}}[v_{a_{\mb g}b_{\mb h}}^{c_{\mb g \mb h}}]$ for vertices
with two incoming edges and one outgoing edge.  Such transformations
correspond to a choice of basis state, and two wave functions related to each other via such redefinitions of vertices should be regarded as being equivalent.
The vertex basis redefinitions induce the following transformation on
${}^{\mb g_0}[\cF^{a_{\mb g}b_{\mb h}c_{\mb k}}_{d_{\mb g \mb h \mb k}}]_{e_{\mb g \mb h},f_{\mb h \mb k}}$,
${}^{\mb g_0}[\cO_{c_{\mb g \mb h}}^{a_{\mb g}b_{\mb h}}]$,${^{\mb g_0}}[\cY^{a_{\mb g}b_{\mb h}}_{c_{\mb g \mb h}}]$:
\begin{align}
\label{ftrans1}
^{\mb g_0}[\cO_{c_{\mb g \mb h}}^{a_{\mb g}b_{\mb h}}]
&\to
{^{\mb g_0}}[v_{a_{\mb g}b_{\mb h}}^{c_{\mb g \mb h}}]\,
{^{\mb g_0}}[v^{a_{\mb g}b_{\mb h}}_{c_{\mb g \mb h}}]\, {^{\mb g_0}}[\cO_{c_{\mb g \mb h}}^{a_{\mb g}b_{\mb h}}] ,
\nonumber\\
{^{\mb g_0}}[\cY^{a_{\mb g}b_{\mb h}}_{c_{\mb g \mb h}}]
&\to
({^{\mb g_0}}[v^{a_{\mb g}b_{\mb h}}_{c_{\mb g \mb h}}])^* \,
({^{\mb g_0}}[v^{c_{\mb g \mb h}}_{a_{\mb g}b_{\mb h}}])^* \,
{^{\mb g_0}}[\cY^{a_{\mb g}b_{\mb h}}_{c_{\mb g \mb h}}],
\end{align}
and
\begin{align}
\label{ftrans2}
{^{\mb g_0}}[\cF^{a_{\mb g}b_{\mb h}c_{\mb k}}_{d_{\mb g \mb h \mb k}}]_{e_{\mb g \mb h}, f_{\mb h \mb k}} &\to
{^{\mb g_0}}[v^{a_{\mb g}b_{\mb h}}_{e_{\mb g \mb h}}]\,
{^{\mb g_0}}[v^{e_{\mb g \mb h}c_{\mb k}}_{d_{\mb g \mb h \mb k}}]
({^{\mb g_0}}[v^{b_{ \mb h}c_{\mb k}}_{f_{\mb h \mb k}}])^*
({^{\mb g_0}}[v^{a_{ \mb g}f_{\mb h\mb k}}_{d_{\mb g\mb h \mb k}}])^*\,
 {^{\mb g_0}}[\cF^{a_{\mb g}b_{\mb h}c_{\mb k}}_{d_{\mb g \mb h \mb k}}]_{e_{\mb g \mb h},f_{\mb h \mb k}} .
\end{align}
We can use the above ``gauge'' degree of freedom to set
\begin{align}
^{\mb g_0}[\cO_{c_{\mb g \mb h}}^{a_{\mb g}b_{\mb h}}] > 0.\label{guage}
\end{align}

\subsection{Consistency relations}
\label{sec:appmoves:cons}

\subsubsection{Dual $F$ move and a relation between $O$ move and $F$ move}
\label{sec:appmoves:OY}
First, we discuss constraints on the $O$ moves, and argue that a particular choice of $O$ moves satisfy the consistency equations. We notice that a fixed-point wave function can have two ways of reduction:
\begin{align}
&\Psi\left(\tikzpicmidline{
  \draw (0, .3) node [above] {\small $\ag d{ghk}$} -- (0, 0) [string];
  \draw (0, 0) ..controls (-.4, 0) and (-.6, -.4).. (-.6, -.6)
  -- (-.6, -1.2) [string] node [midway, left] {\small $\ag ag$};
  \draw (-.6, -1.2) .. controls (-.6, -1.6) and (-.4, -1.8).. (0, -1.8) [string]
        node [midway, left] {\small $\ag e{gh}$};
  \draw (0, 0) ..controls (.4, 0) and (.6, -.4).. (.6, -.6) [string]
        node [midway, right] {\small $\ag f{hk}$};
  \draw (.6, -.6) -- (.6, -1.2) node [midway, right] {\small $\ag ck$}
        .. controls (.6, -1.6) and (.4, -1.8).. (0, -1.8) [string];
  \draw (.6, -.6) -- (-.6, -1.2)
  node [midway, below] {\small $\ag bh$} [string];
  \draw (0, -1.8) -- (0, -2.1) node [below] {\small $\ag d{ghk}$} [string];
  \node at (-.6, .3) [group] {\small $\mb g_0$};
}\right)
=
{^{\mb g_0}}[\cF^{a_{\mb g }b_{\mb h}c_{\mb k}}_{d_{\mb g \mb h \mb k}}]_{ e_{\mb g \mb h}, f_{\mb h \mb k}}
\Psi\left(\tikzpicmidline{
  \draw (0, .3) node [above] {\small $\ag d{ghk}$} -- (0, 0) [string];
  \draw (0, 0) ..controls (-.4, 0) and (-.6, -.4).. (-.6, -.6)
  -- (-.6, -1.2) [string] node [midway, left] {\small $\ag ag$}
  .. controls (-.6, -1.6) and (-.4, -1.8).. (0, -1.8) [string];
  \draw (0, 0) ..controls (.4, 0) and (.6, -.4).. (.6, -.6) [string]
        node [midway, right] {\small $\ag f{hk}$};
  \draw (.6, -.6) -- (.6, -1.2) node [midway, right] {\small $\ag ck$} [string];
  \draw (.6, -1.2).. controls (.6, -1.6) and (.4, -1.8).. (0, -1.8) [string]
  node [midway, right] {\small $\ag f{hk}$};
  \draw (.6, -.6) ..controls (.3, -.7).. (.3, -.8)
  -- (.3, -1) node [midway, left] {\small $\ag bh$}
  ..controls (.3, -1.1)..  (.6, -1.2) [string];
  \draw (0, -1.8) -- (0, -2.1) node [below] {\small $\ag d{ghk}$} [string];
  \node at (-.6, .3) [group] {\small $\mb g_0$};
}\right)
\nonumber\\
&=
{^{\mb g_0}}[\cF^{a_{\mb g }b_{\mb h}c_{\mb k}}_{d_{\mb g \mb h \mb k}}]_{e_{\mb g \mb h};f_{\mb h \mb k}}
{^{\mb g_0}}[\cO_{f_{\mb h \mb k}}^{b_{\mb h}c_{\mb k}}]
{^{\mb g_0}}[\cO_{d_{\mb g \mb h \mb k}}^{a_{\mb g}f_{\mb h \mb k}}]
\Psi \left(\tikzpicmidline{\LineD}\right)
\end{align}
\begin{align}
&\Psi\left(\tikzpicmidline{
  \draw (0, .3) node [above] {\small $\ag d{ghk}$} -- (0, 0) [string];
  \draw (0, 0) ..controls (-.4, 0) and (-.6, -.4).. (-.6, -.6)
  -- (-.6, -1.2) [string] node [midway, left] {\small $\ag ag$};
  \draw (-.6, -1.2) .. controls (-.6, -1.6) and (-.4, -1.8).. (0, -1.8) [string]
        node [midway, left] {\small $\ag e{gh}$};
  \draw (0, 0) ..controls (.4, 0) and (.6, -.4).. (.6, -.6) [string]
        node [midway, right] {\small $\ag f{hk}$};
  \draw (.6, -.6) -- (.6, -1.2) node [midway, right] {\small $\ag ck$}
        .. controls (.6, -1.6) and (.4, -1.8).. (0, -1.8) [string];
  \draw (.6, -.6) -- (-.6, -1.2)
  node [midway, below] {\small $\ag bh$} [string];
  \draw (0, -1.8) -- (0, -2.1) node [below] {\small $\ag d{ghk}$} [string];
  \node at (-.6, .3) [group] {\small $\mb g_0$};
}\right)
=
{^{\mb g_0}}[\t\cF^{a_{\mb g }b_{\mb h}c_{\mb k}}_{d_{\mb g \mb h \mb k}}]_{ e_{\mb g \mb h}, f_{\mb h \mb k}}
\Psi\left(\tikzpicmidline{
  \draw (0, .3) node [above] {\small $\ag d{ghk}$} -- (0, 0) [string];
  \draw (0, 0) ..controls (.4, 0) and (.6, -.4).. (.6, -.6)
  -- (.6, -1.2) [string] node [midway, right] {\small $\ag ck$}
  .. controls (.6, -1.6) and (.4, -1.8).. (0, -1.8) [string];
  \draw (0, 0) ..controls (-.4, 0) and (-.6, -.4).. (-.6, -.6) [string]
        node [midway, left] {\small $\ag e{gh}$};
  \draw (-.6, -.6) -- (-.6, -1.2)
  node [midway, left] {\small $\ag ag$} [string];
  \draw (-.6, -1.2).. controls (-.6, -1.6) and (-.4, -1.8).. (0, -1.8) [string]
  node [midway, left] {\small $\ag e{gh}$};
  \draw (-.6, -.6) ..controls (-.3, -.7).. (-.3, -.8)
  -- (-.3, -1) node [midway, right] {\small $\ag bh$}
  ..controls (-.3, -1.1)..  (-.6, -1.2) [string];
  \draw (0, -1.8) -- (0, -2.1) node [below] {\small $\ag d{ghk}$} [string];
  \node at (-.6, .3) [group] {\small $\mb g_0$};
}\right)
\nonumber\\
&\simeq
{^{\mb g_0}}[\t\cF^{a_{\mb g }b_{\mb h}c_{\mb k}}_{d_{\mb g \mb h \mb k}}]_{e_{\mb g \mb h}, f_{\mb h \mb k}}
{^{\mb g_0}}[\cO_{e_{\mb g \mb h}}^{a_{\mb g}b_{\mb h},\mu}] {^{\mb g_0}}[\cO_{d_{\mb g\mb h \mb k}}^{e_{\mb g \mb h}c_{\mb k}}]
\Psi\left(\tikzpicmidline{\LineD}\right).
\end{align}

This allows us to obtain the following condition
\begin{align}
{^{\mb g_0}}[\t\cF^{a_{\mb g }b_{\mb h}c_{\mb k}}_{d_{\mb g \mb h \mb k}}]_{e_{\mb g \mb h},f_{\mb h \mb k}}
= {^{\mb g_0}}[\cF^{a_{\mb g }b_{\mb h}c_{\mb k}}_{d_{\mb g \mb h \mb k}}]_{e_{\mb g \mb h}, f_{\mb h \mb k}}
{^{\mb g_0}}[\cO_{f_{\mb h \mb k}}^{b_{\mb h}c_{\mb k}}]\,
{^{\mb g_0}}[\cO_{d_{\mb g \mb h \mb k}}^{a_{\mb g}f_{\mb h \mb k}}]\,
{^{\mb g_0}}[\cO_{e_{\mb g \mb h}}^{a_{\mb g}b_{\mb h},\mu}]^{-1}{^{\mb g_0}}[\cO_{d_{\mb g\mb h \mb k}}^{e_{\mb g \mb h}c_{\mb k}}]^{-1}
\end{align}
We require ${^{\mb g_0}}[\t\cF^{a_{\mb g }b_{\mb h}c_{\mb k}}_{d_{\mb g \mb h \mb k}}]_{e_{\mb g \mb h},f_{\mb h \mb k}}$ to be unitary, which leads to

\begin{align}
&\sum_{e_{\mb g \mb h}}
\frac{
{^{\mb g_0}}[\cF^{a_{\mb g }b_{\mb h}c_{\mb k}}_{d_{\mb g \mb h \mb k}}]_{e_{\mb g \mb h},f'_{\mb h \mb k}}^*\,
{^{\mb g_0}}[\cF^{a_{\mb g }b_{\mb h}c_{\mb k}}_{d_{\mb g \mb h \mb k}}]_{ e_{\mb g \mb h},f_{\mb h \mb k}}
}{
({^{\mb g_0}}[\cO_{e_{\mb g \mb h}}^{a_{\mb g}b_{\mb h},\mu}]{^{\mb g_0}}[\cO_{d_{\mb g\mb h \mb k}}^{e_{\mb g \mb h}c_{\mb k}}])^2}
=
\frac{
\del_{f_{\mb h \mb k}f'_{\mb h \mb k}}
}{
({^{\mb g_0}}[\cO_{f_{\mb h \mb k}}^{b_{\mb h}c_{\mb k}}]
{^{\mb g_0}}[\cO_{d_{\mb g \mb h \mb k}}^{a_{\mb g}f_{\mb h \mb k}}])^2
}
,
\label{FOcondition}
\end{align}

The above condition and the unitary condition in Eq.~\eqref{Onorm} can be satisfied by the following ansatz
\begin{align}
	^{\mb g_0}[\cO_{c_{\mb g \mb h}}^{a_{\mb g}b_{\mb h}}] ={}^{\mb 1}[\cO_{c_{\mb g \mb h}}^{a_{\mb g}b_{\mb h},\al}]\equiv[\cO_{c_{\mb g \mb h}}^{a_{\mb g}b_{\mb h}}]=\frac{1}{\cal{D}}\sqrt{\frac{d_{a_{\mb g}}d_{b_{\mb h}}}{ d_{c_{\mb g \mb h}}}}\del_{c_{\mb g \mb h}}^{a_{\mb g}b_{\mb h}},\label{DefO}
\end{align}
where $\del_{c_{\mb g \mb h}}^{a_{\mb g}b_{\mb h}} =1$ for $N_{c_{\mb g \mb h}}^{a_{\mb g}b_{\mb h}} >0$ and $\del_{c_{\mb g \mb h}}^{a_{\mb g}b_{\mb h}} =0$ for
$N_{c_{\mb g \mb h}}^{a_{\mb g}b_{\mb h}} =0$, and $\mathcal{D}=\sqrt{\sum_{\ag ag \in \mathcal{C}_G}d_{\ag ag}^2}$ is the total quantum dimension of $\cal C_G$.
From \eqn{Onorm}, we find that $d_{a_\mb{g}}$ satisfy
\begin{align}
\label{Nddd}
\sum_{a_{\mb g}b_{\mb h}} N_{c_{\mb g \mb h}}^{a_{\mb g}b_{\mb h}}d_{a_{\mb g}}d_{b_{\mb h}} = \mathcal{D}^2d_{c_{\mb g \mb h}}.
\end{align}
The solution of such an equation gives rise to the so-called quantum dimension $d_{a_{\mb g}}$.

Although the equation Eq. \eqref{Nddd} may look unfamiliar, it can in fact be derived from the more familiar relation $d_a d_b =\sum_c N^{ab}_c d_c$ (we will omit the group labels since they do not play any roles in this derivation):
\begin{equation}
	\mathcal{D}^2d_c=\sum_{a}d_a^2 d_c=\sum_a d_a d_{\bar{a}} d_c=\sum_a d_a\sum_b N^{\bar{a}c}_b d_b= \sum_{a,b}N^{ab}_c d_a d_b.
	\label{}
\end{equation}
Here we have used $d_a= d_{\bar{a}}, N^{ab}_c=N^{\bar{a}c}_b$.

We notice that the form of the $O$ moves in Eq.~\eqref{DefO} differs from Eq.~\eqref{eq:locrel2} by a factor of $\cal D^{-1}$. As mentioned before, this is due to the different Hilbert spaces the two moves map to. The moves in Eq.~\eqref{eq:locrel2} are between two states on the \emph{same} lattice, while the $O$ moves in Eq.~\eqref{PhiO} map between wave functions defined on two different graphs. The right-hand side of Eq.~\eqref{PhiO} has fewer degrees of freedom, since the bubble containing two edges, along with the physical states on them, is removed from the Hilbert space. Hence, the amplitudes of the wave function on the right need to be scaled by a factor of $\cal D$, in order to keep the total amplitude normalized. Such a scaling factor is not needed in Eq.~\eqref{eq:locrel2}. In fact, this result in the main text can be derived from the $O$ moves, by relating both sides of Eq.~\eqref{eq:locrel2} to the right-hand side of Eq.~\eqref{PhiO} using the $O$ moves.

\subsubsection{A relation between $O$ move and $Y$ move}

We find that the following wave function has two ways of reduction:
\begin{align}
\label{iOiOOi}
{^{\mb g_0}}[\cY_{c_{\mb g \mb h}}^{a_{\mb g}b_{\mb h}}]
\Psi\left(\tikzpicmidline{\OmD}\right)
=
\Psi\left(\tikzpicmidline{\OmS}\right)
\nonumber
 =
{^{\mb g_0}}[\cO_{c_{\mb g \mb h}}^{a_{\mb g}b_{\mb h}}]
\Psi\left(\tikzpicmidline{\Line}\right),
\end{align}

\begin{align}
{^{\mb g_0}}[\cY_{c_{\mb g \mb h}}^{a_{\mb g }b_{\mb h}}]
\Psi \left(\tikzpicmidline{\OmD}\right)
=
{^{\mb g_0}}[\cY_{c_{\mb g \mb h}}^{a_{\mb g}b_{\mb h}}]
({^{\mb g_0}}[\cO_{c_{\mb g \mb h}}^{a_{\mb g}b_{\mb h}}])^2
\Psi \left(\tikzpicmidline{\Line}\right)
\end{align}
The two reductions should agree, which leads to the condition
\begin{align}
\label{YO1}
{^{\mb g_0}}[\cY_{c_{\mb g \mb h},{\bt\ga}}^{a_{\mb g}b_{\mb h}}]=
{^{\mb g_0}}[\cO_{c_{\mb g \mb h}}^{a_{\mb g}b_{\mb h}}]^{-1}.
\end{align}

\subsubsection{$H$ move and an additional constraint between $O$ move and $F$ move}
Let us consider a new type of move -- $H$ move:
\begin{align}
\label{Hwave1}
\Psi
\left(\tikzpicmidline{\HmL}\right)
\simeq
\sum_{ f_{\mb g \mb h \mb k}}
{^{\mb g_0 }}[\mathcal{H}^{a_{\mb g \mb h}b_{\mb k}}_{c_{\mb g}d_{\mb h \mb k}}]_{ e_{\mb h}, f_{\mb g \mb h \mb k}}
\Psi
\left(\tikzpicmidline{\HmR}\right)
\end{align}
In the following, we will show how to compute the coefficients ${^{\mb g_0 }}[\mathcal{H}^{a_{\mb g \mb h}b_{\mb k}}_{c_{\mb g}d_{\mb h \mb k}}]_{e_{\mb h}, f_{\mb g \mb h \mb k}}$.

First, by applying the $Y$ move, we have:
\begin{align}
\label{Hwave2}
\Psi
\left(\tikzpicmidline{\HmL}\right)
\simeq
\sum_{ f_{\mb g \mb h \mb k}}
{^{\mb g_0}}[\mathcal{Y}^{c_{\mb g}d_{\mb h \mb k}}_{f_{\mb g \mb h \mb k}}]
\Psi
\left(\tikzpicmidline{\Hc}\right).
\end{align}

Next, by applying an inverse $F$ move, we obtain:
\begin{align}
\label{Hwave3}
\Psi
\left(\tikzpicmidline{\Hc}\right)
=
\sum_{ a^\prime_{\mb g \mb h}}
{{^{\mb g_0}}[\mathcal{F}^{c_{\mb g} e_{\mb h} b_{\mb k}}_{f_{\mb g \mb h \mb k}}]^*_{a'_{\mb g \mb h}, d_{\mb h \mb k}}}
\Psi
\left(\tikzpicmidline{\Hd}\right).
\end{align}

Finally, by applying the $O$ move, we end up with:
\begin{align}
\label{Hwave4}
\Psi
\left(\tikzpicmidline{\Hd}\right)
=
{^{\mb g_0}}[\mathcal{O}^{c_{\mb g}e_{\mb h}}_{a_{\mb g \mb h}}]\delta_{a_{\mb g \mb h}a'_{\mb g \mb h}}
\Psi
\left(\tikzpicmidline{\HmR}\right).
\end{align}

All together, we find:
\begin{align}
{^{\mb g_0 }}[\mathcal{H}^{a_{\mb g \mb h}b_{\mb k}}_{c_{\mb g}d_{\mb h \mb k}}]_{e_{\mb h},f_{\mb g \mb h \mb k}}
={^{\mb g_0}}[\mathcal{Y}^{c_{\mb g}d_{\mb h \mb k}}_{f_{\mb g \mb h \mb k}}]\,
{{^{\mb g_0}}[\mathcal{F}^{c_{\mb g} e_{\mb h} b_{\mb k}}_{f_{\mb g \mb h \mb k}}]^*_{a_{\mb g \mb h},  d_{\mb h \mb k}}} {^{\mb g_0}}[\mathcal{O}^{c_{\mb g}e_{\mb h}}_{a_{\mb g \mb h}}]
\end{align}

With the ansatz Eq. \eqref{DefO}, we can further simplify the above expressions as:
\begin{align}
{^{\mb g_0 }}[\mathcal{H}^{a_{\mb g \mb h}b_{\mb k}}_{c_{\mb g}d_{\mb h \mb k}}]_{e_{\mb h}, f_{\mb g \mb h \mb k}}= \sqrt{\frac{d_{f_{\mb g \mb h \mb k}}d_{e_{\mb h}}}{d_{d_{\mb h \mb k}} d_{a_{\mb g \mb h}}}}{{^{\mb g_0}}[\mathcal{F}^{c_{\mb g} e_{\mb h} b_{\mb k}}_{f_{\mb g \mb h \mb k}}]^*_{ a_{\mb g \mb h};  d_{\mb h \mb k}}}
\end{align}
and
\begin{align}
\sum_{f_{\mb g \mb h \mb k}}d_{f_{\mb g \mb h \mb k}} {^{\mb g_0}}[\mathcal{F}^{c_{\mb g} e'_{\mb h} b_{\mb k}}_{f_{\mb g \mb h \mb k}}]_{ a_{\mb g \mb h},  d_{\mb h \mb k}}{{^{\mb g_0}}[\mathcal{F}^{c_{\mb g} e_{\mb h} b_{\mb k}}_{f_{\mb g \mb h \mb k}}]^*_{a_{\mb g \mb h},  d_{\mb h \mb k}}}   =\frac{d_{a_{\mb g \mb h}}d_{d_{\mb h \mb k}}}{d_{e_{\mb h}}}
\delta_{e_{\mb h}e^\prime_{\mb h}}.
\end{align}

\subsection{Summary}

To summarize, all the conditions
form a set of
non-linear equations whose variables are $N^{a_{\mb g}b_{\mb h}}_{c_{\mb g \mb h}}$,
${^{\mb g_0}}[\cF^{a_{\mb g }b_{\mb h}c_{\mb k}}_{d_{\mb g\mb h \mb k}}]_{e_{\mb g \mb h}, f_{\mb h \mb k}}$,
$d_{a_{\mb g}}$,
let us collect those conditions and list them below
\begin{align}
\label{Feq}
&\bullet\
\sum_{e_{\mb g \mb h}} N^{a_{\mb g}b_{\mb h}}_{e_{\mb g\mb h}} N^{e_{\mb g \mb h}c_{\mb k}}_{d_{\mb g \mb h \mb k}}=\sum_{f_{\mb h \mb k}} N^{b_{\mb h }c_{\mb k}}_{f_{\mb h \mb k}}N^{a_{\mb g}f_{\mb h \mb k}}_{d_{\mb g \mb h \mb k}}.
\nonumber\\
&
\bullet\ \sum_{f_{\mb h \mb k}}
{^{\mb g_0}}[\cF^{a_{\mb g }b_{\mb h}c_{\mb k}}_{d_{\mb g\mb h \mb k}}]_{e_{\mb g \mb h}^\prime, f_{\mb h \mb k}}
{^{\mb g_0}}[\cF^{a_{\mb g }b_{\mb h}c_{\mb k}}_{d_{\mb g\mb h \mb k}}]^*_{ e_{\mb g \mb h}, f_{\mb h \mb k}}
=\del_{e_{\mb g \mb h},e_{\mb g\mb h}^\prime},
\nonumber\\
&\bullet\
\sum_{n_{\mb h \mb k}}
{^{\mb g_0}}[\cF^{a_{\mb g }b_{\mb h}c_{\mb k}}_{m_{\mb g \mb h \mb k }}]_{f_{\mb g \mb h },n_{\mb h \mb k }}
{^{\mb g_0}}[\cF^{a_{\mb g }n_{\mb h \mb k}d_{\mb l}}_{e_{\mb g \mb h \mb k \mb l}}]_{m_{\mb g \mb h \mb k }, p_{ \mb h \mb k \mb l}}
{^{\mb g_0 \mb g}}[\cF^{b_{\mb h }c_{\mb k}d_{\mb l}}_{p_{ \mb h \mb k \mb l}}]_{ n_{\mb h \mb k },q_{\mb k \mb l}}
 =
{^{\mb g_0}}[\cF^{f_{\mb g \mb h }c_{ \mb k}d_{\mb l}}_{e_{\mb g \mb h \mb k \mb l}}]_{ m_{\mb g \mb h \mb k }, q_{\mb k \mb l}}
{^{\mb g_0}}[\cF^{a_{\mb g }b_{ \mb h}q_{\mb k \mb l}}_{e_{\mb g \mb h \mb k \mb l}}]_{f_{\mb g \mb h },p_{ \mb h \mb k \mb l}}.
\nonumber\\
&\bullet\    \sum_{a_{\mb g}b_{\mb h}} d_{a_{\mb g}}d_{b_{\mb h}} N_{c_{\mb g \mb h}}^{a_{\mb g}b_{\mb h}}= \mathcal{D}^2d_{c_{\mb g \mb h}},
\nonumber\\
&\bullet\
\sum_{f_{\mb g \mb h \mb k}}d_{f_{\mb g \mb h \mb k}} {^{\mb g_0}}[\mathcal{F}^{c_{\mb g} e'_{\mb h} b_{\mb k}}_{f_{\mb g \mb h \mb k}}]_{ a_{\mb g \mb h},  d_{\mb h \mb k}}{{^{\mb g_0}}[\mathcal{F}^{c_{\mb g} e_{\mb h} b_{\mb k}}_{f_{\mb g \mb h \mb k}}]^*_{a_{\mb g \mb h},  d_{\mb h \mb k}}}   =\frac{d_{a_{\mb g \mb h}}d_{d_{\mb h \mb k}}}{d_{e_{\mb h}}}
\delta_{e_{\mb h}e^\prime_{\mb h}}.
\end{align}
\twocolumngrid

\section{Classification of Group Extensions of Fusion Categories}
\label{sec:eno}
We summarize the categorical classification of (unitary) group extensions of fusion categories. The materials below are distilled from \Ref{ENO2009}.

Given a fusion category $\cal{C}$, \Ref{ENO2009} defined the Brauer-Picard 3-group $\doubleunderline{\text{BrPic}(\cal{C})}$ as the following: there is a single object, the fusion category $\cal{C}$, $1$-morphisms are the $\cal{C}$-bimodule categories, $2$-morphisms are equivalences of such bimodule categories and $3$-morphisms are the isomorphisms of such equivalences.  It can be truncated to a 2-group $\underline{\text{BrPic}(\cal{C})}$, by forgetting the $3$-morphisms, and further down to a group $\text{BrPic}(\cal{C})$, i.e., the group of equivalence classes of such bimodule categories. A central result of \Ref{ENO2009} is the following theorem:

\noindent\textbf{Theorem 1.1}. $\underline{\text{BrPic}(\cal{C})}$ is equivalent to $\underline{\text{EqBr}(\mathcal{Z}(\cal{C}) )}$.

$\underline{\text{EqBr}(\mathcal{Z}(\cal{C}) )}$ is the braided tensor autoequivalence of the Drinfeld center of $\cal{C}$. Consequently, $\text{BrPic}(\cal{C})$ is isomorphic to $\text{EqBr}(\cal{C})$. The latter is called $\text{Aut}(\mathcal{Z}(\cal{C}))$ using the notation of \Ref{SET1}. An explicit description of braided autoequivalences for a modular tensor category in terms of concrete algebraic data can also be found in \Ref{SET1}.

$\doubleunderline{\text{BrPic}(\cal{C})}$ plays an important rule in the classification of extensions of $\cal{C}$, due to the following theorem:

\noindent\textbf{Theorem 7.7}. The equivalence classes of $G$-extensions of $\mathcal{C}$ are in bijection with homotopy classes of maps from the classifying space $BG$ to the classifying space of $\doubleunderline{\text{BrPic}(\cal{C})}$.

Define $\mathcal{A}$ as the group of Abelian anyons in $\cal{Z}(\cal{C})$.  \Ref{ENO2009} gave an explicit description of the extension:

\noindent\textbf{Theorem 1.3}.
Equivalence classes of extensions of $\cal{C}$ by $G$ are parametrized by triplets $(\rho, \coho{w}, \alpha)$. Here $\rho$ is a group homomorphism $\rho: G\rightarrow\text{BrPic}(\cal{C})$, with vanishing obstruction class in $\cal{H}^3[G, \cal{A}]$. Then $\coho{w}$ belongs to a certain torsor over $\cal{H}^2_\rho[G, \cal{A}]$. Notice that the action on $\cal{A}$ is naturally induced from $\rho$ by the group homomorphism between $\text{BrPic}(\cal{C})$ and $\text{Aut}\big(\cal{Z}(\cal{C})\big)$. $\coho{w}$ must satisfy an obstruction-vanishing condition, where the obstruction class belongs to $\cal{H}^4[G, \mathrm{U}(1)]$. Lastly, $\alpha$ belongs to a torsor over $\mathcal{H}^3[G, \mathrm{U}(1)]$.

According to \Ref{ENO2009}, the same triplets also classify equivalence classes of $G$-crossed braided extensions of the modular tensor category $\mathcal{Z}(\cal{C})$. As elaborated in \Ref{SET1}, $G$-symmetry-enriched phases of $\cal{Z}(\cal{C})$ are exactly described by such $G$-crossed braided extensions.
We therefore conclude that the classification of $G$-extension of $\mathcal{C}$ is basically the classification of $G$-enriched phases of $\cal{Z}(\cal{C})$.

\section{Parametrization of $3$-Cocycles in Group Extensions}
\label{sec:param3}
The $3$-cocycle condition of $\cal{G}$ reads
\begin{equation}
	\frac{\omega(a_\mb{g}\times b_\mb{h}, c_\mb{k}, d_\mb{l})\omega(a_\mb{g},b_\mb{h},c_\mb{k}\times d_\mb{l})}{\omega(a_\mb{g}, b_\mb{h}, c_\mb{k})\omega(a_\mb{g},b_\mb{h}\times c_\mb{k}, d_\mb{l})\omega^{s(\mb{g})}(b_\mb{h}, c_\mb{k}, d_\mb{l})}=1.
	\label{}
\end{equation}
Plug in the explicit parametrization:
\begin{multline}
	\frac{\chi_d(\mb{gh,k})\chi_{cd\nu(\mb{k,l})}(\mb{g,h})}{\chi_c(\mb{g,h})\chi_{d}(\mb{g,hk})\chi_d^{s(\mb{g})}(\mb{h,k})}=\\
	\frac{\alpha^{s(\mb{g})}(\mb{h}, \mb{k}, \mb{l})\alpha(\mb{g}, \mb{h}, \mb{k})\alpha(\mb{g},\mb{hk}, \mb{l})}{\alpha(\mb{gh}, \mb{k}, \mb{l})\alpha(\mb{g},\mb{h},\mb{kl})}.
	\label{eqn:xx1}
\end{multline}

First set $\mb{l}=1$. The right-hand side vanishes because of our normalization $\alpha(1, *, *)=\alpha(*,1,*)=\alpha(*,*,1)=1$. The left-hand side becomes
\begin{equation}
	{\chi_d(\mb{gh,k})\chi_{cd}(\mb{g,h})}={\chi_c(\mb{g,h})\chi_{d}(\mb{g,hk})\chi_d^{s(\mb{g})}(\mb{h,k})}.
	\label{eqn:xx2}
\end{equation}
If we further set $c=1$ and use the normalization $\chi_1=1$, we find $\chi_d$ satisfies the $2$-cocycle condition of $G$:
\begin{equation}
	{\chi_d(\mb{gh,k})\chi_{d}(\mb{g,h})}={\chi_{d}(\mb{g,hk})\chi_d^{s(\mb{g})}(\mb{h,k})}.
	\label{}
\end{equation}
Eq. \eqref{eqn:xx2} then implies that $\chi_{cd}(\mb{g,h})=\chi_c(\mb{g,h})\chi_d(\mb{g,h})$.

Finally, Eq. \eqref{eqn:xx1} is reduced to
\begin{equation}
	\chi_{\nu(\mb{k,l})}(\mb{g,h})=\frac{\alpha^{s(\mb{g})}(\mb{h}, \mb{k}, \mb{l})\alpha(\mb{g}, \mb{h}, \mb{k})\alpha(\mb{g},\mb{hk}, \mb{l})}{\alpha(\mb{gh}, \mb{k}, \mb{l})\alpha(\mb{g},\mb{h},\mb{kl})}
	\label{}
\end{equation}

\section{Fractionalization in Symmetry-Enriched Gauge Theories}
\label{sec:frac_gauge_theory}
In the main text we construct a large family of $G$-symmetry-enriched Abelian gauge theories $D(N)$, by first specifying a central extension $\mathcal{G}$ of $N$ by $G$, and a $3$-cocycle $\omega$ of the form Eq. \eqref{eq:w=ca}. In this section, we determine the symmetry fractionalization class for unitary $G$, by gauging the symmetry $G$.

By construction, gauging the symmetry group $G$ yields the Dijkgraaf-Witten gauge theory $D^\omega(\mathcal{G})$ (also known as the twisted quantum double of $\mathcal{G}$). We will first review the anyon content of $D^\omega(\mathcal{G})$~\cite{YTHuTQD, SET1}.
The quasiparticles in the quantum double are labeled by $([g], \pi_g)$ where $[g]$ denotes a conjugacy class of $\mathcal{G}$ with a representative element $g$, and $\pi_g$ is an irreducible representation of the centralizer group $C_g$. However, $\pi_g$ is not necessarily a linear representation; they are generally projective, with the factor sets given by:
\begin{equation}
	\eta_g(h,k)=\frac{\omega(h,k,g)\omega(g,h,k)}{\omega(h,g,k)}, \text{for }h,k \in C_g.
	\label{}
\end{equation}

Now we specialize to the case where $\mathcal{G}$ is a central extension of $N$ by $G$, and parametrize the group elments as in Sec. \ref{sec:group-ext}. The form of $3$-cocycles given in Eq. \eqref{eq:w=ca} reads:
\begin{equation}
	\omega(a_\mb{g},b_\mb{h},\ag ck)=\chi_c(\mb{g,h})\alpha(\mb{g,h,k}).
	\label{}
\end{equation}

For our purpose, a particularly important conjugacy class is $[\ag a1]$ for $a\in N$, and the centralizer group is just $\mathcal{G}$ since $a$ is a central element.  The factor set is given by
 \begin{equation}
	 \eta_{\ag a1}(\ag bh, \ag ck)=\chi_a(\mb{h,k}).
	 \label{}
 \end{equation}
We now explicitly construct all such projective representations $\pi$. By definition, $\pi(\ag bh)\pi(\ag ck)=\chi_a(\mb{h,k})\pi( [bc\nu(\mb{h,k})]_\mb{hk})$. Setting $b=1,\mb{k}=1$, we get
 \begin{equation}
	 \pi(\ag 1h)\pi(\ag c1)=\pi( c_\mb{h})
	 \label{eqn:pich}
 \end{equation}
 Notice that $\pi(\ag c1)$ gives a linear representation of $N$, so $\pi(\ag c1)$ must be a character of $N$. We will denote $\pi(\ag c1)=\lambda(c)$. From Eq. \eqref{eqn:pich} we obtain
\begin{equation}
	\begin{split}
	\pi(\ag bh)\pi(\ag ck)&=\pi(\ag 1h)\pi(\ag 1k)\lambda(b)\lambda(c)\\
	&=\chi_a(\mb{h,k})\pi( 1_\mb{hk}) \lambda({bc\nu(\mb{h,k})}),
	\end{split}
	\label{}
\end{equation}
which gives
\begin{equation}
	\pi(\ag 1h)\pi(\ag 1k)=\chi_a(\mb{h,k})\lambda({\nu(\mb{h,k})})\pi( 1_\mb{hk}).
	\label{eqn:gauging-proj-rep-G}
\end{equation}
In other words, $\pi(\ag 1g)$ is a projective representation of $G$.

Therefore, we can construct all irreducible projective representations of $\mathcal{G}$ with factor set $\eta_{\ag a1}$ in the following way: choose a character $\lambda$ of $N$, and find all the irreducible projective representations of $G$, denoted by $\tilde{\pi}$ (which will be $\pi(\ag 1h)$), with the factor set $\tilde{\eta}(\mb{h,k})=\chi_a(\mb{h,k})\lambda(\nu(\mb{h,k}))$. Then $\pi(\ag bh)=\tilde{\pi}(\mb{h})\lambda(b)$ is an irreducible representations of $\mathcal{G}$. We will label such a quasiparticle by a tuple $([a], \lambda, \tilde{\pi})$. The set of these quasiparticles can be regarded as ``charges'' in the gauged theory since they do not carry any $G$ fluxes.

We can also understand how anyons in the SET are promoted into quasiparticles in the gauged theory. Mathematically this is done by a procedure called equivariantization (for a more detailed account, see \Ref{SET1}). If an anyon type $a$ is not permuted by the symmetry group $G$, after equivariantization, we attach to $a$ a projective representation of $G$, with the factor set the same as the projective phases of $G$ actions on $a$. Given the fractionalization $[\coho{w}]\in \mathcal{H}^2[G, \mathcal{A}]$,  the projective phases are given by $\eta_a(\mb{g,h})=M_{a,\cohosub{w}(\mb{g,h})}$. For the Abelian gauge theory $D(N)$, we have $\mathcal{A}=N\times \hat{N}$. We can represent the fractionalization class as $\coho{w}=(\coho{w}_e, \coho{w}_m)$ where $\coho{w}_e\in \cal{H}^2[G, N], \coho{w}_m\in \cal{H}^2[G, \hat{N}]$. For a quasiparticle $(a,\lambda)$ in $D(N)$, we can then easily compute the projective phases:
\begin{equation}
	\eta_{(a,\lambda)}(\mb{g,h}) = M_{(a,\lambda), \cohosub{w}(\mb{g,h})}=\lambda(\coho{w}_e(\mb{g,h}))[\coho{w}_m(\mb{g,h})](a).
	\label{}
\end{equation}
Comparing with Eq. \eqref{eqn:gauging-proj-rep-G}, we should identify:
\begin{equation}
	\nu(\mb{g,h})\equiv \coho{w}_e(\mb{g,h}), \chi_a(\mb{g,h})\equiv [\coho{w}_m(\mb{g,h})](a).
	\label{}
\end{equation}

\bibliography{quantum_double}
\end{document}